\documentclass[final,5p,times,twocolumn]{elsarticle}

\usepackage{lineno}

\usepackage{graphicx}% Include figure files
\usepackage{dcolumn}% Align table columns on decimal point
\usepackage{microtype}
\usepackage[hyperindex=true,colorlinks=true]{hyperref}
\hypersetup{citecolor=[rgb]{0.125,0.679,0.196}} %To get readable green cite link
\usepackage{url}
\usepackage{amsmath}
\usepackage{amssymb}
\usepackage{braket}
\usepackage{mathrsfs} %for mathscr
\usepackage{bbm} %for double-bar numbers
\usepackage{stmaryrd} % for barred bracket (set of consecutive integers)
\usepackage{slashed}
\usepackage{bm} %for nice bold math symbol
\usepackage{color} %for textcolor
\usepackage{physics}
\usepackage{empheq}
\usepackage{simplewick} %for wick contractions

\DeclareMathOperator{\GL}{GL}
\newcommand*{\transpose}{^{\mkern-1.5mu\mathsf{T}}} % for a nice transpose symbol
\def\mean#1{\left< #1 \right>}
\makeatletter
\newsavebox\myboxA
\newsavebox\myboxB
\newlength\mylenA
\newcommand*\xoverline[2][0.75]{%
    \sbox{\myboxA}{$\m@th#2$}%
    \setbox\myboxB\null% Phantom box
    \ht\myboxB=\ht\myboxA%
    \dp\myboxB=\dp\myboxA%
    \wd\myboxB=#1\wd\myboxA% Scale phantom
    \sbox\myboxB{$\m@th\overline{\copy\myboxB}$}%  Overlined phantom
    \setlength\mylenA{\the\wd\myboxA}%   calc width diff
    \addtolength\mylenA{-\the\wd\myboxB}%
    \ifdim\wd\myboxB<\wd\myboxA%
       \rlap{\hskip 0.5\mylenA\usebox\myboxB}{\usebox\myboxA}%
    \else
        \hskip -0.5\mylenA\rlap{\usebox\myboxA}{\hskip 0.5\mylenA\usebox\myboxB}%
    \fi}
\makeatother
\usepackage{xparse}
\NewDocumentCommand{\evalat}{sO{\big}mm}{%
  \IfBooleanTF{#1}
   {\mleft. #3 \mright|_{#4}}
   {#3#2|_{#4}}%
}

\usepackage{feynmp-auto}
\newcommand{\Farrow}[6]{%
        \fmfcmd{style_def farrow#1
            expr p = drawarrow subpath (9/24, 15/24) of p shifted #6 #2 withpen pencircle scaled 0.4;
            label.#3(btex #4 etex, point 0.5 of p shifted #6 #2);
            enddef;
        }
    \fmf{farrow#1,tension=0}{#5}
}

\def\orcid#1{\href{https://orcid.org/#1}{\!\includegraphics[keepaspectratio,width=0.7em]{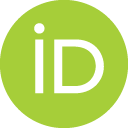}}}

\journal{Annals of Physics}

\begin{document}

\begin{frontmatter}

\title{Nambu-Covariant Many-Body Theory I: Perturbative Approximations}

%% use optional labels to link authors explicitly to addresses:
%% \author[label1,label2]{}
%% \affiliation[label1]{organization={},
%%             addressline={},
%%             city={},
%%             postcode={},
%%             state={},
%%             country={}}
%%
%% \affiliation[label2]{organization={},
%%             addressline={},
%%             city={},
%%             postcode={},
%%             state={},
%%             country={}}

\author[a1,a2]{M.~Drissi \orcid{0000-0001-9472-6280}\corref{cor}}
\ead{mdrissi@triumf.ca}

\author[a3,a2]{A.~Rios \orcid{0000-0002-8759-3202} }

\author[a4,a5,a2]{C.~Barbieri \orcid{0000-0001-8658-6927} }

\affiliation[a1]{organization={TRIUMF},
        addressline={4004 Wesbrook Mall},
        city={Vancouver},
        postcode={V6T 2A3},
        state={British Columbia},
        country={Canada}}

\affiliation[a2]{organization={Department of Physics, University of Surrey},
        city={Guildford},
        postcode={GU2 7XH},
        country={United Kingdom}}

\affiliation[a3]{
        organization={Departament de Fsíca Quàntica i Astrofísica,
            Institut de Ciències del Cosmos (ICCUB), Universitat de Barcelona}, 
        addressline=\hbox{Martí Franquès 1},
        postcode={E08028},
        city={Barcelona},
        country={Spain}}

\affiliation[a4]{organization={Dipartimento di Fisica "Aldo Pontremoli", Università degli studi di Milano},
            addressline={Via Celoria 16}, 
            city={Milano},
            postcode={20133}, 
            %state={},
            country={Italy}}

\affiliation[a5]{organization={INFN, Sezione di Milano},
            addressline={Via Celoria 16}, 
            city={Milano},
            postcode={20133}, 
            %state={},
            country={Italy}}

\cortext[cor]{Corresponding author}

\begin{abstract}
Symmetry-breaking considerations play an important role
in allowing reliable and accurate predictions of complex systems in quantum many-body simulations.
The general theory of perturbations in symmetry-breaking phases
is nonetheless intrinsically more involved than in the unbroken phase 
due to non-vanishing anomalous Green's functions
or anomalous quasiparticle interactions.
In the present paper, we develop a formulation of
many-body theory at non-zero temperature
which is explicitly covariant with respect to a group containing Bogoliubov
transformations. Based on the concept of Nambu tensors,
we derive a factorisation of standard Feynman diagrams
that is valid for a general Hamiltonian.
The resulting factorised amplitudes are indexed over the set of
un-oriented Feynman diagrams with fully antisymmetric vertices. 
We argue that, within this framework,
the design of symmetry-breaking many-body approximations is simplified.
\end{abstract}

\begin{keyword}
%% keywords here, in the form: keyword \sep keyword, up to a maximum of 6 keywords
Quantum many-body theory \sep Symmetry-breaking \sep Perturbation theory \sep Superfluidity
%% PACS codes here, in the form: \PACS code \sep code

%% MSC codes here, in the form: \MSC code \sep code
%% or \MSC[2008] code \sep code (2000 is the default)

\end{keyword}

\end{frontmatter}

\tableofcontents

%\linenumbers

%% main text

\section{Introduction}

Modern descriptions of complex quantum physical systems are largely based on
their decomposition into a set of idealised sub-parts or particles.
The correlations between the individual motion of these particles
are usually described in terms of so-called many-body interactions,
obtained by subtracting the Hamiltonian of free particles
from the Hamiltonian describing the targeted complex quantum physical system.
The quantum theory of many-body systems provides tools to perform this decomposition
conveniently for a wide variety of particles, 
together with means
of computing their correlations and associated many-body observables,
whether approximately or exactly.
How to conveniently devise many-body approximations
when there is a mismatch between the symmetries of
the free system and those of the correlated
many-body system? This is a key question in quantum many-body theory
which we aim to addressing in the present paper.

Since the work of Bardeen, Cooper
and Schrieffer (BCS)~\cite{Bardeen1957a,Bardeen1957b}, the theory of perturbations
over a reference state breaking particle-number symmetry has been formulated in
several ways~\cite{Bogoliubov1958a,Gorkov1958,Soma2011,Signoracci2015}. 
Shortly after the developments of the BCS theory,
Bogoliubov~\cite{Bogoliubov1958a} and Valatin~\cite{Valatin1958}
showed it to be equivalent to a reformulation of many-body theory
in terms of quasiparticle creation and annihilation operators.
This reformulation however required the introduction of
anomalous vertices in the associated perturbation theory~\cite{Bogoliubov1958a,Tolmachev1958,Bogoliubov1958b}. 
Soon after, Gorkov emphasised that one could keep working with
traditional single-particle creation and annihilation operators 
to the price of introducing anomalous Green's functions~\cite{Gorkov1958}.
This amounts to exchanging the complexity associated to anomalous vertices with 
the one associated to anomalous lines in the diagrammatics of perturbation theory.
Later, Anderson pinpointed a two dimensional
space structure as a useful organising principle to mitigate the
growing complexity due to the occurrence of anomalous contributions~\cite{Anderson1958}.
Nambu took this idea further and reformulated the many-body problem
in terms of the so-called Nambu fields~\cite{Nambu1960}.
These extended fields respect the usual canonical anticommutation rules,
and standard perturbation theory can be applied straightforwardly.
The result is an oriented diagrammatic approach, free of anomalous lines, where
propagators are $2\times2$ matrices.
We refer the reader to Ref.~\cite{Mattuck1976} for more details on the oriented
diagrammatics in the formalism of Nambu.
Among subsequent developments in diagrammatic approaches dealing explicitly
with symmetry-breaking, we highlight here the pioneering work 
by De Dominicis and Martin on superfluidity~\cite{DeDominicis1964a,DeDominicis1964b}.
In their work, perturbative contributions were expressed
in terms of diagrams with un-oriented lines
and totally antisymmetric vertices
so that any further development is simplified.
Later, a similar formalism was reintroduced by Kleinert,
in his work on collective excitations~\cite{Kleinert1981,Kleinert1982},
as well as by Haussmann, in his work on the BCS-BEC
crossover~\cite{Haussmann1993,Haussmann1999}.
In this context, the present work can be seen as rooting the specific formalisms developed
in~\cite{DeDominicis1964a,DeDominicis1964b,Kleinert1981,Kleinert1982,Haussmann1993,Haussmann1999}
into the formalism of Nambu tensors, thus extending them to the case of a general Hamiltonian
expressed in a general field basis.

Symmetry breaking considerations
have been understood to be essential
in the description of physical phenomena for more than a century~\cite{Curie1894,Anderson1972}.
In practice, those considerations are of importance to efficiently produce
accurate and reliable predictions of a wide range of physical phenomena.
For example, they are essential to describe BCS electron superconductors, superfluid helium,
the BCS-BEC crossover of ultracold atoms, or for exciton physics.
Nuclear systems are iconic in this respect.
Theoretical descriptions of nuclei regularly break the symmetries of
translation, rotation and particle number~\cite{RingSchuck1980}.
Such considerations are all the more important for infinite homogeneous matter
where experimental data are lacking.
In this case, the reliability of \emph{ab initio} theoretical predictions
is critical for our understanding of neutron-star structure~\cite{Pines1985,Takatsuka1993}.

The main physics motivation behind the formal developments presented here is 
the description of nuclear systems and particularly
of superfluid neutron matter. 
For the last $50$ years, Hartree-Fock-Bogoliubov (HFB) calculations, or their
extensions, have predicted superfluid gaps with a fluctuating magnitude
depending on the inter-nucleonic interaction and on the additional
many-body corrections taken into
account~\cite{Tamagaki1970,Hoffberg1970,Takatsuka1993,Baldo1998,Dean2003,Schwenk2004,Khodel2004,Ding2016,Drischler2017,Rios2017a}.
For example, the question of whether a ${}^{3}PF_2$ gap contributes
significantly to the structure and dynamics of homogeneous neutron matter,
and if so at what densities and temperatures,
is still an open problem~\cite{Dean2003,Yakovlev2004b}.
Simultaneously,
some models of neutron star
cooling have shown a better agreement with observations 
when assuming the ${}^{3}PF_2$ gap to be small enough~\cite{Page1992,Yakovlev2004a,Yakovlev2004b,Blaschke2013}.
The potential tension between neutron star cooling
observations and \emph{ab initio} nuclear estimates of the superfluid pairing gaps
signals the need for clear and quantifiable predictions of the phase diagram of
homogeneous neutron matter. 

This goal requires developing an approach that can be used
to estimate many-body corrections, similarly to what is done routinely in the normal phase~\cite{Hebeler2015,Rios2020,Drischler2019,Drischler2021}. 
Ideally, one would like to include non-perturbative
diagrammatic summations in the description,
to quantify and clarify the importance of the so-called screening corrections,
one of the longstanding issues in the field~\cite{Schwenk2004,Shen2005}. 
In addition, two- and three-body interactions should be considered.
Finally, the approach should be formulated at non-zero temperature in order to
explore extensively the phase diagram.

One possibility along these lines would be to start from 
HFB computations at non-zero temperature and use previously
existing diagrammatic methods to treat the spontaneous breaking of
particle-number symmetry beyond mean-field. Similar approaches have 
proved very effective for finite nuclei~\cite{Soma2020,Soma2020b}.
However, devising high-accuracy approximations with two- and three-body
interactions becomes increasingly difficult already at zero temperature
in a symmetry-conserving framework~\cite{Barbieri2017lnp,Carbone2013b,Raimondi2018}.
Perturbative diagrammatic approaches in the symmetry-breaking
case are even more cumbersome (and hence error-prone) than their
symmetry-conserving counterparts.
Such complications become increasingly relevant if one aims at describing
not one single observable (say, the energy), but rather the equilibrium
dynamics of the system through many-body Green's functions. 
Some of this complexity has been addressed by means of 
automated diagrammatic generation tools~\cite{Stevenson2003,Arthuis2019,Drischler2019,Tichai2020}.
While these are useful for avoiding redundant derivations,
automated frameworks fall short when one is interested
in summing an infinite set of Feynman diagrams and further developments
are required for each type of diagrammatic summation.
For example, many-body approximations based on
Gorkov self-consistent Green's functions~\cite{Soma2011,Soma2014,Barbieri2022}, 
Bogoliubov coupled cluster~\cite{Signoracci2015,Duguet2016}
or random-phase approximations~\cite{Anderson1958},
sum specific sets of infinite diagrams in a particle-number
symmetry-broken phase.

To tackle this recurrent problem, we introduce a reformulation of the
quantum many-body problem in the present paper.
The present work can be seen as the natural continuation
of the pioneering work of De Dominicis and Martin~\cite{DeDominicis1964a,DeDominicis1964b},
Kleinert~\cite{Kleinert1981,Kleinert1982} and Haussmann~\cite{Haussmann1993,Haussmann1999}. 
These developments provided a solid starting point but also 
introduced restrictive hypotheses, either on the choice
of the working field basis or the type of interaction. %, to simplify some formal steps.
The success of the previous formalisms are shown to be underpinned by
the algebraic structure of Nambu tensors which we introduce in the present paper
as an extension of the standard single-particle tensor
algebra~\cite{Sinanoglu1984,Head-Gordon1998}.
This allows us to lift the previous restrictive hypotheses,
thus extending previous developments to the case of a general Hamiltonian and field basis.
The resulting Nambu-covariant formalism has a natural diagrammatic representation in terms of un-oriented Feynman diagrams\footnote{Throughout this work, we use ``un-oriented" to specify
Feynman diagrams with plain lines, i.e.\ without
any line orientation.}.
A key difference with respect to more standard approaches
is the introduction of fully antisymmetric interaction vertices. 
We show that each un-oriented Feynman diagram encloses
a sum of standard (e.g. Gorkov) Feynman diagrams with anomalous propagators 
or vertices. Moreover, the approach provides a perturbative expansion of 
many-body Green's functions in terms of Nambu tensors that are, in particular,
contravariant with respect to Bogoliubov transformations.
For future references, we refer to this formulation of perturbation theory
as Nambu-Covariant Perturbation Theory (NCPT).
Whereas here we focus mainly on NCPT, we 
analyse the properties of self-consistent Green's functions and provide 
formal developments for summation schemes in a follow-up work,
hereafter referred to as Part~II~\cite{part2}.

The advantages of the Nambu-covariant reformulation,
compared to more standard approaches,
are threefold. First, we gain clarity on formal aspects.
The exact properties of many-body
Green's functions and other amplitudes associated to un-oriented Feynman
diagrams can be expressed in a more compact, less cumbersome manner. 
At a given order of perturbation theory, the number
of un-oriented Feynman diagrams is also substantially reduced
compared to other approaches, thus
mitigating as much as possible
the factorial growth in diagram number. 
In turn, the resulting formalism is less error-prone.
Moreover,
any many-body approximation obtained as a truncation on the set of un-oriented
Feynman diagrams is guaranteed to be independent from the field basis
used in practical implementations. 
In other words, there is no need for a re-derivation of many-body equations 
when working with two different field bases related
by a Bogoliubov transformation. 
We expect that these formal results should be useful in numerical
implementations and  associated benchmarking tests.

Second, on numerical aspects, we expect the numerical code resulting
from a direct 
implementation of formulae expressed in the Nambu-covariant formalism
to be more efficient computationally.
The formalism provides compact and factorised
expressions, thus facilitating the implementation
and the maintenance of source codes.
As we discuss later, from a computational efficiency perspective,
the equations derived in this formalism expose more clearly a source of
parallelisation.
The formalism reduces the number of Feynman diagrams whose evaluation
transparently translates into a Nambu tensor network.
Compared to the evaluation of a multitude of single-particle tensor networks,
we expect a greater gain when using massively parallel hardware with
algorithms specifically designed for this kind of
architecture~\cite{Dongarra2014,Gates2019}.

Last, we stress that the Nambu-covariant formalism may also be useful
in the development of automated pipelines. The formalism does not only
reduce the number of diagrams, which would no doubt speed up automated
diagrammatic generation tools, but also removes any consideration in terms
of orientations. We expect this to bring a substantial advantage in terms of
memory processing and practical implementation.

The formalism developed in this paper is entirely equivalent
to any of the previous formulations. In principle, 
a perfectly efficient numerical implementation might not benefit from it.
Factorised and simplified formal many-body equations can also be derived in previous formalisms.
Our claim is, however, that the Nambu-covariant formalism presented here
provides a key to uncover sources of formal simplifications
and generalisations. 
We also expect it will lead to new numerical optimisations in the implementation
of many-body approximations.
Hopefully, this formalism can benefit other many-body practitioners.
To help the reader navigating Part~I and Part~II of this work, we now
proceed to describe its global organisation.

In Part~I, we present the key and foundational aspects of
the Nambu-covariant formalism and its application
to perturbation theory. First, we introduce the essential
ideas of Nambu tensors and their relation
to Bogoliubov transformations in Sec.~\ref{sec:tensor}.
Second, the resulting NCPT, manifestly covariant with respect to
Bogoliubov transformations, is discussed in Sec.~\ref{sec:covPT}.
We introduce many-body Green's functions as Nambu tensors and 
explore their perturbative expansion in terms of un-oriented Feynman diagrams.
We provide explicit Feynman rules for the time and energy representations,
and give an additional set of diagrammatic rules to perform Matsubara sums.
Illustrative examples where the Feynman rules are applied on diagrams
up to third order in the perturbative expansion of the one-body Green's function
are provided in Sec.~\ref{subsec:Examples}. Additional examples
can also be found in Part~II where self-consistently dressed propagators
and vertices are considered.
Third, we explicitly show the connection of this approach to previously
existing formalisms, namely the Gorkov~\cite{Soma2011}
and Bogoliubov~\cite{Signoracci2015} ones, in Sec.~\ref{subsec:ConnectStandardFormalism}.
Finally, we summarise the key points of Part~I in Sec.~\ref{sec:conclusions} and
provide an outlook of future works based on this formalism.

In Part~II, we discuss the application of the Nambu-covariant formalism
to the theory of self-consistent Green's function (SCGF).
To this end, we first derive exact properties of the Nambu-covariant propagator
in Sec.~II.2. We then introduce its associated self-energy
in Sec.~II.3.
Nambu-covariant many-body approximations which are self-consistent in the propagator
are then detailed. As an example of such approximations,
we derive explicitly the HFB equation for a general many-body interaction.
Finally, we introduce Nambu-covariant many-body
approximations which are self-consistent in the two-body interaction in Sec.~II.4.
As an example of such many-body approximation we derive explicitly the self-consistent ladder approximation.
Last, we summarise in Sec.~II.5 the main new results obtained about SCGF theory thanks to
its Nambu-covariant reformulation.

\section{Nambu tensor algebra}\label{sec:tensor}

In this section, we introduce the notations that underpin the
Nambu-covariant formalism. 
We discuss Nambu fields and define general Nambu tensors in terms
of their transformation properties under a general change of basis. 
We provide illustrative examples of such tensors at the end of this section.

\subsection{Definitions}
We consider a many-body system of fermions.
The Fock space $\mathscr{F}$ of the many-fermion system is spanned by the tensor products
of a one-body Hilbert space $\mathscr{H}_1$ of the states of a single fermion.
Let us define a single-particle basis $\mathcal{B} \equiv \Set{\ket{b}}$ of $\mathscr{H}_1$.
Indices $b,c,\dots$ are used to denote states within $\mathcal{B}$.

Since we do not assume the basis to be orthogonal,
it is convenient to introduce the associated dual basis
$\bar{\mathcal{B}} \equiv \Set{\bra{\bar{b}}}$ such that
$\mathcal{B}$ and $\bar{\mathcal{B}}$ verify the biorthogonality condition
\begin{align}
    \Braket{\bar{b} | c} &= \delta_{bc} \ ,
\end{align}
where $\delta_{bc}$ denotes the usual Kronecker symbol.
The dual space of $\mathscr{H}_1$ is denoted as $\mathscr{H}_1^\dagger$.
For any basis $\mathcal{B}$ of $\mathscr{H}_1$, we define the Hermitian
conjugated basis $\mathcal{B}^\dagger \equiv \Set{\bra{b}}$
of $\mathscr{H}_1^\dagger$. In the special case where $\mathcal{B}$
is orthonormal, we have $\bar{\mathcal{B}} = \mathcal{B}^\dagger$.

The creation and annihilation operators associated to $\mathcal{B}$
are denoted as $\bar{a}_b$ and $a_b$, respectively. 
Here, we chose the bar notation used in Ref.~\cite{Balian1969} for the dual basis\footnote{Note that alternative notations exist such, as the one used in Refs.~\cite{Sinanoglu1984,Head-Gordon1998}.}.
We stress that, in general,
$\bar{a}_b = a^\dagger_{\bar{b}} \neq a^\dagger_{b}$~\cite{Balian1969}.
Creation and annihilation operators verify the canonical anticommutation relations
\begin{subequations}\label{CAR_sp}
\begin{align}
    \Set{ \bar{a}_b , \bar{a}_c } &= 0 \ , \label{CAR_sp1} \\
    \Set{ a_b , a_c } &= 0 \ , \label{CAR_sp2} \\
    \Set{ \bar{a}_b , a_c } &= \delta_{bc} \ .
\end{align}
\end{subequations}

At this point, considering tensors over $\mathscr{H}_1$ and
$\mathscr{H}_1^\dagger$ would give us the standard single-particle
tensor algebra, which has been studied and applied in the context of
quantum chemistry in Refs.~\cite{Sinanoglu1984,Head-Gordon1998}.
For instance, let us consider the tensor product space
$\mathscr{H}_1^{\otimes p} \otimes \left(\mathscr{H}_1^\dagger\right)^{\otimes q}$
of type $(p,q)$ single-particle tensors.
A change of single-particle basis modifies the coordinates of a type
$(p,q)$ single-particle tensor according to the standard tensor product
representation of the linear group $\GL(\mathscr{H}_1)$.
If we consider the Fock space $\mathscr{F}$, the associated representation of
$\GL(\mathscr{H}_1)$ characterising a change of single-particle basis
can be decomposed into the sum of representations over the $N$-body Hilbert space
$\mathscr{H}_N \equiv \mathscr{H}_1^{\otimes N}$.
This is a consequence of the stability of $\mathscr{H}_N$ with respect to
a change of single-particle basis.
For example, if $t_{bc}$ are the components of an element of $\mathscr{H}_2$
(i.e.\ of a $(2,0)$ single-particle tensor), and $U$ is the invertible
matrix representing a change of single-particle basis,
the new components after changing the single-particle basis read
\begin{equation}
    t'_{bc} \equiv \sum_{de} \ U^{-1}_{bd} \ U^{-1}_{ce} \ t_{de} \ .
\end{equation}
In practice, working with tensors over $\GL(\mathscr{H}_1)$ 
allows one to keep track of how a change of single-particle basis
affects a set of components. Tensors also provide a powerful organising tool
to classify contributions to observables, which must necessarily be 
invariant with respect to a
change of single-particle basis. Tensorial considerations can also be used to 
guide physically motivated approximations~\cite{Sinanoglu1984,Head-Gordon1998}. 
Unfortunately, the practical advantages of single-particle tensor algebra
cannot be carried over to the larger group of linear canonical transformations, namely Bogoliubov transformations~\cite{Blaizot1986}.
In particular, the sub-spaces $\mathscr{H}_N$ are no longer stable
with respect to Bogoliubov transformations.

We explore here a more convenient tensor algebra
that arises at the price of extending $\mathscr{H}_1$
to a vector space of double dimension.
Such a doubled-dimension vector space was already introduced
in the work of Anderson~\cite{Anderson1958} and Nambu~\cite{Nambu1960}
on symmetry-broken systems.
Instead, we find to be more convenient to work in a second quantisation formulation.
In this case, we work with the vector space
\begin{equation}\label{FieldVectorSpace}
    \mathscr{H}^f \cong \mathrm{Span} \set{ \bar{a}_b } \oplus \mathrm{Span} \set{ a_c }\ .
\end{equation}
We will refer to this space as the \emph{field vector space}, $\mathscr{H}^f$.
For a given complete set of creation and annihilation operators,
we define the \emph{canonical field basis} $\mathcal{B}^f$ of $\mathscr{H}^f$ as
\begin{equation}\label{BasicExampleFieldBasis}
    \mathcal{B}^f \equiv \Set{\bar{a}_b} \cup \set{ a_c } \ .
\end{equation}
In this case, it is convenient to index the elements of $\mathcal{B}^f$
over a global index $\mu \equiv (b,l)$, where $b$ denotes a state in the space
$\mathscr{H}_1$ and $l \in \{ 1, 2\}$ is a Nambu index. This index labels a state of
$\mathcal{B}$ ($l=1$) or of $\bar{\mathcal{B}}$ ($l=2$).
More generally, a \emph{field basis} of $\mathscr{H}^f$ is defined by a set of
Nambu fields $\Set{\mathrm{A}_\mu}$ that spans the whole field vector space $\mathscr{H}^f$.
In the particular case of $\mathcal{B}^f$ given in Eq.~\eqref{BasicExampleFieldBasis}
we have
\begin{subequations}\label{CovNambuFieldsDef}
\begin{align}
    \mathrm{A}_{(b, 1)} &\equiv \bar{a}_b \ , \\
    \mathrm{A}_{(b, 2)} &\equiv a_{b} \ .
\end{align}
\end{subequations}

Given a set of Nambu fields $\mathrm{A}_\mu$, we define the covariant components of
the metric tensor $g_{\mu\nu}$ by the anticommutator
\begin{equation}
     g_{\mu\nu} \equiv \Set{ \mathrm{A}_\mu , \mathrm{A}_\nu } \ ,
\end{equation}
and its contravariant components as
\begin{equation}
    g^{\mu\nu} \equiv (g^{-1})_{\mu\nu} \ .
\end{equation}
By definition, indices in the Nambu fields
can be raised and lowered using the metric tensor, i.e.\
\begin{subequations}\label{RaiseLowIndexFields}
\begin{align}
  \mathrm{A}^{\mu} \equiv \sum_{\nu} g^{\mu\nu} \ \mathrm{A}_{\nu} \ , \\
  \mathrm{A}^{\mu} \equiv \sum_{\nu} {g^{\mu}}_\nu \ \mathrm{A}^{\nu} \ , \\
  \mathrm{A}_{\mu} \equiv \sum_{\nu} g_{\mu\nu} \ \mathrm{A}^{\nu} \ , \\
  \mathrm{A}_{\mu} \equiv \sum_{\nu} {g_{\mu}}^\nu \ \mathrm{A}_{\nu} \ .
\end{align}
\end{subequations}
Consequently, the canonical anticommutation rules are encapsulated in
the metric tensor according to
\begin{subequations}\label{CAR}
\begin{align}
    \Set{ \mathrm{A}^\mu , \mathrm{A}^\nu } &= g^{\mu\nu} \ , \label{CAR_UU}\\
    \Set{ \mathrm{A}^\mu , \mathrm{A}_\nu } &= {g^{\mu}}_{\nu} \ , \label{CAR_UD}\\
    \Set{ \mathrm{A}_\mu , \mathrm{A}^\nu } &= {g_{\mu}}^{\nu} \ , \\
    \Set{ \mathrm{A}_\mu , \mathrm{A}_\nu } &= g_{\mu\nu} \ .
\end{align}
\end{subequations}
For example, in the canonical field basis $\mathcal{B}^f$ defined in Eq.~\eqref{BasicExampleFieldBasis},
the components of the metric tensor simply reads 
\begin{equation}\label{basicExampleMetric}
    g^{\mu\nu} = \delta_{\mu\bar{\nu}} \ ,
\end{equation}
where $\bar{\mu} \equiv (b,\bar{l})$ and $\bar{l}$ is defined on Nambu indices by $\bar{1}=2$
and $\bar{2}=1$.

The above expressions indicate that the Nambu formalism
is underpinned by an implicit tensor algebra structure. 
We stress that, although the metric and Nambu fields expressed in
the canonical basis $\mathcal{B}^f$ are quite simple,
this is no longer the case in a general basis of $\mathscr{H}^f$. 
Working with the general notations $g_{\mu\nu}$ and $\mathrm{A}^{\mu}$ will be particularly useful
to obtain formulae independently of the choice of a particular field basis.
Next, we make explicit the resulting algebra of so-called Nambu tensors
and we relate it to the group of Bogoliubov transformations.

\subsection{Nambu tensors}\label{subsec:NambuTensors}
Nambu tensors are defined as elements of the tensor algebra $T(\mathscr{H}^f)$
built over the vector space $\mathscr{H}^f$.
Intrinsically, i.e.\ without mentioning any basis,
a Nambu tensor $t$ of type $(p,q)$ is a multilinear form over $p$ times
the Cartesian product of $\mathscr{H}^f$ and $q$ times its dual.
Equivalently, one can work directly on the coordinates of a Nambu $(p,q)$-tensor $t$,
which are written as
\begin{equation}
  {t^{\mu_1 \dots \mu_p}}_{\nu_1 \dots \nu_q} \ .
\end{equation}
For these coordinates to be a $(p,q)$  type Nambu tensor, they must transform 
according
to the standard tensor product representation of $\GL(\mathscr{H}^f)$
on $T(\mathscr{H}^f)$. 
In other words, when changing basis, the new coordinates must read
\begin{multline}\label{ChgBasisNambuTensorCoord}
  {{t'}^{\mu_1 \dots \mu_p}}_{\nu_1 \dots \nu_q}
  \equiv
    \sum_{\substack{\lambda_1 \dots \lambda_p \\ \kappa_1 \dots \kappa_q}}
    {\left(\mathcal{W}^{-1}\right)^{\mu_1}}_{\lambda_1}
    \dots
    {\left(\mathcal{W}^{-1}\right)^{\mu_p}}_{\lambda_p} \\
    \times{t^{\lambda_1 \dots \lambda_p}}_{\kappa_1 \dots \kappa_q} \ 
    {\mathcal{W}^{\kappa_1}}_{\nu_1}
    \dots
    {\mathcal{W}^{\kappa_q}}_{\nu_q} \ ,
\end{multline}
where $\mathcal{W}$ is an invertible matrix representing the change of basis
of $\mathscr{H}^f$.

Nambu tensors, as defined above, are of great use when considering
general Bogoliubov transformations.
A Bogoliubov change of basis of the Fock space $\mathscr{F}$ is equivalent to
a linear transformation of Nambu fields~\cite{Balian1969}. In other words, a Nambu
field transforms according to
\begin{subequations}
\label{NambuTransforms}
\begin{align}
 {\mathrm{A}'}^\mu
    &= \sum_{\nu} {\left(\mathcal{W}^{-1}\right)^\mu}_\nu \ \mathrm{A}^\nu \ , \\
  \mathrm{A}'_{\mu}
    &= \sum_{\nu} {\mathcal{W}^\nu}_\mu \ \mathrm{A}_\nu \ ,
\end{align}
\end{subequations}
where ${\mathrm{A}'}^{\mu}$ and $\mathrm{A}'_{\mu}$
are the new Nambu fields\footnote{Note that in a general field basis
${\mathcal{B}^f}'$, the Nambu fields are general linear combinations of
creation and annihilation operators, unlike the specific case given in
Eqs.~\eqref{CovNambuFieldsDef}.}
and
${\mathcal{W}^\mu}_\nu$ are the elements of a
$g$-orthogonal matrix that verifies
\begin{equation}\label{g-Orthogonality}
    \sum_{\lambda\kappa}  g_{\lambda\kappa} \ 
    {\mathcal{W}^\lambda}_\mu \ {\mathcal{W}^\kappa}_\nu
    = g_{\mu\nu} \ .
\end{equation}
This $g$-orthogonality condition is equivalent to restricting linear transformations
of Nambu fields to those conserving the canonical anticommutation
rules in Eqs.~\eqref{CAR}.
Mathematically speaking, the group of Bogoliubov changes of basis of
$\mathscr{F}$ is isomorphic to the orthogonal group
$\mathrm{O}(\mathscr{H}^f, g)$, i.e.\ the group
of basis changes of $\mathscr{H}^f$ which preserve the metric tensor $g$.

As it is common in the theory of tensor algebra,
a Nambu $(p,q)$-tensor will be said to have $p$
contravariant and $q$ covariant indices.
Contravariant and covariant indices
are respectively represented by upper and lower indices.
We note that covariance (contravariance) is here to be understood 
with respect to a change of basis of $\mathscr{H}^f$.
Since
\begin{equation}\label{Nambu_SubGroups}
    \GL(\mathscr{H}_1) \subset \mathrm{O}(\mathscr{H}^f, g) 
                      \subset \GL(\mathscr{H}^f) \ ,
\end{equation}
the covariance (contravariance) remains valid for single-particle and Bogoliubov
transformations.
Let us stress that $\GL(\mathscr{H}^f)$ also contains non-canonical
transformations which modify the components of the metric $g_{\mu\nu}$
i.e.\ the inclusions in Eq.~\eqref{Nambu_SubGroups} are strict.
An example of such transformation will be discussed in
Sec.~\ref{subsec:ConnectStandardFormalism}.

Whenever there is possible ambiguity, we refer to
single-particle, Bogoliubov and Nambu covariance (contravariance) to distinguish
between the specific group of transformations.
In the remainder of this work,
we will be mostly concerned with Nambu tensors.
We stress that such tensors are the cornerstone allowing us
to easily prove our equations to be either contravariant or covariant
with respect to $\GL(\mathscr{H}^f)$ and, as a sub-case,
to any Bogoliubov transformation.

\subsection{Elementary examples}
Having defined Nambu tensors formally in the previous subsection,
we now provide a series of examples illustrating where those Nambu tensors
appear in many-body theory and how to manipulate them.

\subsubsection{Basic Nambu tensors}
We start by looking at a key tensor, the metric. 
We have so far introduced four metric objects, i.e.\ 
$g^{\mu\nu}$, ${g^{\mu}}_{\nu}$, ${g_{\mu}}^{\nu}$ and $g_{\mu\nu}$. 
In our notation, these represent the
coordinates of tensors of type $(2,0)$, $(1,1)$, $(1,1)$ and $(0,2)$, respectively.
Another basic Nambu tensor is made of the matrix elements of a change of basis of $\mathscr{H}^f$,
${\mathcal{W}^{\mu}}_{\nu}$. In this case it defines a $(1,1)$-tensor.

More generally, any $k$-body operator $O$ can be represented by a $(p,q)$-tensor,
so long as $p+q=2k$. 
For instance, in the case where $p=k$ and $q=k$, the mixed $(k,k)$
representation of $O$ reads
\begin{equation}\label{Okk_tensor}
  O \equiv
    \sum_{\substack{\mu_1 \dots \mu_k \\ \nu_1 \dots \nu_k}}
      {o^{\mu_1 \dots \mu_k}}_{\nu_1 \dots \nu_k} \
      \mathrm{A}_{\mu_1} \dots \mathrm{A}_{\mu_k}
      \mathrm{A}^{\nu_1} \dots \mathrm{A}^{\nu_k} \ ,
\end{equation}
where ${o^{\mu_1 \dots \mu_k}}_{\nu_1 \dots \nu_k}$ are the coordinates of a $(k,k)$-tensor. 
Equivalently, a fully covariant representation of the same operator $O$ reads
\begin{equation}
  O \equiv
    \sum_{\mu_1 \dots \mu_{2k}}
      o_{\mu_1 \dots \mu_{2k}} \
      \mathrm{A}^{\mu_1} \dots \mathrm{A}^{\mu_{2k}} \ ,
\end{equation}
where $o_{\mu_1 \dots \mu_{2k}}$ are now the coordinates of a $(0,2k)$-tensor.
Finally, the fully contravariant representation reads
\begin{equation}
  O \equiv
    \sum_{\mu_1 \dots \mu_{2k}}
      o^{\mu_1 \dots \mu_{2k}} \
      \mathrm{A}_{\mu_{1}} \dots \mathrm{A}_{\mu_{2k}} \ ,
\end{equation}
where $o^{\mu_1 \dots \mu_{2k}}$ are coordinates of a $(2k,0)$-tensor.
We can use the index raising and lowering operations in Eqs.~\eqref{RaiseLowIndexFields} 
to relate the coordinates of the different types of tensors,
\begin{subequations}\label{RaiseLowIndex}
\begin{align}
    o_{\mu_1 \dots \mu_{2k}}
    &= \sum_{\alpha_1 \dots \alpha_k}
        g_{\mu_1\alpha_1} \dots g_{\mu_k\alpha_k} \
        {o^{\alpha_1 \dots \alpha_k}}_{\mu_{k+1} \dots \mu_{2k}} \ ,
    \label{RaiseLowIndex_cov} \\
    o^{\mu_1 \dots \mu_{2k}}
    &= \sum_{\alpha_1 \dots \alpha_k}
        {o^{\mu_1 \dots \mu_k}}_{\alpha_1 \dots \alpha_k} \
        g^{\alpha_1\mu_{k+1}} \dots g^{\alpha_k\mu_{2k}} \ .
\end{align}
\end{subequations}

\subsubsection{Building new Nambu tensors}
We now turn our attention to a series of additional tensor operations that will be 
necessary in our derivations. In particular, we discuss here 
transpositions, linear combinations, tensor products and tensor contractions.

Let us first start with transpositions, which
essentially correspond to a different reordering of the indices.
For example, in the case of the $(1,1)$- and $(0,2)$-tensors of
coordinates ${t^{\mu}}_{\nu}$ and $s_{\mu\nu}$,
the only possible transpositions read
\begin{subequations}\label{DefTranspose}
\begin{align}
    {\left(t\transpose\right)_{\mu}}^{\nu} &\equiv {t^{\nu}}_{\mu} \ , \label{DefTranspose_11} \\
    \left(s\transpose\right)_{\mu\nu} &\equiv s_{\nu\mu} \ . \label{DefTranspose_02}
\end{align}
\end{subequations}
Note that the contravariant or covariant character of the indices is
kept by transpositions.
For example, using this transposition together with the raising and lowering of indices
given in Eqs.~\eqref{RaiseLowIndex}, the g-orthogonality
condition in Eq.~\eqref{g-Orthogonality} takes the familiar form
\begin{equation}
    \sum_{\lambda}
        {(\mathcal{W}\transpose)^\mu}_\lambda \ {\mathcal{W}^\lambda}_\nu
    = {g^{\mu}}_{\nu} \ .
\end{equation}

We now turn our attention to linear combinations. 
The space of tensors of a fixed type is a vector space.
As such, tensors of the same type can be linearly combined while keeping
the tensorial structure intact.
For example, the (anti)symmetrisation of a tensor gives back a tensor of the same type.
We note, however, 
that contravariant and covariant indices must be (anti)symmetrised separately.
Consider, for instance, the coordinates
\begin{equation}\label{partialAntisymSymEx}
  {o^{[\mu_1 \dots \mu_k]}}_{(\nu_1 \dots \nu_k)} \, ,
\end{equation}
which define a new $(k,k)$-tensor based on the original components of Eq.~(\ref{Okk_tensor}).
The bracketed indices correspond to the %standard 
shorthand notations for (anti)symmetrisation
\begin{align}
    t_{[\mu_1 \dots \mu_p] \mu_{p+1}\dots}
        &\equiv \frac{1}{p!}
        \sum_{\sigma \in S_p} \epsilon(\sigma) \ t_{\mu_{\sigma(1)} \dots \mu_{\sigma(p)} \mu_{p+1}\dots} \ , \label{AntisymNotation}\\
    t_{(\mu_1 \dots \mu_p) \mu_{p+1}\dots}
        &\equiv \frac{1}{p!}
        \sum_{\sigma \in S_p} \ t_{\mu_{\sigma(1)} \dots \mu_{\sigma(p)} \mu_{p+1}\dots} \ , \label{SymNotation}
\end{align}
where $S_p$ is the symmetric group of order $p$ and $\epsilon(\sigma)$
the signature of the permutation $\sigma$.
Eq.~\eqref{partialAntisymSymEx} thus corresponds to a new $(k,k)$ tensor which is 
(anti)symmetric in its (contravariant) covariant indices.  
In contrast, if $t$ is a $(1,1)$-tensor, the quantity
\begin{equation}
    \frac{{t^{\mu}}_{\nu} + {t^{\nu}}_{\mu} }{2}
\end{equation}
does \emph{not} define a new tensor. The $\nu$ index is covariant in the first term but
contravariant in the second one. As a result, the sum of both behaves
neither covariantly nor contravariantly with respect to a change of field basis.

Finally, new tensors can also be built via tensor networks, i.e.\
via a combination of tensor products and tensor contractions of previously existing tensors.
For example, let $t^{\mu_1\nu_1}$ and ${s_{\mu_2}}^{\nu_2}$ be the coordinates of
two tensors of type $(2,0)$ and $(1,1)$, respectively.
The coordinates of their tensor product
\begin{equation}
    {{r^{\mu_1\nu_1}}_{\mu_2}}^{\nu_2} \equiv t^{\mu_1\nu_1} \   {s_{\mu_2}}^{\nu_2}
\end{equation}
define a tensor of type $(3,1)$.
A contraction of the two original tensors can be defined as
\begin{equation}
    {r'}^{\mu_1\nu_2} \equiv \sum_{\lambda} t^{\mu_1\lambda} \ {s_{\lambda}}^{\nu_2} \, ,
\end{equation}
and yields coordinates of a tensor of type $(2,0)$.

In the following, most derivations will start from tensors obtained from a set
of operators. New tensors will be built by applying transpositions, linear combinations
or tensor networks.
Ultimately,
observables will necessarily be $(0,0)$-tensors, 
insuring Nambu invariance.
In standard many-body perturbation theory, theoretical predictions
are independent of the choice of the single-particle basis of $\mathscr{H}_1$. 
In the NCPT developed in Sec.~\ref{sec:covPT}, this independence of
theoretical predictions is extended to the choice of basis of
$\mathscr{H}^f$, thus including Bogoliubov transformations.

\subsection{Quadratic Hamiltonian}
As a first concrete example, let us relate the standard expression
of a quadratic Hamiltonian to its fully covariant representation.
For a given choice of single-particle basis $\mathcal{B}$,
a quadratic Hamiltonian $H_0$ reads, in general,
\begin{equation}\label{Quadra_Omega_Gorkov}
  H_0 \equiv \frac{1}{2} \sum_{bc} U^{11}_{bc}\  \bar{a}_b a_c
                    + U^{22}_{bc}\  a_{b} \bar{a}_{c}
                    + U^{12}_{bc}\  \bar{a}_b \bar{a}_{c}
                    + U^{21}_{bc}\  a_{b} a_c\ ,
\end{equation}
where $U^{l_b l_c}_{bc}$ are complex numbers.
In terms of a type $(1,1)$-tensor ${U^{\mu}}_{\nu}$,
$H_0$ reads
\begin{align}\label{Quadra_Omega_Nambu}
    H_0 &=
        \frac{1}{2}\sum_{\mu\nu} {U^{\mu}}_{\nu} \ 
            \mathrm{A}_{\mu} \mathrm{A}^{\nu} \ .
\end{align}
For convenience, we can choose to work with Nambu fields
related to the same single-particle basis $\mathcal{B}$, i.e.\
we work with the basis defined in Eq.~\eqref{BasicExampleFieldBasis}.
In this case, ${U^{\mu}}_{\nu}$ and $U^{l_b l_c}_{bc}$ are
simply related according to
\begin{align}
  {U^{(b,l_b)}}_{(c,l_c)} &= U^{l_bl_c}_{bc}
      \ , \label{Quadra_Omega_NambuGorkov_Relation}
\end{align}
and any other representation, say $U_{\mu\nu}$, can be obtained with the metric tensor
using $U_{\mu\nu}= \sum_{\lambda} g_{\mu\lambda} \ {U^{\lambda}}_{\nu}$. Explicitly,
using Eq.~\eqref{basicExampleMetric},
\begin{equation}
    U_{(b,l_b)(c,l_c)} =  U^{\bar{l}_bl_c}_{bc} \ .
\end{equation}
For completeness, similar relations for $k$-body operators
are described in~\ref{app:MatElts}.

Now that $U_{\mu\nu}$ has been related to traditional matrix elements
$U^{l_b l_c}_{bc}$, let us specify the symmetry properties of $U_{\mu\nu}$.
Decomposing $U_{\mu\nu}$ into its symmetric, $U_{(\mu\nu)}$,
and antisymmetric, $U_{[\mu\nu]}$, parts,
$H_0$ reads
\begin{equation}
    H_0 =
        \frac{1}{4} \sum_{\mu} {U^{\mu}}_{\mu}
         + \frac{1}{2} \sum_{\mu\nu} U_{[\mu\nu]} \mathrm{A}^{\mu} \mathrm{A}^{\nu} \ ,
\end{equation}
where Eqs.~\eqref{CAR_UU} and~\eqref{RaiseLowIndex} have been used.
For simplicity, we assume the term proportional to the identity
vanishes, i.e.\ $H_0$ is purely quadratic.
This is equivalent to assuming $U_{\mu\nu}$ to be antisymmetric and traceless, namely
\begin{subequations}\label{Symmetries_CovU}
\begin{align}
    U_{\mu\nu} &= - U_{\nu\mu} \ , \\
    \sum_{\mu} {U^{\mu}}_{\mu} &= 0 \ .
\end{align}
\end{subequations}
We stress that the antisymmetrisation of $U_{\mu\nu}$ was trivially obtained
by working in a fully covariant representation of $H_0$.
We will take full advantage of the idea that fully covariant
representations can be easily antisymmetrised while keeping
their tensor character intact in Sec.~\ref{sec:covPT}, where fully
antisymmetric vertices associated to $k$-body interactions will be introduced.

\section{Nambu-covariant perturbation theory}\label{sec:covPT}

The general formalism of Nambu tensors introduced
in Sec.~\ref{sec:tensor} allows us to present the perturbation theory
of many-body Green's functions in a Nambu-covariant fashion.
We refer to this specific formulation of perturbation theory as NCPT.
To obtain such formulation, we first introduce many-body Green's functions as Nambu tensors.
Second, their perturbative expansion is expressed in terms of contributions
respecting the associated co- and contravariance, i.e.\ any perturbative contribution
is a tensor of the same type as the Green's function being expanded.
These perturbative contributions are shown to be indexed over a set of
un-oriented Feynman diagrams obtained by a set of diagrammatic rules, which 
we describe explicitly.
Third, one-particle irreducible (1PI) contributions to the one-body Green's function
are given up to third order as an example of the versatility of the formalism.
Those examples, given in Sec.~\ref{subsec:Examples}, together with additional ones detailed
in Part~II~\cite{part2} are also meant as illustrative examples of the Feynman rules
discussed in this section.

We consider in this section a physical system of
many-fermions at equilibrium. 
The system is in a statistical ensemble at inverse temperature $\beta$, 
described by the Hamiltonian $H$.
This Hamiltonian describes fermionic interactions as well as the statistical
ensemble. For generality, and anticipating
applications in nuclear systems, we consider two-, three- and 
up to $k_{\text{max}}$-body interactions.

We define the perturbation theory for a partitioning of the Hamiltonian, 
$ H \equiv H_0 + H_1$,
where the unperturbed Hamiltonian, $H_0$, is purely quadratic
and non-necessarily Hermitian. In terms of Nambu fields and tensors,
\begin{subequations}\label{GeneralPartitionPT}
  \begin{align}
    H_0 &\equiv
      \frac{1}{2} \sum_{\mu\nu} U_{\mu\nu} \mathrm{A}^\mu \mathrm{A}^\nu \ , \\
    H_1 &\equiv
      \sum^{k_{\text{max}}}_{k=0} \ \frac{1}{(2k)!}
                        \sum_{\mu_1 \dots \mu_{2k}}
                              v^{(k)}_{\mu_1 \dots \mu_{2k}} \
                              \mathrm{A}^{\mu_1} \dots \mathrm{A}^{\mu_{2k}} \ .
  \end{align}
\end{subequations}
Here, and in the following, we use the fully covariant representation
of operators and contravariant Nambu fields.

\subsection{Many-body Green's functions}
We write the exact and the unperturbed (statistical) density matrices of the system,
together with their partition functions, as
\begin{subequations}
\begin{align}
    \rho &\equiv \frac{1}{Z} e^{-\beta H} \ , \\
    Z &\equiv \Tr\left( e^{-\beta H} \right) \ , \\
    \rho_0 &\equiv \frac{1}{Z_0} e^{-\beta H_0} \ , \\
    Z_0 &\equiv \Tr\left( e^{-\beta H_0} \right) \ .
\end{align}
\end{subequations}
We use the notation $\mean{ \dots }$ and $\mean{ \dots }_0$ for the ensemble averages
with respect to the exact and unperturbed density matrices, respectively.
Using the imaginary-time formalism, we define the contravariant $k$-body Green's function
as\footnote{Throughout this work we assume natural units where
$\hbar = c = k_\text{B} = 1$.}
\begin{equation}\label{DefkBodyGF}
    (-1)^k \mathcal{G}^{\mu_1 \dots \mu_{2k}}(\tau_1, \dots, \tau_{2k})
    \equiv
    \mean{\mathrm{T}\left[ \mathrm{A}^{\mu_1}(\tau_{1}) \dots \mathrm{A}^{\mu_{2k}}(\tau_{2k}) \right]} \ , 
\end{equation}
Here, the (imaginary) time evolution depends on the complete Hamiltonian, $H$, and 
$\mathrm{T}\left[ \dots \right]$ denotes the time-ordering
from right to left when increasing imaginary-time, $\tau$.
At any fixed time, the contravariant $k$-body Green's function is a $(2k,0)$-tensor.
We could study any other $(p,q)$-tensor (with $p+q=2k$)
obtained by raising or lowering indices in the fully-contravariant $k$-body Green's function.
Working on fully contravariant tensors, however, will be convenient
due to their higher degree of symmetry.

Let us also emphasise the simplicity of the definition of a fully contravariant $k$-body Green's function.
Depending on the Nambu indices contained in the global $\mu$ indices, we can recover
from this expression 
the normal as well as any other possible anomalous components of the Green's function.
In particular, for the one-body Green's function, one anomalous term
appears~\cite{Gorkov1958}. This motivates the definition of a $2\times2$ matrix
propagator, encompassing both normal and anomalous terms.
In a way, Eq.~\eqref{DefkBodyGF} is the fully contravariant tensor generalisation
of the more traditional one-body matrix Green's function.
We discuss this connection in more detail in Sec.~\ref{subsubsec:GorkovConnection}, where we
explicitly relate our approach to the standard Gorkov formulation
exposed in Ref.~\cite{Soma2011}. 

Using the fact that $H_1$ contains only terms with an even number of Nambu fields,
the contravariant $k$-body Green's function can be re-expressed as
\begin{multline}
    (-1)^{k} \ \mathcal{G}^{\mu_1 \dots \mu_{2k}}(\tau_1, \dots, \tau_{2k})
    \ = \\
    \frac{
       \left<
        \mathrm{T}
        \left[
          e^{- \int^{\beta}_{0}\mathrm{d}s \ H_1(s)} \
          {\mathrm{A}^{\mu_1}}(\tau_1) \
          \dots \
          {\mathrm{A}^{\mu_{2k}}}(\tau_{2k})
         \right]
        \right>_0
    }{
        \left<
           \mathrm{T} e^{-\int^{\beta}_{0}\mathrm{d}s \ H_1(s)}
        \right>_0
     } \ .
\end{multline}
The time dependence is now with respect to $H_0$.
Expanding the exponential and permuting sums and integrals with the ensemble average,
we find the perturbative expansion of the contravariant $k$-body Green's function:
\begin{align}\label{kBodyGF_PT}
    (-1)^{k} \ \mathcal{G}&^{\mu_1 \dots \mu_{2k}}(\tau_1, \dots, \tau_{2k})
    \ = \nonumber \\
    \sum^{+\infty}_{n=0} \frac{(-1)^n}{n!}
    &\int^{\beta}_0 \mathrm{d}s_1 \dots
    \int^{\beta}_0 \mathrm{d}s_n \nonumber \\
        &\frac{
                \mean{
                \mathrm{T}
                \left[
                    H_1(s_1) \dots H_1(s_n)
                    \mathrm{A}^{\mu_1}(\tau_1)
                    \dots
                    \mathrm{A}^{\mu_{2k}}(\tau_{2k})
                \right]
                }_0
            }{
                \mean{\mathrm{T} e^{-\int^{\beta}_{0}\mathrm{d}s \ H_1(s)}}_0
            } \ .
\end{align}
This expression is analogous to the perturbative expansion in the normal, symmetry-conserving 
case~\cite{stefanucci_van_leeuwen_2013}, but replacing creation and annihilation operators by (contravariant) 
Nambu fields.

The integral at a given order in the perturbative expansion of
the contravariant $k$-body Green's function can be computed as usual
via a statistical time-dependent Wick's theorem.
The proof of Wick's theorem can be adapted straightforwardly
to the case of Nambu fields following Ref.~\cite{Gaudin1960a}.
The statistical time-dependent Wick's theorem for Nambu fields reads
\begin{multline}\label{TimeDepStatWickTheorem}
  \mean{\mathrm{A}^{\mu_1}(\tau_{1}) \dots \mathrm{A}^{\mu_{2p}}(\tau_{2p})}_0
  = \\
  \sum_{ \mathcal{P}=\Set{\Set{i,j}} }
  \epsilon(\mathcal{P}) \prod_{\substack{\Set{i,j} \in \mathcal{P}\\ i<j}}
  \mean{\mathrm{A}^{\mu_i}(\tau_i) \mathrm{A}^{\mu_j}(\tau_j)}_0 \ ,
\end{multline}
where the sum is over the set of pairings $\mathcal{P}$ of the first $2p$
positive integers and $\epsilon(\mathcal{P})$ is the standard sign factor
associated to a pairing, $\mathcal{P}$.
The different time-orderings in Eq.~\eqref{kBodyGF_PT} are taken into account
by using the unperturbed time-ordered contravariant propagator
\begin{equation}\label{UnpertContravProp}
    -\mathcal{G}^{(0)\mu\nu}(\tau,\tau')
        \equiv
    \mean{\mathrm{T}\left[ \mathrm{A}^{\mu}(\tau) \mathrm{A}^{\nu}(\tau') \right]}_0
\end{equation}
and by cancelling out any double-counting in the time integration
with an appropriate symmetry factor.
As a consequence of the symmetry of the contravariant propagator,
the contravariant $k$-body Green's function is decomposed
into a sum of amplitudes indexed over the set of \emph{un-oriented Feynman diagrams}.
We now provide the resulting Feynman rules in the time and in the energy representations.
For more details on the exact properties of the contravariant
$k$-body Green's functions, we refer the reader to Part~II of this work.

\subsection{Feynman rules in the time representation}\label{subsec:FeynRulesTime}
Before expressing Feynman rules in the Nambu-covariant framework,
let us define a notation for partially antisymmetrised tensor coordinates.
These appear in the algebraic expression of un-oriented Feynman diagrams,
in particular at the level of interaction vertices $v^{(k)}_{\mu_1 \dots \mu_k}$.
In general, each vertex of a diagram will correspond to
the totally antisymmetric part of the associated interaction. 
In this case, using the standard notation given in
Eq.~\eqref{AntisymNotation}, the totally antisymmetric vertex reads
\begin{equation}\label{FullyAntisymVertexSym}
  v^{(k)}_{[\mu_1 \dots \mu_{2k}]}
  \equiv
    \frac{1}{(2k)!} \sum_{\sigma \in S_{2k}} \epsilon(\sigma)\
        v^{(k)}_{\mu_{\sigma(1)}\dots \mu_{\sigma(2k)}} \ .
\end{equation}
An important subtlety that arises in NCPT has to do with the interaction matrix elements in
diagrams involving tadpoles. 
In the case of a vertex with $p$ tadpoles, one needs to perform
a sum over a subset of the permutations of the indices.
We will denote such partial antisymmetrisation as
\begin{multline}\label{PartialAntisymVertexSym}
  v^{(k)}_{[\mu_1 \dots \dot{\mu}_{x} \dots \dot{\mu}_{y} \dots \mu_{2k}]}
  \equiv
    \frac{2^p p!}{(2k)!} \sum_{\sigma \in S_{2k}/S^{p}_2 \times S_p} \!\!\! \epsilon(\sigma)\
        v^{(k)}_{\mu_{\sigma(1)}\dots
                \dot{\mu}_{x} \dots
                \dot{\mu}_{y} \dots
                \mu_{\sigma(2k)}} \, ,
\end{multline}
where the sum is to be understood as running only over the permutations $\sigma$
that \emph{do not} exchange two indices within a dotted pair \emph{nor}
two different dotted pairs of indices. We denote the subset of these specific permutations as
$S_{2k}/S^{p}_2 \times S_p$.

Having established our notations, we now proceed to give the Feynman 
rules in the time domain. 
The $n^{\text{th}}$ order contribution to
$(-1)^{k} \ \mathcal{G}^{\mu_1 \dots \mu_{2k}}(\tau_1, \dots, \tau_{2k})$
is obtained as follows:
\begin{enumerate}
  \item Draw all topologically distinct un-oriented unlabelled linked diagrams
   $\mathscr{G}_n$ with $n$ vertices and $2k$ external lines.
   Two diagrams are topologically
   equivalent if one is obtained from the other by a continuous deformation.
  \item Assign a label $\tau_1 \dots \tau_n$ to the vertices and compute
  the symmetry factor $S$ which is the number of permutations of vertex labels
  leaving invariant the labelled diagram (up to a continuous deformation).
  \item Assign a global index $\mu$ to every half-lines of
  $\mathscr{G}_n$. For each line joining $(\mu,\tau)$ and $(\nu,\tau')$,
  multiply by a factor $-\mathcal{G}^{(0)\mu\nu}(\tau,\tau')$.
  In the case of a tadpole, the factor reads $-\mathcal{G}^{(0)\mu\nu}(\tau+\eta,\tau)$
  with $\eta \to 0^+$.
  \item For each $k_i$-body vertex with indices $\mu_1 \dots \mu_{2k_i}$,
  multiply by a factor $v^{(k_i)}_{[\mu_1 \dots \mu_{2k_i}]}$ where indices
  belonging to a same tadpole are dotted in the same way, according to Eq.~\eqref{PartialAntisymVertexSym}.
  \item Sum over global indices $\mu$ and integrate over
  $\tau$ on $[0,\beta]$.
  \item Multiply by a factor
  $\frac{(-1)^{n+L}}{S \times 2^{T} \prod_{l=2}^{l_{\text{max}}}(l!)^m}$
  where $L$ is the number of loops (excluding tadpoles), $m$ the number of
  $l$-tuple equivalent lines (with $l \in \llbracket 2,l_{\text{max}} \rrbracket$)
  and $T$ the number of tadpoles.
\end{enumerate}
In the above rules, we choose to work with a time conventionally flowing from the bottom
to the top of the diagram.
The reading convention of a line is from top to bottom while
the writing of the algebraic expression is from left to right.
For vertices and tadpoles, indices are to be read clockwise from the angle
$+\pi$ towards $-\pi$ and the writing is still from left to right.
In the case of no external legs, i.e.\ $k=0$, we obtain the $n^{\text{th}}$ order
contribution to $\ln{\frac{Z}{Z_0}}$ rather than to a Green's function.

Just like in standard versions of perturbation theory,
the linked-cluster theorem allows us to consider only linked diagrams.
The resulting Feynman amplitude
$\mathcal{A}^{\mu_1 \dots \mu_{2k}}(\tau_{\mu_1}, \dots, \tau_{\mu_{2k}})$
associated with the un-oriented Feynman diagram $\mathscr{G}_{n}$,
which contributes to
$(-1)^{k} \ \mathcal{G}^{\mu_1 \dots \mu_{2k}}(\tau_1, \dots, \tau_{2k})$,
reads generically
\begin{multline}\label{GenericFeynmanAmplitudeTime}
    \mathcal{A}^{\mu_1 \dots \mu_{2k}}(\tau_{\mu_1}, \dots, \tau_{\mu_{2k}})
    =
    \frac{(-1)^{n+L}}{S \times 2^{T} \prod_{l=2}^{l_{\text{max}}}(l!)^m} \\
    \times \sum_{\lambda\dots\lambda}
        v^{(k_1)}_{[\lambda \dots \lambda]}
        \dots
        v^{(k_n)}_{[\lambda \dots \lambda]}
        \int^{\beta}_{0} \mathrm{d}\tau_1 \dots \mathrm{d}\tau_n \
        \prod_{e \in I} -\mathcal{G}^{(0)\lambda\lambda}(\tau_{i},\tau_{j}) \\
        \times \prod_{e \in E_{\text{in}}} -\mathcal{G}^{(0)\lambda\mu}(\tau_{i},\tau_{\mu})
        \prod_{e \in E_{\text{out}}} -\mathcal{G}^{(0)\mu\lambda}(\tau_{\mu},\tau_{j})
        \ ,
\end{multline}
where $\lambda$ and $\mu$ denote respectively generic global indices for internal
and (incoming or outgoing) external lines\footnote{Diagrams are un-oriented here,
so ``outgoing" or ``incoming" lines are to be understood with respect to the time flow.}.
Labels $k_i$ characterise the $k$-body type of vertex $i$ and $\tau_i$ denotes the
corresponding time label. The set of internal, incoming external and outgoing
external lines are respectively denoted by $I$, $E_\text{in}$ and $E_\text{out}$.
In Eq.~\eqref{GenericFeynmanAmplitudeTime} the tadpole case is not explicitly
taken into account for the sake of conciseness.

We now make several observations about the above Feynman rules.
In the case where the vertex is free of tadpoles, only its totally antisymmetric
part contributes to the Feynman amplitude.
This generalises what was noticed by
De Dominicis and Martin~\cite{DeDominicis1964a,DeDominicis1964b},
Kleinert~\cite{Kleinert1981,Kleinert1982} and
Haussmann~\cite{Haussmann1993,Haussmann1999} to the case of
a generic $k$-body interaction and a general set of Nambu fields.
In the case where the vertex is contracted with a set of tadpoles,
it is actually a specific partial antisymmetrisation that contributes
to the Feynman amplitude. 
Let us point out that this subtle case of partial
antisymmetrisation is neither mentioned
in Refs.~\cite{DeDominicis1964a,DeDominicis1964b,Kleinert1981,Kleinert1982,Haussmann1993,Haussmann1999} nor, to our knowledge, elsewhere.
The occurrence of totally and partially antisymmetric vertices is the consequence
of a factorisation of several Feynman diagrams. This factorisation property, in turn, 
arises because the Feynman amplitudes are expressed in terms of:
\begin{itemize}
    \item a sum over pairings, thanks to
    Wick's theorem~\eqref{TimeDepStatWickTheorem};
    \item a sum over single-particle and Nambu indices,
    thanks to our decomposition of $H_0$ and $H_1$
    as polynomials in Nambu fields in Eqs.~\eqref{GeneralPartitionPT};
    \item fully covariant vertices, which can be antisymmetrised
    while keeping their tensor character intact.
\end{itemize}
More details on how totally and partially antisymmetric vertices arise
in the above Feynman rules are given in~\ref{app:AntisymVertices}.
The factorisation property is further discussed in connection with standard
Gorkov and Bogoliubov formalisms in Sec.~\ref{subsec:ConnectStandardFormalism}.

We stress that NCPT requires an \emph{extension of the Hugenholtz antisymmetrisation of vertices}. 
In standard Hugenholtz diagrams,
vertices are only antisymmetric with respect to permutations of outgoing
and incoming half-lines, separately. Here, in contrast, the antisymmetrisation is complete. 
This antisymmetrisation is found \emph{empirically} to hold in the standard zero-temperature Gorkov formalism of Refs.~\cite{Soma2011} but it requires an \emph{ad hoc} selection of contributions.
There, several oriented Feynman diagrams contributing to the self-energy at second order were combined to get an expression in Eqs.~(79) of Ref.~\cite{Soma2011}. 
This combination lead to the introduction of the so-called 
$\mathcal{C}$ and $\mathcal{D}$ objects, defined
in Eqs.~(78) of Ref.~\cite{Soma2011} and that satisfy proper Pauli statistics. Similar summations of antisymmetrised terms are found at higher order of the Gorkov-Feynman expansion~\cite{Barbieri2022}.
The NCPT developed here shows why such property is not a coincidence. 
Moreover, our approach generalises these considerations to any set of Feynman diagrams, 
and to the non-zero temperature case.

Finally, we stress that the amplitude associated to an un-oriented Feynman diagram, 
Eq.~\eqref{GenericFeynmanAmplitudeTime},
is essentially a tensor network of covariant vertices and contravariant propagators.
As a consequence, the amplitude associated to an un-oriented Feynman diagram is a tensor
of the same nature as the $k$-body Green's function to which it contributes.

\subsection{Feynman rules in the energy representation}\label{subsec:FeynRulesEnergy}
Similarly to other formalisms of perturbation theory, the time-independence of
the partitioning in Eq.~\eqref{GeneralPartitionPT} allows us to simplify the Feynman
amplitudes when working in the energy representation.
Before stating the Feynman rules in the energy representation, however,
we need to specify the conventions we are using to define the propagators in
their energy representation.
In thermal equilibrium, the unperturbed contravariant one-body propagator only depends
on the time difference, i.e.\
\begin{equation}
    \mathcal{G}^{(0)\mu\nu}(\tau+\tau',\tau')
  = \mathcal{G}^{(0)\mu\nu}(\tau,0)
  \equiv \mathcal{G}^{(0)\mu\nu}(\tau) \ .
\end{equation}
Furthermore, the propagator is a $\beta$-quasiperiodic function, i.e.\ 
\begin{align}
    \mathcal{G}^{(0)\mu\nu}(\tau + \beta) &= - \mathcal{G}^{(0)\nu\mu}(\tau) \ ,
\end{align}
and fulfils the antisymmetry property 
\begin{align}
\mathcal{G}^{(0)\mu\nu}(\tau) = - \mathcal{G}^{(0)\nu\mu}(-\tau) \, .
\end{align}
As a consequence, the Fourier transform of the unperturbed contravariant
propagator yields the energy representation
\begin{align}
    \mathcal{G}^{(0)\mu\nu}(\omega_m)
        &\equiv \int_{0}^{\beta} \mathrm{d}\tau \ e^{i\omega_m \tau} \
                            \mathcal{G}^{(0)\mu\nu}(\tau) \ , \\
    \mathcal{G}^{(0)\mu\nu}(\tau)
        &= \frac{1}{\beta} \sum_{\omega_m}
                              \ e^{-i\omega_m \tau} \  \mathcal{G}^{(0)\mu\nu}(\omega_m) \ ,
\end{align}
where $\omega_m \equiv (2m+1) \frac{\pi}{\beta}$ are fermionic Matsubara frequencies.

These considerations are easily extended to $k$-body propagators. 
Energy conservation dictates that 
contravariant $k$-body Green's functions are Fourier transformed
to their energy representation leaving out a factor
$\beta \delta_{\omega_{\text{in}}, \omega_{\text{out}}}$,
\begin{align}
    &\mathcal{G}^{\mu_1 \dots \mu_{2k}}(\omega_{m_1}, \dots, \omega_{m_{2k}})
      \ \beta \delta_{\omega_{\text{in}}, \omega_{\text{out}}}
        \nonumber \\
    &\equiv  \int_{0}^{\beta} \mathrm{d}\tau_1 \dots \mathrm{d}\tau_{2k} \
            e^{+ i\omega_{m_\text{in}} \tau_{m_\text{in}} + \dots} \
            e^{- i\omega_{m_{\text{out}}} \tau_{m_{\text{out}}} - \dots}
            \nonumber \\
    &\phantom{\equiv \int_{0}^{\beta} \mathrm{d}\tau_1 \dots \mathrm{d}\tau_{2k} \ }
            \times \mathcal{G}^{\mu_1 \dots \mu_{2k}}(\tau_1, \dots, \tau_{2k}) \ , \label{DefEnergykBodyGF}\\
    &\mathcal{G}^{\mu_1 \dots \mu_{2k}}(\tau_1, \dots, \tau_{2k})
        \nonumber \\
    &= \frac{1}{\beta^{2k}} \sum_{\omega_{m_1} \dots \omega_{m_{2k}}}
            e^{- i\omega_{m_\text{in}} \tau_{m_\text{in}} - \dots} \
            e^{+ i\omega_{m_{\text{out}}} \tau_{m_{\text{out}}} + \dots}
            \nonumber \\
    &\phantom{\frac{1}{\beta^{2k}} \sum_{\omega_{m_1} \dots \omega_{m_{2k}}}}
            \times \mathcal{G}^{\mu_1 \dots \mu_{2k}}(\omega_{m_1}, \dots, \omega_{m_{2k}}) \
                   \beta \delta_{\omega_{\text{in}}, \omega_{\text{out}}} \ .
\end{align}
Here, $\omega_{\text{in}}$ and $\omega_{\text{out}}$
denote respectively the sum of the external incoming and outgoing energies of the Green's function\footnote{The
external energy flow is left to be fixed by convention.}.
Energies flowing in and out are denoted generically as
$\omega_{m_\text{in}}$ and $\omega_{m_\text{out}}$, respectively.
Their associated times are also generically denoted as $\tau_{m_\text{in}}$
and $\tau_{m_\text{out}}$, respectively.

With the above definitions for the energy representation of contravariant
many-body Green's functions, the $n^{\text{th}}$ order contribution to
$(-1)^{k} \mathcal{G}^{\mu_1 \dots \mu_{2k}}(\omega_{m_1}, \dots, \omega_{m_{2k}})$
is obtained as follows:
\begin{enumerate}
\item Draw all topologically distinct un-oriented unlabelled linked diagrams
   $\mathscr{G}_n$ with $n$ vertices and $2k$ external lines.
   Two diagrams are topologically equivalent if one is obtained
   from the other by a continuous deformation.
\item Assign a label $1 \dots n$ to the vertices and
  compute the symmetry factor $S$ which is the number of permutations
  of vertex labels leaving invariant the labelled diagram
  (up to a continuous deformation).
\item Choose a spanning forest of $\mathscr{G}_n$\footnote{By definition, a spanning
  forest of the diagram $\mathscr{G}_n$ is a sub-diagram without cycles
  which is maximal in the sense that no line can be added to the sub-diagram
  without creating a cycle. It necessarily contains all the vertices of
  $\mathscr{G}_n$.}. For each internal line $l_i$ \emph{not} in the forest,
  assign an independent Matsubara frequency $\omega_{l_i}$.
  Remaining internal lines are assigned the linear combination of
  $\omega_{l_i}$ and external $\omega_{m_j}$ determined by conservation of energy
  at each vertex. The Matsubara frequency thus associated to any line $e$
  is denoted generically as $\omega_e$.
\item  Assign a global index $\mu$ to any half-line of $\mathscr{G}_n$.
  For each line $e$, joining half-lines $\mu$ to $\nu$,
  multiply by a factor $-\mathcal{G}^{(0)\mu\nu}(\omega_e)$.
  In the case of a tadpole, the factor reads
  $-\mathcal{G}^{(0)\mu\nu}(\omega_e)e^{- i\omega_e \eta}$ with $\eta \to 0^+$.
\item For each $k_i$-body vertex with indices $\mu_1 \dots \mu_{2k_i}$
  multiply by a factor $v^{(k_i)}_{[\mu_1 \dots \mu_{2k_i}]}$ where indices
  belonging to a same tadpole are dotted in the same way, according to Eq.~\eqref{PartialAntisymVertexSym}.
\item Sum over global indices $\mu$ as $\sum_{\mu}$ and over
  each independent Matsubara frequencies $\omega_{l_i}$ (defined for the chosen
  spanning forest) as $\frac{1}{\beta}\sum_{\omega_{l_i}}$.
\item Multiply by a factor
  $\frac{(-1)^{n+L} \beta^{C - 1}}{S \times 2^T \prod_{l=2}^{l_{\text{max}}}(l!)^m}$
  where $C$ is the number of connected components, $L$ the number of loops (excluding tadpoles),
  $m$ the number of $l$-tuple equivalent lines (with $l \in \llbracket 2,l_{\text{max}} \rrbracket$)
  and $T$ the number of tadpoles.
\end{enumerate}
In the above rules, the energy flow convention is left to be specified.
We discuss this choice further in Sec.~\ref{subsec:MatsubaraSum}, where
Matsubara sums are performed explicitly.
The reading convention of a line is going against
the energy flow orientation while the writing of the associated propagator is from left to right.
The reading of vertices is the same as for the Feynman rules in
the time representation.
The factor $\beta^{C}$ appears from the change of representation
from time to energy~\cite{Blaizot1986}, and the remaining $\beta^{-1}$
from our choice to factorise the global term
$\beta \delta_{\omega_{\text{in}}, \omega_{\text{out}}}$
out of the definition of the energy representation of the $k$-body Green's function.
% In the case of $k=0$, we obtain the $n^{\text{th}}$ order contribution
% to $\frac{1}{\beta}\ln{\frac{Z}{Z_0}}$ rather than to a Green's function.

Following the above diagrammatic rules, the resulting Feynman amplitude
$\mathcal{A}^{\mu_1 \dots \mu_{2k}}(\omega_{m_1}, \dots, \omega_{m_{2k}})$
associated to the un-oriented Feynman diagram $\mathscr{G}_{n}$
reads generically
\begin{multline}\label{GenericFeynmanAmplitudeEnergy}
    \mathcal{A}^{\mu_1 \dots \mu_{2k}}(\omega_{m_1}, \dots, \omega_{m_{2k}})
    =
    \frac{(-1)^{n+L} \beta^{C-1}}{S \times 2^{T} \prod_{l=2}^{l_{\text{max}}}(l!)^m}
    \\
    \times \sum_{\lambda\dots\lambda}
        v^{(k_1)}_{[\lambda \dots \lambda]}
        \dots
        v^{(k_n)}_{[\lambda \dots \lambda]}
        \frac{1}{\beta^L}\sum_{\omega_{l_1} \dots \omega_{l_L}}
        \prod_{e \in I} -\mathcal{G}^{(0)\lambda\lambda}(\omega_e) \\
        \times  \prod_{e \in E_{\text{in}}}
        -\mathcal{G}^{(0)\lambda\mu}(\omega_{m_\text{in}})
        \prod_{e \in E_{\text{out}}} 
        -\mathcal{G}^{(0)\mu\lambda}(\omega_{m_\text{out}})
        \ ,
\end{multline}
where $\lambda$ and $\mu$ denote respectively generic global indices for internal
and (incoming or outgoing) external lines\footnote{Diagrams are un-oriented here,
so ``outgoing" or ``incoming" lines are to be understood with respect to the energy flow.}.
For each vertex $i$, the label $k_i$ indicates that this is a $k_i$-body interaction.
Independent Matsubara frequencies that are summed over
are denoted by $\omega_{l_i}$.
External incoming and outgoing Matsubara frequencies are generically denoted as 
$\omega_{m_\text{in}}$ and $\omega_{m_\text{out}}$, respectively.
Internal Matsubara frequencies, which are linear combinations of $\omega_{l_i}$,
$\omega_{m_\text{in}}$ and $\omega_{m_\text{out}}$, are generically
denoted as $\omega_e$. Again, $I$ denotes the set of internal lines,
$E_\text{in}$ the set of incoming external lines and $E_\text{out}$
the set of outgoing external lines.
In Eq.~\eqref{GenericFeynmanAmplitudeEnergy} the tadpole case
is not explicitly taken into account for the sake of conciseness. 

Let us stress that it is not the first time
that un-oriented diagrams occur in formal many-body theory work.
Those do appear sometimes in classical textbooks, such as Ref.~\cite{Blaizot1986}.
However, to the best of our knowledge un-oriented diagrams
have always been either restricted to Majorana fields, or
were to be understood as a shorthand notation for summing over amplitudes
associated to all compatible oriented diagrams, such as in Ref.~\cite{Nozieres1964}.
Such shorthand notation appears already in the work of Nambu on superconductivity~\cite{Nambu1960}, for instance. 
Making the distinction between this common shorthand notation and the NCPT diagrammatics
obtained here is essential. In the former approach, e.g.\ see the diagrammatic rules
given in Chap.~7 of~\cite{Nozieres1964}, the dependence on Nambu indices of the
Feynman amplitude arises from two different sources. On the one hand, the amplitude is the result of a
sum of several single-particle tensor networks whose values
depend on the kind of propagators at stake, i.e.\ on Nambu indices.
On the other hand, there is an additional effect on
the symmetry factor of the amplitude.
For example, in Ref.~\cite{Nozieres1964}, a factor
$\frac{1}{2}$ arises for each anomalous tadpole, but not for normal tadpoles.
The factor due to equivalent lines also typically depends on Nambu indices.
Such non-trivial dependence on the Nambu indices obscures
the study of Nambu-covariance of the amplitudes contributing
to the many-body Green's functions, and precludes transparent
formal and numerical developments.

In contrast, in the Feynman rules of NCPT, given in Secs.~\ref{subsec:FeynRulesTime}
and~\ref{subsec:FeynRulesEnergy}, the symmetry factor
arises solely from the topology of the un-oriented diagram.
There is no explicit dependence of the symmetry factor on Nambu indices.
With these rules, the amplitude of an un-oriented diagram
is thus clearly decoupled into a Nambu tensor network, and a symmetry (and sign) factor.

\subsection{Gaudin's summation rules}\label{subsec:MatsubaraSum}
Feynman amplitudes, as obtained by applying the rules of the previous subsection,
provide a decomposition of Green's functions
in terms of tensor networks \emph{and} of sums over Matsubara frequencies,
as explicitly shown in Eq.~\eqref{GenericFeynmanAmplitudeEnergy}.
The sums over Matsubara frequencies can be performed exactly by considering
all different orderings of vertices of a given diagram. Such decomposition
gives up to $n!$ terms, with $n$ the number of vertices of the diagram.
An alternative method is based on applying the residue theorem until
all Matsubara sums have been replaced by complex integrals.

A third, more convenient approach to tackle the Matsubara sums consists in decomposing
a given Feynman amplitude into a sum of contributions, one for each spanning
forest of a diagram, as shown by Gaudin in Ref.~\cite{Gaudin1965}.
Compared to the residue theorem, only diagrammatic considerations are necessary.
Compared to vertex ordering, the number of spanning forests grows much more slowly
with $n$.
For example, the number of spanning forests for a connected diagram
with only two-body vertices is bounded from above by $4^{n-1} \ll n!$.
In Ref.~\cite{Gaudin1965}, Gaudin derived a set of diagrammatic rules
giving the algebraic expression of a Feynman diagram which is obtained
after performing sums over Matsubara frequencies.
His focus was on two-particle-irreducible (2PI) diagrams with dressed propagators at finite $\beta$
in a symmetry-conserving case.
This approach has been exploited several times in a very similar fashion
for different applications~\cite{Guerin1994,Dib1997,Wong2001,Espinosa2004}.
As noted in Ref.~\cite{Blaizot2006}, Gaudin's work on the summation rules
remains relatively unknown, so we recapitulate the general rationale behind these rules
and adapt them to the diagrams of NCPT in the energy representation.
For simplicity, we focus on
connected diagrams without tadpoles nor external lines.
In this case, all spanning forest are connected, i.e.\ they are (by definition) 
spanning trees. The extension to the case of a general diagram is detailed
in~\ref{app:SumGeneralGraph}.

\subsubsection{Rationale}
In general, a contravariant propagator can be expressed in terms of
a spectral function $S^{\mu\nu}(\epsilon)$,
\begin{equation}\label{SpectralRepresentation}
    \mathcal{G}^{\mu\nu}(\omega_m)
        = \int_{-\infty}^{+\infty} \frac{\mathrm{d}\epsilon}{2\pi} \
                \frac{S^{\mu\nu}(\epsilon)}{i\omega_m - \epsilon} \ .
\end{equation}
This so-called spectral representation of a propagator
is discussed, along with other exact properties, in Part~II of this work.
Here we are interested in the unperturbed contravariant propagator
$\mathcal{G}^{(0)\mu\nu}(\omega_m)$, associated to a Hamiltonian $H_0$ which is quadratic
in the Nambu fields. In this case,
the spectral function associated to $\mathcal{G}^{(0)\mu\nu}(\omega_m)$ is expressed as
\begin{equation}
   S^{(0)\mu\nu}(\epsilon)
    = \sum_{n}
            X^{(n) \mu}\bar{X}^{(n) \nu} (2\pi) \ \delta(\epsilon - \epsilon_n) \ , \label{QuadraSpFunctionEnergy}
\end{equation}
where $\epsilon_n$ are quasiparticle energies and $X^{(n) \mu}$,
the associated spectroscopic amplitudes.
By definition, $\epsilon_n$ are the eigenvalues of the matrix made of the
mixed $(1,1)$ coordinates ${U^{\mu}}_{\nu}$.
For simplicity, we assume that these eigenvalues are labelled by
a discrete quantum number $n$ (not to be confused
with the order in perturbation theory) and are non-degenerate.
$X^{(n) \mu}$ and $\bar{X}^{(n)}_{\mu}$ are, respectively,
the coordinates of the associated right and left eigenvectors, i.e.\
\begin{subequations}
\begin{align}
    \sum_{\nu} {U^{\mu}}_{\nu} \ X^{(n) \nu} &= \epsilon_n X^{(n) \mu} \ , \\
    \sum_{\nu} \bar{X}^{(n)}_{\mu} {U^{\mu}}_{\nu} &= \epsilon_n \bar{X}^{(n)}_{\nu}
        \, .
\end{align}
\end{subequations}
These eigenvectors are normalised such that
they form a biorthogonal system, i.e.\
\begin{equation}
    \sum_{\mu} \bar{X}^{(n)}_{\mu} X^{(n') \mu} = \delta_{nn'} \ .
\end{equation}
Note that the antisymmetry of $U_{\mu\nu}$ would in principle provide a relation between the
 left and right eigenvectors.
For clarity, however, we keep a notation that distinguishes them explicitly.

The rationale behind Gaudin's summation rules is the following.
Let $\mathscr{G}_n$ be a connected un-oriented Feynman diagram
with $L$ loops and no tadpoles nor external lines.
The problem is to compute within Eq.~\eqref{GenericFeynmanAmplitudeEnergy}
the following sum
\begin{equation}\label{MatsubSum}
    I\left(\mathscr{G}_n\right) \equiv \frac{1}{\beta^L}\sum_{\omega_{l_1} \dots \omega_{l_L}}
        \prod_{e \in I} -\mathcal{G}^{(0)\lambda\lambda}(\omega_e) \ .
\end{equation}
Replacing each propagator by the corresponding spectral
representation in Eq.~\eqref{SpectralRepresentation}, Eq.~\eqref{MatsubSum} reads
\begin{multline}
    I\left(\mathscr{G}_n\right) =
    \int^{+\infty}_{-\infty}
        \left( \prod_{e \in I }
            S^{(0)\lambda\lambda}(\epsilon_e) \ \mathrm{d}\epsilon_e \right) \\
        \times \frac{1}{\beta^L}\sum_{\omega_{l_1} \dots \omega_{l_L}}
        \left( \prod_{e \in I}
            \frac{1}{\epsilon_e -  i\omega_e}
        \right)
    \ ,
\end{multline}
where $\epsilon_e$ denotes the energy associated to the line $e$.
The problem is thus
reduced to computing
\begin{equation}\label{WeightFactorIntegral}
    \frac{1}{\beta^L}\sum_{\omega_{l_1} \dots \omega_{l_L}}
        \left( \prod_{e \in I }
            \frac{1}{\epsilon_e -  i\omega_e}
        \right) \ ,
\end{equation}
where each $\omega_e$ is a linear combination of $L$ independent Matsubara
frequencies $\omega_{l_i}$, obtained by applying
energy conservation laws at the vertices of $\mathscr{G}_n$.
To compute it, 
the above product is decomposed into partial fractions, 
where each term is associated to one spanning tree $\mathscr{A}$ of $\mathscr{G}_n$.
For each spanning tree, the associated Matsubara sum 
is decoupled and performed analytically.
As a result, $I\left(\mathscr{G}_n\right)$ is decomposed as
\begin{equation}\label{MatsubaraSumDecomposition}
    I\left(\mathscr{G}_n\right)
        =
        \sum_{\mathscr{A} \in \text{Spanning trees}} I\left( \mathscr{A} \right) \ .
\end{equation}
In the case of a dressed propagator, the resulting expression for
$I\left(\mathscr{G}_n\right)$ contains energy integrals of a product
of spectral functions multiplied by the factor in Eq.~\eqref{WeightFactorIntegral}.
In the case of an unperturbed one-body propagator with a spectral function given by
Eq.~\eqref{QuadraSpFunctionEnergy}, the energy integrals simplify
to sums over quasiparticle energies and the spectral functions
$S^{\mu\nu}(\epsilon_e)$ are replaced by the spectroscopic amplitudes
$X^{(n_e) \mu}\bar{X}^{(n_e) \nu}$.
The factor~\eqref{WeightFactorIntegral} remains the same in both cases.
Eventually, $I\left(\mathscr{G}_n\right)$ is plugged back into
Eq.~\eqref{GenericFeynmanAmplitudeEnergy}, which reads generically
\begin{multline}
    \mathcal{A}^{\mu_1 \dots \mu_{2k}}(\omega_{m_1}, \dots, \omega_{m_{2k}})
    = \\
    \frac{(-1)^{n+L} \beta^{C-1}}{S \times 2^{T} \prod_{l=2}^{l_{\text{max}}}(l!)^m} \
    \sum_{\lambda\dots\lambda} \
        v^{(k_1)}_{[\lambda \dots \lambda]}
        \dots
        v^{(k_n)}_{[\lambda \dots \lambda]}\
        \times \
        I\left( \mathscr{G}_n \right)
        \ .
\end{multline}
The sum of Matsubara sums, $I\left( \mathscr{G}_n \right)$, can be evaluated explicitly by following
a set of summation rules that we enumerate below. 

\subsubsection{Summation rules}
We start by defining the complementary diagram, $\mathscr{B}$, of a given spanning tree $\mathscr{A}$ in a diagram  $\mathscr{G}_n$.
The complementary diagram $\mathscr{B}$ is the diagram made of all the vertices
of $\mathscr{G}_n$ and of all the lines of
$\mathscr{G}_n$ that are not present in the set of lines of $\mathscr{A}$. 
Further, we denote the lines belonging to $\mathscr{A}$ as $a,b,\dots$,
lines belonging to $\mathscr{B}$ as $p,q,\dots$, and
a generic line of $\mathscr{G}_n$ by~$e$.
In the following, $\epsilon_{n_e}$ indicates the quasiparticle energy associated to a fermion line $e$.
The function $f(\epsilon)$ is the standard Fermi-Dirac distribution, i.e.\
      $f(\epsilon) \equiv \frac{1}{1 + e^{\beta \epsilon}}$.
      
Let $\mathscr{G}_n$ be a connected diagram of order $n$ without tadpoles
nor external lines. The Matsubara sum $I\left(\mathscr{G}_n\right)$ is obtained
in terms of the spectroscopic amplitudes, $X^{(n_e) \mu}$ and $\bar{X}^{(n_e) \nu}$, and quasiparticle energies, $\epsilon_{n_e}$, as follows:
\begin{enumerate}
  \item Build the set of spanning trees $\mathscr{A}$ of $\mathscr{G}_n$. Associate a complementary diagram $\mathscr{B}$ to each spanning tree. 
  \item Fix an orientation on the diagram, i.e.\ associate a choice of
  direction and an intensity integer $i_e$ to each line such that for each cycle $(p)$,
  associated to the line $p$ of $\mathscr{B}$\footnote{For each
  spanning tree $\mathscr{A}$ and line $p$ in $\mathscr{B}$, the cycle $(p)$ is uniquely defined
  as the one obtained when adding $p$ to $\mathscr{A}$.}, its total orientation
  $N_p$ is different from $0$. The total orientation $N_p$ of a cycle $(p)$ is
  obtained by adding 
  $i_e$ for each line $e$ of the cycle oriented in the same way as $p$,
  and
  by subtracting $i_e$ for each line $e$ of the cycle oriented in
  the opposite way to $p$.
  \item The Matsubara sum $I\left(\mathscr{G}_n\right)$ is the sum of
  the contributions $I\left(\mathscr{A}\right)$ associated to each spanning
  tree $\mathscr{A}$. $I\left(\mathscr{A}\right)$ is the sum over
  quasiparticle energies $\epsilon_{n_e}$ of a product of statistical factors,
  one for each line
  $p$ in $\mathscr{B}$; of energy denominators, one for each line
  $a$ in $\mathscr{A}$; and of spectroscopic amplitudes, one for each line
  $e$ in $\mathscr{G}_n$. The contribution of one spanning tree
  $I(\mathscr{A})$ is obtained as follows:
    \begin{enumerate}
      \item[$3$.a.]  For each line $p$ in $\mathscr{B}$, multiply by
      the statistical factor $f(-\epsilon_{n_p})$ or $-f(\epsilon_{n_p})$,
      depending on whether $N_p$ is positive or negative,  respectively.
      \item[$3$.b.] For each line $a$ in $\mathscr{A}$, multiply by the energy
      denominator obtained as follows : when removing the line $a$ from
      $\mathscr{A}$ the tree is divided in two sub-trees $\mathscr{A}^+$ and
      $\mathscr{A}^-$ defined such that $a$ is oriented from $\mathscr{A}^-$
      towards $\mathscr{A}^+$.
      The denominator is the sum of $\epsilon_{n_e}$ for each line
      connecting $\mathscr{A}^-$ to $\mathscr{A}^+$, positively or negatively
      counted when going in the same or opposite direction as $a$, respectively.
      Note that by definition the denominator contains
      a $+\epsilon_{n_a}$ term.
      \item[$3$.c.] For each line $e$ in $\mathscr{G}_n$ multiply by a factor
      $X^{(n_e) \mu_e} \bar{X}^{(n_e) \nu_e}$ and sum over $n_e$
      which indexes the quasiparticle energies and spectroscopic amplitudes
      associated to the line $e$.
    \end{enumerate}
\end{enumerate}
Reading and writing conventions are the same as in Sec.~\ref{subsec:FeynRulesEnergy}
given the orientation fixed in the summation rule $2$
(without taking into account the chosen intensities).
Examples of application of the above rules are given in Sec.~\ref{subsec:Examples}.
Following Gaudin's summation rules, the Feynman amplitude associated to
$\mathscr{G}_n$ reads
\begin{multline}\label{GenericFeynmanSummedAmplitudeEnergy}
    \mathcal{A}^{\mu_1 \dots \mu_{2k}}(\omega_{m_1}, \dots, \omega_{m_{2k}})
    =
    \frac{(-1)^{n+L} \beta^{C-1}}{S \times 2^{T} \prod_{l=2}^{l_{\text{max}}}(l!)^m} \\
    \times \sum_{\lambda\dots\lambda}
        v^{(k_1)}_{[\lambda \dots \lambda]}
        \dots
        v^{(k_n)}_{[\lambda \dots \lambda]} \\
        \times \sum_{n_e \dots n_e} \
        \prod_{e \in I} X^{(n_e) \lambda} \bar{X}^{(n_e) \lambda} % \\
        \ \sum_{\mathscr{A}} \frac{N\left(\mathscr{B}\right)}{D\left(\mathscr{A}\right)}
        \ ,
\end{multline}
where $n_e$ denotes a generic index for quasiparticle energies and spectroscopic
amplitudes associated to a line $e$.
The notation $\mathscr{A}$ denotes a generic spanning tree, $\mathscr{B}$ its
complementary diagram, $N\left(\mathscr{B}\right)$ the numerator obtained
following summation rule $3.a.$ and $D\left(\mathscr{A}\right)$
the denominator obtained following summation rule $3.b$.
Although not displayed in Eq.~\eqref{GenericFeynmanSummedAmplitudeEnergy},
we stress that $N\left(\mathscr{B}\right)$ and $D\left(\mathscr{A}\right)$
depend on the quasiparticle energies $\epsilon_{n_e}$.

\subsubsection{Discussion}
The summation rules just discussed decompose $I\left(\mathscr{G}_n\right)$
into a sum of contributions $I\left(\mathscr{A}\right)$, one for each spanning tree as in Eq.~\eqref{MatsubaraSumDecomposition}.
When looking at the contribution for one spanning tree $\mathscr{A}$,
several infrared divergences might appear, i.e.\ divergences due to a denominator
(computed in rule $3.b.$) converging to zero.
The problem is even worse in the case of a translation invariant system
where some denominators are \emph{only} evaluated when they are vanishing
because of momentum conservation at vertices and of the unperturbed propagator
being diagonal in momentum.
This was originally the reason why only 2PI diagrams were considered
in Ref.~\cite{Gaudin1965}. Since then, the origin of these divergences
has been pinpointed to stem from the splitting of $I(\mathscr{G}_n)$
into spanning tree contributions, Eq.~\eqref{MatsubaraSumDecomposition}.
When the set of spanning tree contributions are added up,
the infrared divergences cancel out.
This cancellation of infrared divergences is briefly discussed in App.~B
of Ref.~\cite{Blaizot2006} and in Chaps.~2-3 of  Ref.~\cite{Reinosa2004}.
It was shown there that the infrared divergences appearing in Gaudin's formulae
are artificial, and always cancel out when combining several spanning tree
contributions.

Typically, the limit where a denominator, made of a linear combination
of quasiparticle energies, goes to zero is cancelled out by the combination
of several numerators (from different spanning trees) going also to zero -
so that the overall ratio converges to a well-defined finite value.
In practical implementations, numerical instabilities can be avoided
by replacing the ratio by a Taylor expansion of the numerator simplified
by the denominator. In doing so, derivatives of the statistical distributions appear.
We study one of these examples in Sec.~\ref{subsec:Examples}: a diagram at third 
order which is not 2PI. It is explicitly shown there how spanning tree contributions
get their infrared divergences cancelled out.
We also give an asymptotically valid expression around a removable singularity
depending on the derivative of the Fermi-Dirac distribution.
Obtaining explicit rules which systematically remove those artificial infrared
divergences would be interesting to automatically and efficiently generate
expressions free of numerical instabilities.
Such refinements are beyond the scope of this article.

\subsection{Examples}\label{subsec:Examples}
In this section, we give some basic and illustrative applications of the 
Feynman rules of NCPT. Contributions at first, second and third order of
the contravariant one-body Green's function are worked out explicitly.
The factorisation of energy denominators and the cancellation
of infrared divergences is then briefly discussed.

\subsubsection{First order perturbations}
\begin{figure}[t]
  \centering
  $-\mathcal{G}^{\mu\nu}_{(1)}(\omega_m) =$
  \hspace{-0.7cm}\parbox{60pt}{\begin{fmffile}{1stOrderGF_2body}
  \begin{fmfgraph*}(50,100)
    \fmfkeep{1stOrderGF_2body}
  \fmfbottom{i}
  \fmftop{o}
  \fmf{plain, tension=1, tag=1}{v1,o}
  \fmf{plain, tension=0.9, tag=2}{v1,v1}
  \fmf{plain, tension=1, tag=3}{i,v1}
  \fmfv{d.shape=circle,d.filled=full,d.size=3thick,l.angle=180,l.dist=8thick}{v1}
  \fmfposition
  \fmfipath{p[]}
  \fmfiset{p1}{vpath1(__v1,__o)}
  \fmfiset{p2}{vpath2(__v1,__v1)}
  \fmfiset{p3}{vpath3(__i,__v1)}
  \fmfiv{label=$\mu$,l.dist=2thick,l.angle=170}{point 14length(p1)/15 of p1}
  \fmfiv{label=$\lambda$,l.dist=2thick,l.angle=170}{point length(p1)/5 of p1}
  \fmfiv{label=$\lambda_2$,l.dist=1.6thick,l.angle=100}{point 14length(p2)/15 of p2}
  \fmfiv{label=$\lambda_3$,l.dist=1.6thick,l.angle=-100}{point 1length(p2)/15 of p2}
  \fmfiv{label=$\lambda'$,l.dist=1thick,l.angle=170}{point 4length(p3)/5 of p3}
  \fmfiv{label=$\nu$,l.dist=2thick,l.angle=170}{point 1length(p3)/15 of p3}

  \fmfiv{label=$\uparrow \omega_l$,l.dist=1thick,l.angle=0}{point length(p2)/2 of p2}
  \fmfiv{label=$\uparrow \omega_m$,l.dist=1thick,l.angle=0}{point 15length(p1)/15 of p1}
  \fmfiv{label=$\uparrow \omega_m$,l.dist=1thick,l.angle=0}{point 0length(p3)/15 of p3}
\end{fmfgraph*}
\end{fmffile}}
  \hspace{1cm}$+ \dots +$\hspace{-2.2cm}
  \parbox{150pt}{\begin{fmffile}{1stOrderGF_kbody}
  \begin{fmfgraph*}(150,100)
    \fmfkeep{1stOrderGF_kbody}
  \fmfbottom{i}
  \fmftop{o}
  \fmfright{r1,r2,r3}
  \fmf{plain, tension=1, tag=1}{v1,o}
  \fmf{plain, tension=1, tag=4}{i,v1}
  \fmfv{d.shape=circle,d.filled=full,d.size=3thick,l.angle=180,l.dist=8thick}{v1}
  \fmffreeze
  \fmfipath{p[]}
  \fmfiset{p1}{vpath1(__v1,__o)}
  \fmfiset{p2}{vloc(__v1){dir 5} .. tension 1
                .. vloc(__r3) shifted (-7mm,-5mm) .. tension 1
                .. vloc(__v1){dir -100}}
  \fmfiset{p3}{vloc(__v1){dir -80} .. tension 1
                .. vloc(__r1) shifted (-7mm,5mm)  .. tension 1
                .. vloc(__v1){dir 175}}
  \fmfiset{p4}{vpath4(__i,__v1)}
  \fmfiset{p5}{fullcircle scaled .5w shifted (.5w,.5h)}
  \fmfi{plain, tag=2}{p2}
  \fmfi{plain, tag=3}{p3}
  \fmfi{dots, tag=5}{subpath (-length(p5)/25,length(p5)/25) of p5}
  \fmfiv{label=$\mu$,l.dist=2thick,l.angle=170}{point 14length(p1)/15 of p1}
  \fmfiv{label=$\lambda$,l.dist=2thick,l.angle=170}{point length(p1)/5 of p1}
  \fmfiv{label=$\lambda_2$,l.dist=1.4thick,l.angle=20}{point 13length(p2)/15 of p2}
  \fmfiv{label=$\lambda_3$,l.dist=1.4thick,l.angle=90}{point 2length(p2)/15 of p2}
  \fmfiv{label=$\lambda_{2k-2}$,
         l.dist=1thick,l.angle=-170}{point 12length(p3)/15 of p3}
  \fmfiv{label=$\lambda_{2k-1}$,
         l.dist=1thick,l.angle=10}{point 2length(p3)/15 of p3}
  \fmfiv{label=$\lambda'$,l.dist=1thick,l.angle=170}{point 4length(p4)/5 of p4}
  \fmfiv{label=$\nu$,l.dist=2thick,l.angle=170}{point 1length(p4)/15 of p4}

  \fmfiv{label=$\uparrow \omega_{l_1}$,l.dist=1thick,l.angle=0}{point 4length(p2)/10 of p2}
  \fmfiv{label=$\uparrow \omega_{l_{k-1}}$,l.dist=1thick,l.angle=0}{point 6length(p2)/10 of p3}
  \fmfiv{label=$\uparrow \omega_m$,l.dist=1thick,l.angle=0}{point 15length(p1)/15 of p1}
  \fmfiv{label=$\uparrow \omega_m$,l.dist=1thick,l.angle=0}{point 0length(p4)/15 of p4}
\end{fmfgraph*}
\end{fmffile}}
\vspace{0.5cm}
\caption{Labelled diagrams contributing to the propagator at first
order with $2$- up to $k$-body interactions.
The orientation convention for the energy flow is also made explicit.}
\label{fig:1stOrderGF}
\end{figure}
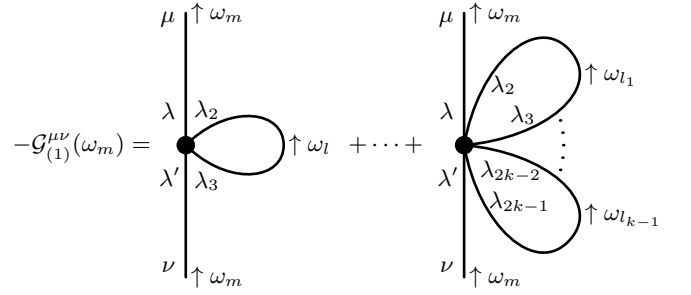

As a first example, we consider the system and partitioning defined by
\begin{align*}
    H_0 &\equiv
      \frac{1}{2} \sum_{\mu\nu} U_{\mu\nu} \mathrm{A}^\mu \mathrm{A}^\nu \ , \\
    H_1 &\equiv
      \sum^{k_{\text{max}}}_{k=2} \ \frac{1}{(2k)!}
                        \sum_{\mu_1 \dots \mu_{2k}}
                              v^{(k)}_{\mu_1 \dots \mu_{2k}} \
                              \mathrm{A}^{\mu_1} \dots \mathrm{A}^{\mu_{2k}} \ .
  \end{align*}
At first order in terms of number of vertices, the set of diagrams contributing
is shown in Fig.~\ref{fig:1stOrderGF}.
If $k_{\text{max}} = 2$, the Feynman amplitude
$\mathcal{A}^{\mu\nu}_{(1)}(\omega_m)$ contributing to
$-\mathcal{G}^{\mu\nu}_{(1)}(\omega_m)$
reads
\begin{align}\label{1stOrderkmax2}
   -\mathcal{A}^{\mu\nu}_{(1)}(\omega_m)
   &=
   \sum_{\lambda \lambda'}
    \mathcal{G}^{(0)\mu\lambda}(\omega_m) \nonumber \\
    &\phantom{=} \times \left[
        \frac{1}{2}
        \sum_{\lambda_2 \lambda_3}
            v^{(2)}_{[\lambda \dot{\lambda}_2 \dot{\lambda}_3 \lambda']}
            \frac{1}{\beta} \sum_{\omega_l} -\mathcal{G}^{(0)\lambda_2\lambda_3}(\omega_l) e^{-i\omega_l \eta}
    \right] \nonumber \\
    &\phantom{=} \times \ \mathcal{G}^{(0)\lambda'\nu}(\omega_m) \ .
\end{align}
Using Eq.~\eqref{tadpoleIdentity}, the Matsubara sum associated
to the tadpole can be performed and Eq.~\eqref{1stOrderkmax2} becomes
\begin{align}\label{1stOrderkmax2Summed}
   -\mathcal{A}^{\mu\nu}_{(1)}(\omega_m)
   &=
   \sum_{\lambda \lambda'}
    \mathcal{G}^{(0)\mu\lambda}(\omega_m) \nonumber \\
    &\phantom{\sum} \times
    \left[
        \frac{1}{2}
        \sum_{\lambda_2 \lambda_3}
            v^{(2)}_{[\lambda \dot{\lambda}_2 \dot{\lambda}_3 \lambda']}
%    \right. \nonumber \\
%    &\phantom{\sum \times \sum v} \times
%    \left.
            \sum_{n_t} f(-\epsilon_{n_t}) X^{(n_t) \lambda_2} \bar{X}^{(n_t) \lambda_3}
    \right] \nonumber \\
    &\phantom{\sum} \times \ \mathcal{G}^{(0)\lambda'\nu}(\omega_m) \ ,
\end{align}
where $n_t$ indexes quasiparticle energies and their associated
spectroscopic amplitudes. 

For a general $k_{\text{max}}$-body interaction, the 
same first-order contribution reads
\begin{align}\label{1stOrder}
   &-\mathcal{A}^{\mu\nu}_{(1)}(\omega_m)
   = \nonumber \\
   &\sum_{\lambda \lambda'}
    \mathcal{G}^{(0)\mu\lambda}(\omega_m)
    \left\{
    \sum^{k_{\text{max}}}_{k=2}
    \left[
        \frac{1}{2^{k-1}(k-1)!}
        \sum_{\lambda_2 \dots \lambda_{2k-1}}
        v^{(k)}_{[\lambda \dot{\lambda}_2 \dot{\lambda}_3
                  \ddot{\lambda}_4 \ddot{\lambda}_5 \dots \lambda']}
    \right. \right. \nonumber \\
    &\left. \left.
        \times \prod^{k-1}_{j = 1}
        \left(
            \frac{1}{\beta} \sum_{\omega_{l_j}} -\mathcal{G}^{(0)\lambda_{2j}\lambda_{2j+1}}(\omega_{l_j})
                                                e^{-i\omega_{l_j} \eta_j}
        \right)
    \right]
    \right\} \times \ \mathcal{G}^{(0)\lambda'\nu}(\omega_m) \ .
\end{align}
In Eq.~\eqref{1stOrder}, the $k-1$ different tadpoles
are denoted with indices dotted a different number of times.
Just like for Eq.~\eqref{1stOrderkmax2Summed}, Matsubara sums
can be explicitly performed so that $-\mathcal{A}^{\mu\nu}_{(1)}(\omega_m)$ reads
\begin{align}\label{1stOrderSummed}
   &-\mathcal{A}^{\mu\nu}_{(1)}(\omega_m)
   = \nonumber \\
   &\sum_{\lambda \lambda'}
    \mathcal{G}^{(0)\mu\lambda}(\omega_m)
    \left\{
    \sum^{k_{\text{max}}}_{k=2}
    \left[
        \frac{1}{2^{k-1}(k-1)!}
        \sum_{\lambda_2 \dots \lambda_{2k-1}}
        v^{(k)}_{[\lambda \dot{\lambda}_2 \dot{\lambda}_3
                  \ddot{\lambda}_4 \ddot{\lambda}_5 \dots \lambda']}
    \right. \right. \nonumber \\
    &\left. \left.
        \times \prod^{k-1}_{j = 1}
        \left(
            \sum_{n_j} f(-\epsilon_{n_j}) X^{(n_j) \lambda_{2j}} \bar{X}^{(n_j) \lambda_{2j+1}}
        \right)
    \right]
    \right\} \mathcal{G}^{(0)\lambda'\nu}(\omega_m) \ ,
\end{align}
where $n_j$ indexes the $j^{\text{th}}$ tadpole quasiparticle energies
and associated spectroscopic amplitudes.

Eqs.~\eqref{1stOrderkmax2} and~\eqref{1stOrder},
when stripped off the contraction with external propagators,
are closely related to the self-consistent HFB self-energies
with up to $k_{\text{max}}$-body interactions.
If Eq.~\eqref{1stOrderkmax2} is only slightly more simple than the HFB equation
with a two-body interaction, Eq.~\eqref{1stOrder} is remarkably simple
and compact compared to what one would expect from the HFB equations
with arbitrary large $k$-body interactions. 
Note, in particular, that different contractions with normal and anomalous lines
would have to be considered explicitly.
In general, the more complex the interaction is (the higher the $k$),
the more powerful is the simplification coming from NCPT.

\subsubsection{Second order perturbations}
In addition to different types of interaction, NCPT also facilitates the development of 
higher-order perturbative approximations. 
To illustrate this point, we consider here another example. This time,
the Hamiltonian and its partitions are defined by
\begin{subequations}
\begin{equation}
  H \equiv H^{\text{HFB}}_0 + H^{\text{HFB}}_1
\end{equation}
where
\begin{align}
    H^{\text{HFB}}_0 &\equiv
      H'_{0}
      +
      \frac{1}{2} \sum_{\mu\nu} U_{\mu\nu} \ \mathrm{A}^\mu \mathrm{A}^\nu \ , \\
    H^{\text{HFB}}_1 &\equiv
    - H'_{0} +
      \frac{1}{4!}
      \sum_{\mu_1\mu_2\mu_3\mu_4}
      v^{(2)}_{\mu_1\mu_2\mu_3\mu_4} \
      \mathrm{A}^{\mu_1} \mathrm{A}^{\mu_2} \mathrm{A}^{\mu_3} \mathrm{A}^{\mu_4} \ .
\end{align}
\end{subequations}
Here, $H'_{0}$ is a quadratic Hamiltonian correction to the
reference Hamiltonian $H_0$ of Eq.~\eqref{GeneralPartitionPT}.
This correction provides the standard HFB partitioning,
which we employ here for conciseness.
In this particular case, any diagram with a tadpole
is cancelled out by the same diagram where the tadpole is replaced by
the quadratic perturbation $-H'_{0}$~\cite{Negele1987}.
More details on the unperturbed propagator associated to the HFB partitioning
are given in Part~II.
We also assume to have only two-body interactions for simplicity.
As a slight abuse of notation, we keep denoting the unperturbed propagator,
the quasiparticle energies and the spectroscopic amplitudes respectively as
$\mathcal{G}^{(0)\mu\nu}(\omega_e)$, $\epsilon_n$ and $X^{(n)\mu}$,
$\bar{X}^{(n)\nu}$, although here they are to be understood as those associated to
the HFB mean-field.

\begin{figure}[t]
  \centering
  $-\mathcal{G}^{\mu\nu}_{(2+3)}(\omega_m) =$
  \hspace{-0.2cm}\parbox{50pt}{\begin{fmffile}{2ndOrderGF}
    \begin{fmfgraph*}(50,100)
       \fmfkeep{2ndOrderGF}
    \fmfbottom{i}
    \fmftop{o}
    \fmf{plain, tension=1.5, tag=1}{i,v1}
    \fmf{plain, tag=3}{v1,v2}
    \fmf{plain, tension=1.5, tag=5}{v2,o}
    \fmffreeze
    \fmf{plain, right=0.7, tag=4}{v1,v2}
    \fmffreeze
    \fmf{plain, right=0.7, tag=2}{v2,v1}
    \fmfv{d.shape=circle,d.filled=full,d.size=3thick}{v1}
    \fmfv{d.shape=circle,d.filled=full,d.size=3thick}{v2}
    \fmflabel{\Large$\mathscr{G}_2$}{i}
\end{fmfgraph*}
\end{fmffile}} \hspace{-0.3cm}$+$
  \hspace{-0.3cm}\parbox{50pt}{\begin{fmffile}{3rdOrderGF_2PI}
  \begin{fmfgraph*}(50,100)
    \fmfkeep{3rdOrderGF_2PI}
    \fmfbottom{i}
    \fmftop{o}
    \fmf{plain, tension=2.5, tag=1}{i,v1}
    \fmf{plain, right=0.75, tag=2}{v1,v2}
    \fmf{plain, right=0.75, tag=3}{v2,v3}
    \fmf{plain, right=0.75, tag=4}{v2,v1}
    \fmf{plain, right=0.75, tag=5}{v3,v2}
    \fmf{plain, tension=2.5, tag=6}{v3,o}
    \fmffreeze
    \fmf{plain, right=0.75, tag=7}{v1,v3}
    \fmfv{d.shape=circle,d.filled=full,d.size=3thick}{v1}
    \fmfv{d.shape=circle,d.filled=full,d.size=3thick}{v2}
    \fmfv{d.shape=circle,d.filled=full,d.size=3thick}{v3}
    \fmflabel{\Large$\mathscr{G}_3$}{i}
\end{fmfgraph*}
\end{fmffile}} $+$
  \hspace{0.1cm}\parbox{50pt}{\begin{fmffile}{3rdOrderGF_2PR}
  \begin{fmfgraph*}(40,125)
    \fmfkeep{3rdOrderGF_2PR}
    \fmfbottom{i1,i2}
    \fmftop{o1,o2}
    \fmf{plain, tag=1}{i1,v1}
    \fmf{plain, tag=2}{v1,o1}
    \fmf{phantom, tension=2}{i2,v2}
    \fmf{plain}{v2,v3}
    \fmf{phantom, tension=2}{v3,o2}
    \fmffreeze
    \fmf{plain, tension=1}{v1,v2}
    \fmf{plain, tension=1}{v3,v1}
    \fmffreeze
    \fmf{plain, right=0.5}{v2,v3}
    \fmffreeze
    \fmf{plain, right=0.5}{v3,v2}
    \fmfv{d.shape=circle,d.filled=full,d.size=3thick}{v1}
    \fmfv{d.shape=circle,d.filled=full,d.size=3thick}{v2}
    \fmfv{d.shape=circle,d.filled=full,d.size=3thick}{v3}
    \fmflabel{\Large$ \mathscr{G}_3'$}{i1}
\end{fmfgraph*}
\end{fmffile}}
\vspace{0.5cm}
\caption{Diagrams contributing to the one-body Green's function at second and third
orders for a two-body interaction and a HFB partitioning.}
\label{fig:2nd3rdOrderGF}
\end{figure}
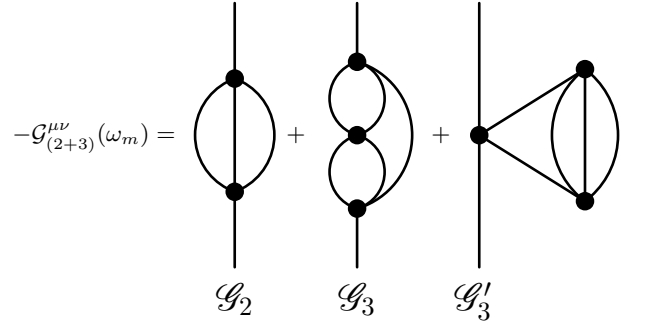

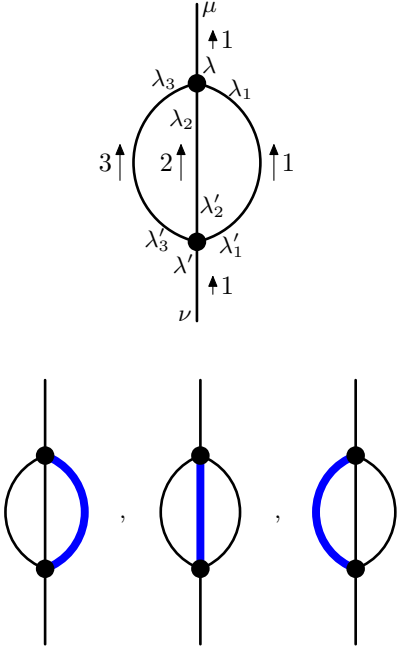
\begin{figure}[t]
  \centering
  \hspace{-0.2cm}\parbox{60pt}{\begin{fmffile}{2ndOrderGF_labelled}
    \begin{fmfgraph*}(60,120)
       \fmfkeep{2ndOrderGF_labelled}
    \fmfbottom{i}
    \fmftop{o}
    \fmf{plain, tension=2, tag=1}{i,v1}
    \fmf{plain, tag=3}{v1,v2}
    \fmf{plain, tension=2, tag=5}{v2,o}
    \fmffreeze
    \fmf{plain, right=0.8, tag=4}{v1,v2}
    \fmffreeze
    \fmf{plain, right=0.8, tag=2}{v2,v1}
    \fmfv{d.shape=circle,d.filled=full,d.size=3thick}{v1}
    \fmfv{d.shape=circle,d.filled=full,d.size=3thick}{v2}
    \fmffreeze
    % Flow arrows
    \Farrow{a}{right}{rt}{$1$}{v1,v2}{29}
    \Farrow{b}{left}{lft}{$2$}{v1,v2}{6}
    \Farrow{c}{left}{lft}{$3$}{v1,v2}{29}
    \Farrow{d}{right}{rt}{$1$}{v2,o}{6}
    \Farrow{e}{right}{rt}{$1$}{i,v1}{6}
    % Vertex labels
    \fmfposition
    \fmfipath{p[]}
    \fmfiset{p1}{vpath1(__i,__v1)}
    \fmfiset{p2}{vpath2(__v2,__v1)}
    \fmfiset{p3}{vpath3(__v1,__v2)}
    \fmfiset{p4}{vpath4(__v1,__v2)}
    \fmfiset{p5}{vpath5(__v2,__o)}

    \fmfiv{label=$\nu$,l.dist=1thick,l.angle=170}{point 1length(p1)/15 of p1}
    \fmfiv{label=$\lambda'$,l.dist=0.5thick,l.angle=170}{point 4length(p1)/5 of p1}
    \fmfiv{label=$\lambda_2'$,l.dist=0.5thick,l.angle=0}{point length(p3)/5 of p3}
    \fmfiv{label=$\lambda_3'$,l.dist=2thick,l.angle=200}{point 19length(p2)/20 of p2}
    \fmfiv{label=$\lambda_1'$,l.dist=1.5thick,l.angle=-80}{point 2length(p4)/14 of p4}

    \fmfiv{label=$\lambda_2$,l.dist=0.5thick,l.angle=170}{point 4length(p3)/5 of p3}
    \fmfiv{label=$\lambda_3$,l.dist=2thick,l.angle=90}{point 2length(p2)/14 of p2}
    \fmfiv{label=$\lambda_1$,l.dist=2thick,l.angle=90}{point 11length(p4)/14 of p4}
    \fmfiv{label=$\lambda$,l.dist=1thick,l.angle=0}{point 2length(p5)/14 of p5}
    \fmfiv{label=$\mu$,l.dist=1thick,l.angle=0}{point 14length(p5)/15 of p5}

\end{fmfgraph*}
\end{fmffile}}

\vspace{0.75cm}

\parbox{50pt}{\begin{fmffile}{2ndOrderGF_Tree1}
    \begin{fmfgraph*}(50,100)
       \fmfkeep{2ndOrderGF_Tree1}
    \fmfbottom{i}
    \fmftop{o}
    \fmf{plain, tension=1.5, tag=1}{i,v1}
    \fmf{plain, tag=3}{v1,v2}
    \fmf{plain, tension=1.5, tag=5}{v2,o}
    \fmffreeze
    \fmf{plain, right=0.7, tag=4, fore=blue, width=3thin}{v1,v2}
    \fmffreeze
    \fmf{plain, right=0.7, tag=2}{v2,v1}
    \fmfv{d.shape=circle,d.filled=full,d.size=3thick}{v1}
    \fmfv{d.shape=circle,d.filled=full,d.size=3thick}{v2}
\end{fmfgraph*}
\end{fmffile}}
,
\parbox{50pt}{\begin{fmffile}{2ndOrderGF_Tree2}
    \begin{fmfgraph*}(50,100)
       \fmfkeep{2ndOrderGF_Tree2}
    \fmfbottom{i}
    \fmftop{o}
    \fmf{plain, tension=1.5, tag=1}{i,v1}
    \fmf{plain, tag=3, fore=blue, width=3thin}{v1,v2}
    \fmf{plain, tension=1.5, tag=5}{v2,o}
    \fmffreeze
    \fmf{plain, right=0.7, tag=4}{v1,v2}
    \fmffreeze
    \fmf{plain, right=0.7, tag=2}{v2,v1}
    \fmfv{d.shape=circle,d.filled=full,d.size=3thick}{v1}
    \fmfv{d.shape=circle,d.filled=full,d.size=3thick}{v2}
\end{fmfgraph*}
\end{fmffile}}
,
\parbox{50pt}{\begin{fmffile}{2ndOrderGF_Tree3}
    \begin{fmfgraph*}(50,100)
       \fmfkeep{2ndOrderGF_Tree3}
    \fmfbottom{i}
    \fmftop{o}
    \fmf{plain, tension=1.5, tag=1}{i,v1}
    \fmf{plain, tag=3}{v1,v2}
    \fmf{plain, tension=1.5, tag=5}{v2,o}
    \fmffreeze
    \fmf{plain, right=0.7, tag=4}{v1,v2}
    \fmffreeze
    \fmf{plain, right=0.7, tag=2 ,fore=blue, width=3thin}{v2,v1}
    \fmfv{d.shape=circle,d.filled=full,d.size=3thick}{v1}
    \fmfv{d.shape=circle,d.filled=full,d.size=3thick}{v2}
\end{fmfgraph*}
\end{fmffile}}
\vspace{0.5cm}
\caption{Top: second-order labelled diagram $\mathscr{G}_2$
contributing to the propagator. The chosen orientations and strengths
are given explicitly. We recall that, by convention, the energy flows positively
when following the chosen orientation convention (without taking into account
the chosen intensities).
Bottom: the three spanning trees of internal lines are shown with bold blue lines.
Thin black lines are part of the corresponding complementary diagrams.}
\label{fig:tree2ndOrder}
\end{figure}

In this setup, only one diagram contributes to the one-body Green's function
at second order, and two, at third order.
We show these three diagrams in Fig.~\ref{fig:2nd3rdOrderGF}.
The Feynman amplitude $\mathcal{A}^{\mu\nu}_{(2)}(\omega_m)$ contributing to
$-\mathcal{G}^{\mu\nu}(\omega_m)$ at second order reads
\begin{align}\label{2ndOrder}
   &-\mathcal{A}^{\mu\nu}_{(2)}(\omega_m)
   = \nonumber \\
   &\sum_{\lambda \lambda'}
    \mathcal{G}^{(0)\mu\lambda}(\omega_m)
    \left\{
        \frac{1}{3!}
        \sum_{\substack{\lambda_1 \lambda_2 \lambda_3 \\ \lambda_1' \lambda_2' \lambda_3'}}
        v^{(2)}_{[\lambda \lambda_1 \lambda_2 \lambda_3]}
        v^{(2)}_{[\lambda_3' \lambda_2' \lambda'_1 \lambda']}
    \right. \nonumber \\
    &\left.
        \times \frac{1}{\beta^2} \sum_{\omega_{l_1} \omega_{l_2}}
        \left(
            \mathcal{G}^{(0)\lambda_{1}\lambda'_{1}}(\omega_{l_1})
            \mathcal{G}^{(0)\lambda_{2}\lambda'_{2}}(\omega_{l_2})
        \right.\right. \nonumber \\
    &\phantom{\times \frac{1}{\beta^2} \sum_{\omega_{l_1} \omega_{l_2}} (\times}
        \left.\left.
            \mathcal{G}^{(0)\lambda_{3}\lambda'_{3}}
                (\omega_{m} - \omega_{l_1} - \omega_{l_2})
        \right)\vphantom{\sum_{\substack{\lambda_1 \\ \lambda_1'}}}
    \right\} \nonumber \\
    &\phantom{\sum \frac{1}{\beta}}
        \times \ \mathcal{G}^{(0)\lambda'\nu}(\omega_m) \ .
\end{align}
The orientation of the energy flow is explicitly shown in the top diagram of Fig.~\ref{fig:tree2ndOrder}.
The Matsubara sum $I\left( \mathscr{G}_2 \right)$ is defined by
\begin{align}
    I\left( \mathscr{G}_2 \right) \equiv
    \frac{-1}{\beta^2} \sum_{\omega_{l_1} \omega_{l_2}}
            &\mathcal{G}^{(0)\lambda_{1}\lambda'_{1}}(\omega_{l_1})
            \mathcal{G}^{(0)\lambda_{2}\lambda'_{2}}(\omega_{l_2}) \nonumber \\
            &\times \mathcal{G}^{(0)\lambda_{3}\lambda'_{3}}
                        (\omega_{m} - \omega_{l_1} - \omega_{l_2}) \ .
\end{align}
Applying Gaudin's summation rules as given in Sec.~\ref{subsec:MatsubaraSum}
and~\ref{app:SumGeneralGraph}, $I\left( \mathscr{G}_2 \right)$ reads
\begin{multline}\label{MatsubaraSummed2ndOrder}
    I\left( \mathscr{G}_2 \right) = \\
    \sum_{n_1 n_2 n_3}
    \frac{f(-\epsilon_{n_3}) f(-\epsilon_{n_2})
        - f(-\epsilon_{n_3}) f(\epsilon_{n_1})
        + f(\epsilon_{n_2})  f(\epsilon_{n_1})}
    {-i \omega_m + \epsilon_{n_1} + \epsilon_{n_2} + \epsilon_{n_3}} \\
    \times
    X^{(n_1)\lambda_1} \bar{X}^{(n_1)\lambda'_1} \
    X^{(n_2)\lambda_2} \bar{X}^{(n_2)\lambda'_2} \
    X^{(n_3)\lambda_3} \bar{X}^{(n_3)\lambda'_3} \ .
\end{multline}
Eq.~\eqref{MatsubaraSummed2ndOrder} is obtained as the sum of the amplitudes
associated to the three spanning trees (bold blue lines) shown in the bottom of Fig.~\ref{fig:tree2ndOrder}.
Note that all three trees
contribute with the same denominator, which is why
only one factorised denominator appears in Eq.~\eqref{MatsubaraSummed2ndOrder}. 

\subsubsection{Third order perturbations}
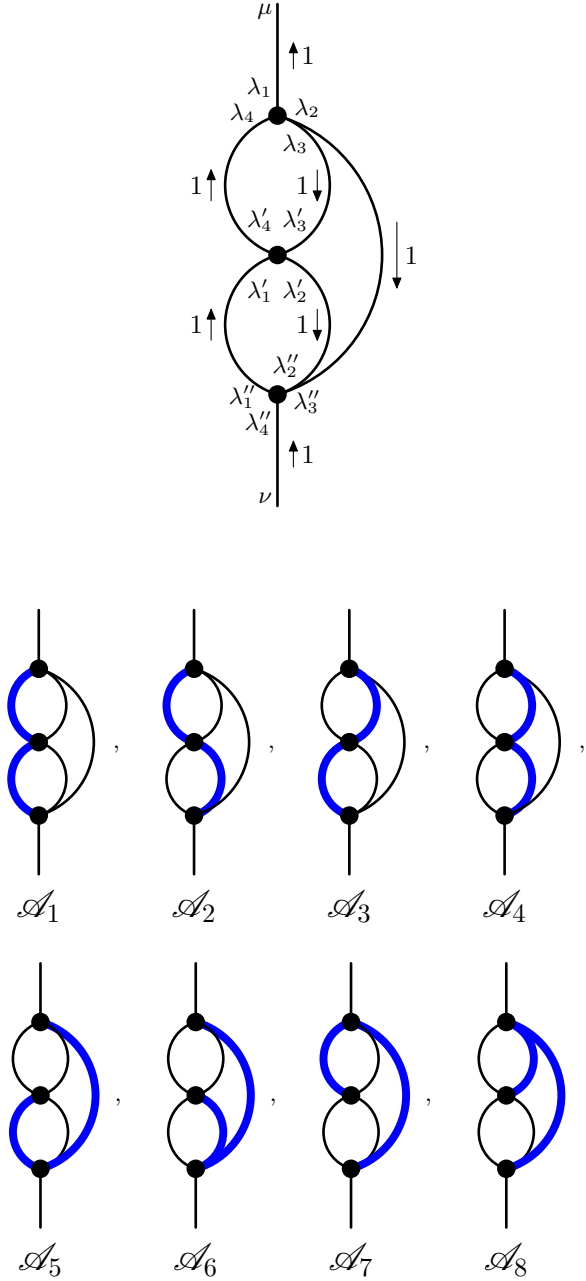
\begin{figure}[tbp]
  \centering
  \parbox{80pt}{\begin{fmffile}{3rdOrderGF_2PI_labelled}
  \begin{fmfgraph*}(50,190)
    \fmfkeep{3rdOrderGF_2PI_labelled}
    \fmfbottom{i}
    \fmftop{o}
    \fmf{plain, tension=2.5, tag=1}{i,v1}
    \fmf{plain, right=0.75, tag=2}{v1,v2}
    \fmf{plain, right=0.75, tag=3}{v2,v3}
    \fmf{plain, right=0.75, tag=4}{v2,v1}
    \fmf{plain, right=0.75, tag=5}{v3,v2}
    \fmf{plain, tension=2.5, tag=6}{v3,o}
    \fmffreeze
    \fmf{plain, right=0.75, tag=7}{v1,v3}
    \fmfv{d.shape=circle,d.filled=full,d.size=3thick}{v1}
    \fmfv{d.shape=circle,d.filled=full,d.size=3thick}{v2}
    \fmfv{d.shape=circle,d.filled=full,d.size=3thick}{v3}
    \fmffreeze
    % Flow arrows
    \Farrow{a}{right}{rt}{$1$}{v3,v1}{45}
    \Farrow{b}{right}{lft}{$1$}{v3,v2}{15}
    \Farrow{c}{right}{lft}{$1$}{v2,v1}{15}
    \Farrow{d}{left}{lft}{$1$}{v1,v2}{25}
    \Farrow{e}{left}{lft}{$1$}{v2,v3}{25}
    \Farrow{f}{right}{rt}{$1$}{i,v1}{6}
    \Farrow{g}{right}{rt}{$1$}{v3,o}{6}
    % Vertex labels
    \fmfposition
    \fmfipath{p[]}
    \fmfiset{p1}{vpath1(__i,__v1)}
    \fmfiset{p2}{vpath2(__v1,__v2)}
    \fmfiset{p3}{vpath3(__v2,__v3)}
    \fmfiset{p4}{vpath4(__v2,__v1)}
    \fmfiset{p5}{vpath5(__v3,__v2)}
    \fmfiset{p6}{vpath6(__v3,__o)}
    \fmfiset{p7}{vpath7(__v1,__v3)}

    \fmfiv{label=$\nu$,l.dist=1thick,l.angle=170}{point 1length(p1)/15 of p1}
    \fmfiv{label=$\lambda_4''$,l.dist=1thick,l.angle=180}{point 12length(p1)/15 of p1}
    \fmfiv{label=$\lambda_3''$,l.dist=1.5thick,l.angle=-90}{point 1length(p7)/15 of p7}
    \fmfiv{label=$\lambda_2''$,l.dist=1thick,l.angle=150}{point 2length(p2)/15 of p2}
    \fmfiv{label=$\lambda_1''$,l.dist=2.5thick,l.angle=-90}{point 12length(p4)/15 of p4}

    \fmfiv{label=$\lambda_4'$,l.dist=3thick,l.angle=90}{point 14length(p5)/15 of p5}
    \fmfiv{label=$\lambda_3'$,l.dist=3thick,l.angle=90}{point 1length(p3)/15 of p3}
    \fmfiv{label=$\lambda_2'$,l.dist=3thick,l.angle=-90}{point 14length(p2)/15 of p2}
    \fmfiv{label=$\lambda_1'$,l.dist=3thick,l.angle=-90}{point 1length(p4)/15 of p4}

    \fmfiv{label=$\lambda_4$,l.dist=3thick,l.angle=90}{point 3length(p5)/15 of p5}
    \fmfiv{label=$\lambda_3$,l.dist=2thick,l.angle=-90}{point 14length(p3)/15 of p3}
    \fmfiv{label=$\lambda_2$,l.dist=2thick,l.angle=90}{point 14length(p7)/15 of p7}
    \fmfiv{label=$\lambda_1$,l.dist=1thick,l.angle=180}{point 3length(p6)/15 of p6}
    \fmfiv{label=$\mu$,l.dist=1thick,l.angle=180}{point 14length(p6)/15 of p6}

\end{fmfgraph*}
\end{fmffile}}

\vspace{1.35cm}

\parbox{50pt}{\begin{fmffile}{3rdOrderGF_Tree1}
    \begin{fmfgraph*}(50,100)
       \fmfkeep{3rdOrderGF_Tree1}
    \fmfbottom{i}
    \fmftop{o}
    \fmf{plain, tension=2.5, tag=1}{i,v1}
    \fmf{plain, right=0.75, tag=2}{v1,v2}
    \fmf{plain, right=0.75, tag=3}{v2,v3}
    \fmf{plain, right=0.75, fore=blue, width=3thin, tag=4}{v2,v1}
    \fmf{plain, right=0.75, fore=blue, width=3thin, tag=5}{v3,v2}
    \fmf{plain, tension=2.5, tag=6}{v3,o}
    \fmffreeze
    \fmf{plain, right=0.75, tag=7}{v1,v3}
    \fmfv{d.shape=circle,d.filled=full,d.size=3thick}{v1}
    \fmfv{d.shape=circle,d.filled=full,d.size=3thick}{v2}
    \fmfv{d.shape=circle,d.filled=full,d.size=3thick}{v3}
    \fmflabel{\Large$ \mathscr{A}_1$}{i}
\end{fmfgraph*}
\end{fmffile}}
,
\parbox{50pt}{\begin{fmffile}{3rdOrderGF_Tree2}
    \begin{fmfgraph*}(50,100)
       \fmfkeep{3rdOrderGF_Tree2}
    \fmfbottom{i}
    \fmftop{o}
    \fmf{plain, tension=2.5, tag=1}{i,v1}
    \fmf{plain, right=0.75, fore=blue, width=3thin, tag=2}{v1,v2}
    \fmf{plain, right=0.75, tag=3}{v2,v3}
    \fmf{plain, right=0.75, tag=4}{v2,v1}
    \fmf{plain, right=0.75, fore=blue, width=3thin, tag=5}{v3,v2}
    \fmf{plain, tension=2.5, tag=6}{v3,o}
    \fmffreeze
    \fmf{plain, right=0.75, tag=7}{v1,v3}
    \fmfv{d.shape=circle,d.filled=full,d.size=3thick}{v1}
    \fmfv{d.shape=circle,d.filled=full,d.size=3thick}{v2}
    \fmfv{d.shape=circle,d.filled=full,d.size=3thick}{v3}
    \fmflabel{\Large$ \mathscr{A}_2$}{i}
\end{fmfgraph*}
\end{fmffile}}
,
\parbox{50pt}{\begin{fmffile}{3rdOrderGF_Tree3}
    \begin{fmfgraph*}(50,100)
       \fmfkeep{3rdOrderGF_Tree3}
    \fmfbottom{i}
    \fmftop{o}
    \fmf{plain, tension=2.5, tag=1}{i,v1}
    \fmf{plain, right=0.75, tag=2}{v1,v2}
    \fmf{plain, right=0.75, fore=blue, width=3thin, tag=3}{v2,v3}
    \fmf{plain, right=0.75, fore=blue, width=3thin, tag=4}{v2,v1}
    \fmf{plain, right=0.75, tag=5}{v3,v2}
    \fmf{plain, tension=2.5, tag=6}{v3,o}
    \fmffreeze
    \fmf{plain, right=0.75, tag=7}{v1,v3}
    \fmfv{d.shape=circle,d.filled=full,d.size=3thick}{v1}
    \fmfv{d.shape=circle,d.filled=full,d.size=3thick}{v2}
    \fmfv{d.shape=circle,d.filled=full,d.size=3thick}{v3}
    \fmflabel{\Large$ \mathscr{A}_3$}{i}
\end{fmfgraph*}
\end{fmffile}}
,
\parbox{50pt}{\begin{fmffile}{3rdOrderGF_Tree4}
    \begin{fmfgraph*}(50,100)
       \fmfkeep{3rdOrderGF_Tree3}
    \fmfbottom{i}
    \fmftop{o}
    \fmf{plain, tension=2.5, tag=1}{i,v1}
    \fmf{plain, right=0.75, fore=blue, width=3thin, tag=2}{v1,v2}
    \fmf{plain, right=0.75, fore=blue, width=3thin, tag=3}{v2,v3}
    \fmf{plain, right=0.75, tag=4}{v2,v1}
    \fmf{plain, right=0.75, tag=5}{v3,v2}
    \fmf{plain, tension=2.5, tag=6}{v3,o}
    \fmffreeze
    \fmf{plain, right=0.75, tag=7}{v1,v3}
    \fmfv{d.shape=circle,d.filled=full,d.size=3thick}{v1}
    \fmfv{d.shape=circle,d.filled=full,d.size=3thick}{v2}
    \fmfv{d.shape=circle,d.filled=full,d.size=3thick}{v3}
    \fmflabel{\Large$ \mathscr{A}_4$}{i}
\end{fmfgraph*}
\end{fmffile}}
,

\vspace{1.15cm}

\parbox{50pt}{\begin{fmffile}{3rdOrderGF_Tree5}
    \begin{fmfgraph*}(50,100)
       \fmfkeep{3rdOrderGF_Tree5}
    \fmfbottom{i}
    \fmftop{o}
    \fmf{plain, tension=2.5, tag=1}{i,v1}
    \fmf{plain, right=0.75, tag=2}{v1,v2}
    \fmf{plain, right=0.75, tag=3}{v2,v3}
    \fmf{plain, right=0.75, fore=blue, width=3thin, tag=4}{v2,v1}
    \fmf{plain, right=0.75, tag=5}{v3,v2}
    \fmf{plain, tension=2.5, tag=6}{v3,o}
    \fmffreeze
    \fmf{plain, right=0.75, fore=blue, width=3thin, tag=7}{v1,v3}
    \fmfv{d.shape=circle,d.filled=full,d.size=3thick}{v1}
    \fmfv{d.shape=circle,d.filled=full,d.size=3thick}{v2}
    \fmfv{d.shape=circle,d.filled=full,d.size=3thick}{v3}
    \fmflabel{\Large$ \mathscr{A}_5$}{i}
\end{fmfgraph*}
\end{fmffile}}
,
\parbox{50pt}{\begin{fmffile}{3rdOrderGF_Tree6}
    \begin{fmfgraph*}(50,100)
       \fmfkeep{3rdOrderGF_Tree6}
    \fmfbottom{i}
    \fmftop{o}
    \fmf{plain, tension=2.5, tag=1}{i,v1}
    \fmf{plain, right=0.75, fore=blue, width=3thin, tag=2}{v1,v2}
    \fmf{plain, right=0.75, tag=3}{v2,v3}
    \fmf{plain, right=0.75, tag=4}{v2,v1}
    \fmf{plain, right=0.75, tag=5}{v3,v2}
    \fmf{plain, tension=2.5, tag=6}{v3,o}
    \fmffreeze
    \fmf{plain, right=0.75, fore=blue, width=3thin, tag=7}{v1,v3}
    \fmfv{d.shape=circle,d.filled=full,d.size=3thick}{v1}
    \fmfv{d.shape=circle,d.filled=full,d.size=3thick}{v2}
    \fmfv{d.shape=circle,d.filled=full,d.size=3thick}{v3}
    \fmflabel{\Large$ \mathscr{A}_6$}{i}
\end{fmfgraph*}
\end{fmffile}}
,
\parbox{50pt}{\begin{fmffile}{3rdOrderGF_Tree7}
    \begin{fmfgraph*}(50,100)
       \fmfkeep{3rdOrderGF_Tree7}
    \fmfbottom{i}
    \fmftop{o}
    \fmf{plain, tension=2.5, tag=1}{i,v1}
    \fmf{plain, right=0.75, tag=2}{v1,v2}
    \fmf{plain, right=0.75, tag=3}{v2,v3}
    \fmf{plain, right=0.75, tag=4}{v2,v1}
    \fmf{plain, right=0.75, fore=blue, width=3thin, tag=5}{v3,v2}
    \fmf{plain, tension=2.5, tag=6}{v3,o}
    \fmffreeze
    \fmf{plain, right=0.75, fore=blue, width=3thin,  tag=7}{v1,v3}
    \fmfv{d.shape=circle,d.filled=full,d.size=3thick}{v1}
    \fmfv{d.shape=circle,d.filled=full,d.size=3thick}{v2}
    \fmfv{d.shape=circle,d.filled=full,d.size=3thick}{v3}
    \fmflabel{\Large$ \mathscr{A}_7$}{i}
\end{fmfgraph*}
\end{fmffile}}
,
\parbox{50pt}{\begin{fmffile}{3rdOrderGF_Tree8}
    \begin{fmfgraph*}(50,100)
       \fmfkeep{3rdOrderGF_Tree8}
    \fmfbottom{i}
    \fmftop{o}
    \fmf{plain, tension=2.5, tag=1}{i,v1}
    \fmf{plain, right=0.75, tag=2}{v1,v2}
    \fmf{plain, right=0.75, fore=blue, width=3thin, tag=3}{v2,v3}
    \fmf{plain, right=0.75, tag=4}{v2,v1}
    \fmf{plain, right=0.75, tag=5}{v3,v2}
    \fmf{plain, tension=2.5, tag=6}{v3,o}
    \fmffreeze
    \fmf{plain, right=0.75, fore=blue, width=3thin, tag=7}{v1,v3}
    \fmfv{d.shape=circle,d.filled=full,d.size=3thick}{v1}
    \fmfv{d.shape=circle,d.filled=full,d.size=3thick}{v2}
    \fmfv{d.shape=circle,d.filled=full,d.size=3thick}{v3}
    \fmflabel{\Large$ \mathscr{A}_8$}{i}
\end{fmfgraph*}
\end{fmffile}}
\vspace{0.95cm}
\caption{
The same as Fig.~\ref{fig:tree2ndOrder} for the diagram $\mathscr{G}_3$. For this diagram,
there are eight distinct spanning trees. 
}
\label{fig:tree3rdOrder2PI}
\end{figure}

The simplifications obtained with NCPT become more
important as one considers higher perturbative orders.
To give a clear illustration,
we derive the Feynman amplitudes contributing to the contravariant
one-body Green's function at third order.
We start with the contribution of the 2PI diagram $\mathscr{G}_3$ of Fig.~\ref{fig:2nd3rdOrderGF}.
The Feynman amplitude $\mathcal{A}^{\mu\nu}_{(3)}(\omega_m)$ contributing to
$-\mathcal{G}^{\mu\nu}(\omega_m)$ 
in this case reads
\begin{align}\label{3rdOrder2PI}
   &-\mathcal{A}^{\mu\nu}_{(3)}(\omega_m)
   =
   -\frac{1}{(2!)^2}
   \sum_{\lambda_1 \lambda_4'}
    \mathcal{G}^{(0)\mu\lambda_1}(\omega_m) \nonumber \\
    &\times \left\{
        \sum_{\substack{\lambda_2 \lambda_3 \lambda_4 \\
                        \lambda_1' \lambda_2' \lambda_3' \lambda_4'\\
                        \lambda_1'' \lambda_2'' \lambda_3''}}
        v^{(2)}_{[\lambda_1 \lambda_2 \lambda_3 \lambda_4]}
        v^{(2)}_{[\lambda_4' \lambda_3' \lambda_2' \lambda_1']}
        v^{(2)}_{[\lambda_1'' \lambda_2'' \lambda_3'' \lambda_4'']} \
        I\left(\mathscr{G}_3\right)
    \right\} \nonumber \\
    &\times \ \mathcal{G}^{(0)\lambda_4''\nu}(\omega_m) \ ,
\end{align}
where $I\left(\mathscr{G}_3\right)$ is the Matsubara sum associated to
$\mathscr{G}_3$, as defined in Eq.~\eqref{MatsubSum}.
The Matsubara sum is computed using Gaudin's summation rules.
There are eight spanning trees within $\mathscr{G}_3$
which are identified in Fig.~\ref{fig:tree3rdOrder2PI}.
% \begin{figure}[t]
The associated numerators and denominators of each contribution read
\begin{subequations}\label{3rdOrderTrees2PI}
\begin{align}
    \mathscr{A}_1 &: \frac{f(-\epsilon_{n_2})f(-\epsilon_{n_3})f(-\epsilon_{n})}
                          {[-i\omega_m + \epsilon_{n_1} - \epsilon_{n_2} - \epsilon_{n}]
                           [-i\omega_m + \epsilon_{n_4} - \epsilon_{n_3} - \epsilon_{n}]} \ , \\
    \mathscr{A}_2 &: \frac{f(-\epsilon_{n_1})f(-\epsilon_{n_3})f(-\epsilon_{n})}
                          {[i\omega_m + \epsilon_{n_2} - \epsilon_{n_1} + \epsilon_{n}]
                           [-i\omega_m + \epsilon_{n_4} - \epsilon_{n_3} - \epsilon_{n}]} \ , \\
    \mathscr{A}_3 &: \frac{f(-\epsilon_{n_2})f(-\epsilon_{n_4})f(-\epsilon_{n})}
                          {[-i\omega_m + \epsilon_{n_1} - \epsilon_{n_2} - \epsilon_{n}]
                           [i\omega_m + \epsilon_{n_3} - \epsilon_{n_4} + \epsilon_{n}]} \ , \\
    \mathscr{A}_4 &: \frac{f(-\epsilon_{n_1})f(-\epsilon_{n_4})\left(-f(\epsilon_{n})\right)}
                          {[i\omega_m + \epsilon_{n_2} - \epsilon_{n_1} + \epsilon_{n}]
                           [i\omega_m + \epsilon_{n_3} - \epsilon_{n_4} + \epsilon_{n}]} \ , \\
    \mathscr{A}_5 &: \frac{f(-\epsilon_{n_4})\left(-f(\epsilon_{n_3})\right)f(-\epsilon_{n_2})}
                          {[\epsilon_{n_1} - \epsilon_{n_2} + \epsilon_{n_3} - \epsilon_{n_4}]
                           [i\omega_m + \epsilon_{n} + \epsilon_{n_3} - \epsilon_{n_4}]} \ , \\
    \mathscr{A}_6 &: \frac{f(-\epsilon_{n_1})f(-\epsilon_{n_4})f(-\epsilon_{n_3})}
                          {[\epsilon_{n_2} - \epsilon_{n_1} + \epsilon_{n_4} - \epsilon_{n_3}]
                           [i\omega_m + \epsilon_{n} + \epsilon_{n_3} - \epsilon_{n_4}]} \ , \\
    \mathscr{A}_7 &: \frac{f(-\epsilon_{n_1})\left(-f(\epsilon_{n_2})\right)f(-\epsilon_{n_3})}
                          {[\epsilon_{n_4} - \epsilon_{n_3} + \epsilon_{n_2} - \epsilon_{n_1}]
                           [i\omega_m + \epsilon_{n} + \epsilon_{n_2} - \epsilon_{n_1}]} \ , \\
    \mathscr{A}_8 &: \frac{f(-\epsilon_{n_1})f(-\epsilon_{n_2})f(-\epsilon_{n_4})}
                          {[\epsilon_{n_3} - \epsilon_{n_4} + \epsilon_{n_1} - \epsilon_{n_2}]
                           [i\omega_m + \epsilon_{n} + \epsilon_{n_2} - \epsilon_{n_1}]} \ .
\end{align}
\end{subequations}
The resulting total Matsubara sum reads, 
\begin{align}\label{MatsubaraSummed3rdOrder2PI}
   &I\left(\mathscr{G}_3\right)
   =
    \sum_{\substack{n_1 \, n_2 \, n_3 \\ n_4 \, n}}
    X^{(n_1)\lambda_1'} \bar{X}^{(n_1)\lambda_1''}\
    X^{(n_2)\lambda_2''} \bar{X}^{(n_2)\lambda_2'}  \nonumber \\
    &\phantom{\sum}\times
        X^{(n_3)\lambda_3'} \bar{X}^{(n_3)\lambda_3}\
        X^{(n_4)\lambda_4} \bar{X}^{(n_4)\lambda_4'}\
        X^{(n)\lambda_3''} \bar{X}^{(n)\lambda_2}\  \nonumber \\
    &\phantom{}\times
    \left\{
        \frac{1}
            {[i\omega_m + \epsilon_{n_2} - \epsilon_{n_1} + \epsilon_{n}]
             [i\omega_m + \epsilon_{n_3} - \epsilon_{n_4} + \epsilon_{n}]}
    \right. \nonumber \\
    &\phantom{ \times (  ( }\times
    \left.
        \left[
            f(-\epsilon_{n}) f(-\epsilon_{n_2}) f(-\epsilon_{n_3})
            -f(\epsilon_{n}) f(-\epsilon_{n_1}) f(-\epsilon_{n_4})
        \right.
    \right. \nonumber \\
    &\phantom{ \times ( ( \times}
    \left.
        \left.
            -f(-\epsilon_{n}) f(-\epsilon_{n_1}) f(-\epsilon_{n_3})
            -f(-\epsilon_{n}) f(-\epsilon_{n_2}) f(-\epsilon_{n_4})
        \right]
    \right. \nonumber \\
    &\phantom{ \times ( }
    \left.
        +
        \frac{1}{\epsilon_{n_2} - \epsilon_{n_1} + \epsilon_{n_4} - \epsilon_{n_3}}
    \right. \nonumber \\
    &\phantom{ \times ( ( }\times
    \left.
        \left[
            \frac{f(-\epsilon_{n_4}) f(\epsilon_{n_3}) f(-\epsilon_{n_2})
                 +f(-\epsilon_{n_4}) f(-\epsilon_{n_3}) f(-\epsilon_{n_1})}
            {i\omega_m + \epsilon_{n} + \epsilon_{n_3} - \epsilon_{n_4}}
        \right.
    \right. \nonumber \\
    &\phantom{\times ( ( \times - }\left.
        \left.
            - \frac{f(-\epsilon_{n_1}) f(\epsilon_{n_2}) f(-\epsilon_{n_3})
                   +f(-\epsilon_{n_1}) f(-\epsilon_{n_2}) f(-\epsilon_{n_4})}
             {i\omega_m + \epsilon_{n} + \epsilon_{n_2} - \epsilon_{n_1}}
        \right]
    \right\}
\end{align}
where denominators have been factorised as done in the second-order case.

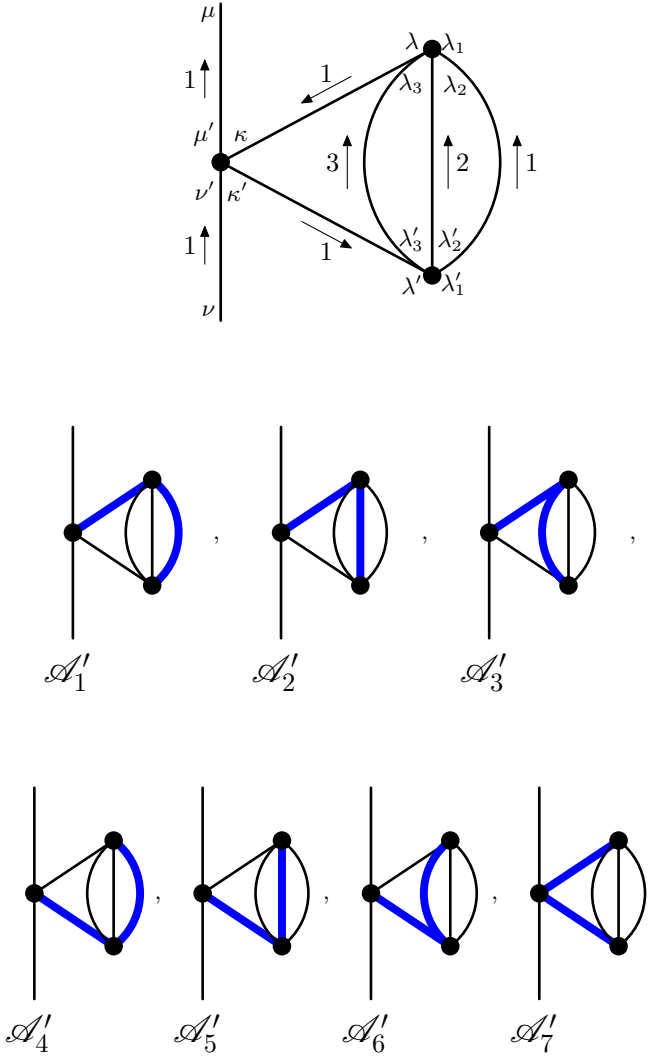
\begin{figure}[t]
  \centering
  \parbox{80pt}{\begin{fmffile}{3rdOrderGF_2PR_labelled}
  \begin{fmfgraph*}(80,150)
    \fmfkeep{3rdOrderGF_2PR_labelled}
    \fmfbottom{i1,i2}
    \fmftop{o1,o2}
    \fmf{plain, tag=1}{i1,v1}
    \fmf{plain, tag=2}{v1,o1}
    \fmf{phantom, tension=5}{i2,v2}
    \fmf{plain, tag=3}{v2,v3}
    \fmf{phantom, tension=5}{v3,o2}
    \fmffreeze
    \fmf{plain, tension=1, tag=4}{v1,v2}
    \fmf{plain, tension=1, tag=5}{v3,v1}
    \fmffreeze
    \fmf{plain, right=0.6, tag=6}{v2,v3}
    \fmffreeze
    \fmf{plain, right=0.6,tag=7}{v3,v2}
    \fmfv{d.shape=circle,d.filled=full,d.size=3thick}{v1}
    \fmfv{d.shape=circle,d.filled=full,d.size=3thick}{v2}
    \fmfv{d.shape=circle,d.filled=full,d.size=3thick}{v3}
    % Flow arrows
    \fmffreeze
    \Farrow{a}{right}{rt}{$1$}{v2,v3}{32}
    \Farrow{b}{right}{rt}{$2$}{v2,v3}{6}
    \Farrow{c}{left}{lft}{$3$}{v2,v3}{32}
    \Farrow{d}{down}{bot}{$1$}{v1,v2}{6}
    \Farrow{e}{up}{top}{$1$}{v3,v1}{6}
    \Farrow{f}{left}{lft}{$1$}{i1,v1}{6}
    \Farrow{g}{left}{lft}{$1$}{v1,o1}{6}
    % Vertex labels
    \fmfipath{p[]}
    \fmfiset{p1}{vpath1(__i1,__v1)}
    \fmfiset{p2}{vpath2(__v1,__o1)}
    \fmfiset{p3}{vpath3(__v2,__v3)}
    \fmfiset{p4}{vpath4(__v1,__v2)}
    \fmfiset{p5}{vpath5(__v3,__v1)}
    \fmfiset{p6}{vpath6(__v2,__v3)}
    \fmfiset{p7}{vpath7(__v3,__v2)}

    \fmfiv{label=$\mu$,l.dist=1thick,l.angle=180}{point 14length(p2)/15 of p2}
    \fmfiv{label=$\mu'$,l.dist=1thick,l.angle=180}{point 2length(p2)/15 of p2}
    \fmfiv{label=$\nu'$,l.dist=1thick,l.angle=180}{point 13length(p1)/15 of p1}
    \fmfiv{label=$\nu$,l.dist=1thick,l.angle=180}{point 1length(p1)/15 of p1}
    \fmfiv{label=$\kappa'$,l.dist=2thick,l.angle=-90}{point 1length(p4)/20 of p4}
    \fmfiv{label=$\kappa$,l.dist=2thick,l.angle=90}{point 14length(p5)/15 of p5}

    \fmfiv{label=$\lambda$,l.dist=2thick,l.angle=90}{point 1length(p5)/15 of p5}
    \fmfiv{label=$\lambda_3$,l.dist=2thick,l.angle=-90}{point 1length(p7)/15 of p7}
    \fmfiv{label=$\lambda_2$,l.dist=2thick,l.angle=0}{point 13length(p3)/15 of p3}
    \fmfiv{label=$\lambda_1$,l.dist=2thick,l.angle=90}{point 14length(p6)/15 of p6}

    \fmfiv{label=$\lambda'$,l.dist=2thick,l.angle=-90}{point 14length(p4)/15 of p4}
    \fmfiv{label=$\lambda_3'$,l.dist=2thick,l.angle=90}{point 14length(p7)/15 of p7}
    \fmfiv{label=$\lambda_2'$,l.dist=1thick,l.angle=0}{point 2length(p3)/15 of p3}
    \fmfiv{label=$\lambda_1'$,l.dist=2thick,l.angle=-90}{point 1length(p6)/15 of p6}

\end{fmfgraph*}
\end{fmffile}}

\vspace{0.5cm}

\parbox{60pt}{\begin{fmffile}{3rdOrderGF_2PR_Tree1}
    \begin{fmfgraph*}(30,100)
       \fmfkeep{3rdOrderGF_2PR_Tree1}
    \fmfbottom{i1,i2}
    \fmftop{o1,o2}
    \fmf{plain, tag=1}{i1,v1}
    \fmf{plain, tag=2}{v1,o1}
    \fmf{phantom, tension=2}{i2,v2}
    \fmf{plain}{v2,v3}
    \fmf{phantom, tension=2}{v3,o2}
    \fmffreeze
    \fmf{plain, tension=1}{v1,v2}
    \fmf{plain, tension=1, fore=blue, width=3thin}{v3,v1}
    \fmffreeze
    \fmf{plain, right=0.5, fore=blue, width=3thin}{v2,v3}
    \fmffreeze
    \fmf{plain, right=0.5}{v3,v2}
    \fmfv{d.shape=circle,d.filled=full,d.size=3thick}{v1}
    \fmfv{d.shape=circle,d.filled=full,d.size=3thick}{v2}
    \fmfv{d.shape=circle,d.filled=full,d.size=3thick}{v3}
    \fmflabel{\Large$ \mathscr{A}_1'$}{i1}
\end{fmfgraph*}
\end{fmffile}}
,
\hspace{2mm}
\parbox{60pt}{\begin{fmffile}{3rdOrderGF_2PR_Tree2}
    \begin{fmfgraph*}(30,100)
       \fmfkeep{3rdOrderGF_2PR_Tree2}
    \fmfbottom{i1,i2}
    \fmftop{o1,o2}
    \fmf{plain, tag=1}{i1,v1}
    \fmf{plain, tag=2}{v1,o1}
    \fmf{phantom, tension=2}{i2,v2}
    \fmf{plain, fore=blue, width=3thin}{v2,v3}
    \fmf{phantom, tension=2}{v3,o2}
    \fmffreeze
    \fmf{plain, tension=1}{v1,v2}
    \fmf{plain, tension=1, fore=blue, width=3thin}{v3,v1}
    \fmffreeze
    \fmf{plain, right=0.5}{v2,v3}
    \fmffreeze
    \fmf{plain, right=0.5}{v3,v2}
    \fmfv{d.shape=circle,d.filled=full,d.size=3thick}{v1}
    \fmfv{d.shape=circle,d.filled=full,d.size=3thick}{v2}
    \fmfv{d.shape=circle,d.filled=full,d.size=3thick}{v3}
    \fmflabel{\Large$ \mathscr{A}_2'$}{i1}
\end{fmfgraph*}
\end{fmffile}}
,
\hspace{2mm}
\parbox{60pt}{\begin{fmffile}{3rdOrderGF_2PR_Tree3}
    \begin{fmfgraph*}(30,100)
       \fmfkeep{3rdOrderGF_2PR_Tree3}
    \fmfbottom{i1,i2}
    \fmftop{o1,o2}
    \fmf{plain, tag=1}{i1,v1}
    \fmf{plain, tag=2}{v1,o1}
    \fmf{phantom, tension=2}{i2,v2}
    \fmf{plain}{v2,v3}
    \fmf{phantom, tension=2}{v3,o2}
    \fmffreeze
    \fmf{plain, tension=1}{v1,v2}
    \fmf{plain, tension=1, fore=blue, width=3thin}{v3,v1}
    \fmffreeze
    \fmf{plain, right=0.5}{v2,v3}
    \fmffreeze
    \fmf{plain, right=0.5, fore=blue, width=3thin}{v3,v2}
    \fmfv{d.shape=circle,d.filled=full,d.size=3thick}{v1}
    \fmfv{d.shape=circle,d.filled=full,d.size=3thick}{v2}
    \fmfv{d.shape=circle,d.filled=full,d.size=3thick}{v3}
    \fmflabel{\Large$ \mathscr{A}_3'$}{i1}
\end{fmfgraph*}
\end{fmffile}}
,

\vspace{1.25cm}

\parbox{45pt}{\begin{fmffile}{3rdOrderGF_2PR_Tree4}
    \begin{fmfgraph*}(30,100)
       \fmfkeep{3rdOrderGF_2PR_Tree3}
    \fmfbottom{i1,i2}
    \fmftop{o1,o2}
    \fmf{plain, tag=1}{i1,v1}
    \fmf{plain, tag=2}{v1,o1}
    \fmf{phantom, tension=2}{i2,v2}
    \fmf{plain}{v2,v3}
    \fmf{phantom, tension=2}{v3,o2}
    \fmffreeze
    \fmf{plain, tension=1, fore=blue, width=3thin}{v1,v2}
    \fmf{plain, tension=1}{v3,v1}
    \fmffreeze
    \fmf{plain, right=0.5, fore=blue, width=3thin}{v2,v3}
    \fmffreeze
    \fmf{plain, right=0.5}{v3,v2}
    \fmfv{d.shape=circle,d.filled=full,d.size=3thick}{v1}
    \fmfv{d.shape=circle,d.filled=full,d.size=3thick}{v2}
    \fmfv{d.shape=circle,d.filled=full,d.size=3thick}{v3}
    \fmflabel{\Large$ \mathscr{A}_4'$}{i1}
\end{fmfgraph*}
\end{fmffile}}
,
\hspace{2mm}
\parbox{45pt}{\begin{fmffile}{3rdOrderGF_2PR_Tree5}
    \begin{fmfgraph*}(30,100)
       \fmfkeep{3rdOrderGF_2PR_Tree5}
    \fmfbottom{i1,i2}
    \fmftop{o1,o2}
    \fmf{plain, tag=1}{i1,v1}
    \fmf{plain, tag=2}{v1,o1}
    \fmf{phantom, tension=2}{i2,v2}
    \fmf{plain, fore=blue, width=3thin}{v2,v3}
    \fmf{phantom, tension=2}{v3,o2}
    \fmffreeze
    \fmf{plain, tension=1, fore=blue, width=3thin}{v1,v2}
    \fmf{plain, tension=1}{v3,v1}
    \fmffreeze
    \fmf{plain, right=0.5}{v2,v3}
    \fmffreeze
    \fmf{plain, right=0.5}{v3,v2}
    \fmfv{d.shape=circle,d.filled=full,d.size=3thick}{v1}
    \fmfv{d.shape=circle,d.filled=full,d.size=3thick}{v2}
    \fmfv{d.shape=circle,d.filled=full,d.size=3thick}{v3}
    \fmflabel{\Large$ \mathscr{A}_5'$}{i1}
\end{fmfgraph*}
\end{fmffile}}
,
\hspace{2mm}
\parbox{45pt}{\begin{fmffile}{3rdOrderGF_2PR_Tree6}
    \begin{fmfgraph*}(30,100)
       \fmfkeep{3rdOrderGF_2PR_Tree6}
    \fmfbottom{i1,i2}
    \fmftop{o1,o2}
    \fmf{plain, tag=1}{i1,v1}
    \fmf{plain, tag=2}{v1,o1}
    \fmf{phantom, tension=2}{i2,v2}
    \fmf{plain}{v2,v3}
    \fmf{phantom, tension=2}{v3,o2}
    \fmffreeze
    \fmf{plain, tension=1, fore=blue, width=3thin}{v1,v2}
    \fmf{plain, tension=1}{v3,v1}
    \fmffreeze
    \fmf{plain, right=0.5}{v2,v3}
    \fmffreeze
    \fmf{plain, right=0.5, fore=blue, width=3thin}{v3,v2}
    \fmfv{d.shape=circle,d.filled=full,d.size=3thick}{v1}
    \fmfv{d.shape=circle,d.filled=full,d.size=3thick}{v2}
    \fmfv{d.shape=circle,d.filled=full,d.size=3thick}{v3}
    \fmflabel{\Large$ \mathscr{A}_6'$}{i1}
\end{fmfgraph*}
\end{fmffile}}
,
\hspace{2mm}
\parbox{45pt}{\begin{fmffile}{3rdOrderGF_2PR_Tree7}
    \begin{fmfgraph*}(30,100)
       \fmfkeep{3rdOrderGF_2PR_Tree7}
    \fmfbottom{i1,i2}
    \fmftop{o1,o2}
    \fmf{plain, tag=1}{i1,v1}
    \fmf{plain, tag=2}{v1,o1}
    \fmf{phantom, tension=2}{i2,v2}
    \fmf{plain}{v2,v3}
    \fmf{phantom, tension=2}{v3,o2}
    \fmffreeze
    \fmf{plain, tension=1, fore=blue, width=3thin}{v1,v2}
    \fmf{plain, tension=1, fore=blue, width=3thin}{v3,v1}
    \fmffreeze
    \fmf{plain, right=0.5}{v2,v3}
    \fmffreeze
    \fmf{plain, right=0.5}{v3,v2}
    \fmfv{d.shape=circle,d.filled=full,d.size=3thick}{v1}
    \fmfv{d.shape=circle,d.filled=full,d.size=3thick}{v2}
    \fmfv{d.shape=circle,d.filled=full,d.size=3thick}{v3}
    \fmflabel{\Large$ \mathscr{A}_7'$}{i1}
\end{fmfgraph*}
\end{fmffile}}
\vspace{1cm}
\caption{
The same as Fig.~\ref{fig:tree2ndOrder} for the diagram $\mathscr{G}_3'$. For this diagram,
there are seven distinct spanning trees. 
}
\label{fig:tree3rdOrder2PR}
\end{figure}

The second diagram contributing at third order is not 2PI (it is of a non-skeleton type) and
is indicated by $\mathscr{G}_3'$ in Fig.~\ref{fig:2nd3rdOrderGF}.
The associated Feynman amplitude $\mathcal{A}^{\mu\nu}_{(3')}(\omega_m)$ reads
\begin{align}\label{3rdOrder2PR}
   &-\mathcal{A}^{\mu\nu}_{(3')}(\omega_m)
   =
   -\frac{1}{2\times3!}\sum_{\mu' \nu'}
    \mathcal{G}^{(0)\mu\mu'}(\omega_m) \nonumber \\
    &\times
    \left\{
        \sum_{\substack{\lambda_1 \lambda_2 \lambda_3 \\
                        \lambda_1' \lambda_2' \lambda_3' \\
                        \lambda \lambda' \kappa \kappa'}}
        v^{(2)}_{[\mu' \kappa \kappa' \nu']}
        v^{(2)}_{[\lambda \lambda_1 \lambda_2 \lambda_3]}
        v^{(2)}_{[\lambda_3' \lambda_2' \lambda_1' \lambda']} \
        I\left(\mathscr{G}_3'\right)
    \right\} \nonumber \\
    &\times \ \mathcal{G}^{(0)\nu'\nu}(\omega_m) \ ,
\end{align}
where $I\left(\mathscr{G}_3'\right)$ is the associated Matsubara sum.  
Again, we use Gaudin's summation rules to compute it based on the 
seven spanning trees of $\mathscr{G}_3'$ shown in
Fig.~\ref{fig:tree3rdOrder2PR}.
The associated numerators and denominators read
\begin{subequations}\label{3rdOrderTrees2PR}
\begin{align}
    \mathscr{A}_1' &: \frac{f(-\epsilon_{n_3}) f(-\epsilon_{n_2}) f(-\epsilon_{n'})}
                          {[\epsilon_{n_1} + \epsilon_{n_2} + \epsilon_{n_3} - \epsilon_{n'}]
                           [\epsilon_{n} - \epsilon_{n'}]} \ , \\
    \mathscr{A}_2' &: \frac{f(-\epsilon_{n_3}) \left(-f(\epsilon_{n_1})\right) f(-\epsilon_{n'})}
                          {[\epsilon_{n_1} + \epsilon_{n_2} + \epsilon_{n_3} - \epsilon_{n'}]
                           [\epsilon_{n} - \epsilon_{n'}]} \ , \\
    \mathscr{A}_3' &: \frac{\left(-f(\epsilon_{n_2})\right) \left(-f(\epsilon_{n_1})\right) f(-\epsilon_{n'})}
                          {[\epsilon_{n_1} + \epsilon_{n_2} + \epsilon_{n_3} - \epsilon_{n'}]
                           [\epsilon_{n} - \epsilon_{n'}]} \ , \\
    \mathscr{A}_4' &: \frac{f(-\epsilon_{n_3}) f(-\epsilon_{n_2}) f(-\epsilon_{n})}
                          {[\epsilon_{n_1} + \epsilon_{n_2} + \epsilon_{n_3} - \epsilon_{n}]
                           [\epsilon_{n'} - \epsilon_{n}]} \ , \\
    \mathscr{A}_5' &: \frac{f(-\epsilon_{n_3}) \left(-f(\epsilon_{n_1})\right) f(-\epsilon_{n})}
                          {[\epsilon_{n_1} + \epsilon_{n_2} + \epsilon_{n_3} - \epsilon_{n}]
                           [\epsilon_{n'} - \epsilon_{n}]} \ , \\
    \mathscr{A}_6' &: \frac{\left(-f(\epsilon_{n_2})\right) \left(-f(\epsilon_{n_1})\right) f(-\epsilon_{n})}
                          {[\epsilon_{n_1} + \epsilon_{n_2} + \epsilon_{n_3} - \epsilon_{n}]
                           [\epsilon_{n'} - \epsilon_{n}]} \ , \\
    \mathscr{A}_7' &: \frac{f(-\epsilon_{n_3}) f(-\epsilon_{n_2}) f(-\epsilon_{n_1})}
                          {[\epsilon_{n} - \epsilon_{n_1} - \epsilon_{n_2} - \epsilon_{n_3}]
                           [\epsilon_{n'} - \epsilon_{n_1} - \epsilon_{n_2} - \epsilon_{n_3}]} \ .
\end{align}
\end{subequations}
The resulting Matsubara sum, after factorisation, reads
\begin{align}\label{MatsubaraSummed3rdOrder2PR}
   I&\left(\mathscr{G}_3'\right)
   =
    \sum_{\substack{n_1 \, n_2 \,  n_3 \\   n \, n'}}
        X^{(n)\kappa} \bar{X}^{(n)\lambda}\ 
        X^{(n_1)\lambda_1} \bar{X}^{(n_1)\lambda_1'}
        \nonumber \\
    &\phantom{\sum}
        \times 
        X^{(n_2)\lambda_2} \bar{X}^{(n_2)\lambda_2'}\ 
        X^{(n_3)\lambda_3} \bar{X}^{(n_3)\lambda_3'}\ 
        X^{(n')\lambda'} \bar{X}^{(n')\kappa'}\ \nonumber \\
    &\phantom{\sum}\times
    \left\{ \vphantom{\frac{f(-\epsilon_{n_3})}{\epsilon_{n}}}
    \left[
        f(-\epsilon_{n_3}) f(-\epsilon_{n_2})
        - f(-\epsilon_{n_3}) f(\epsilon_{n_1})
        + f(\epsilon_{n_2}) f(\epsilon_{n_1})
    \right]
    \right. \nonumber\\
    &\phantom{\sum ((( f(\epsilon) }
    \times
    \left[
    \frac{
        \frac{f(-\epsilon_{n'})}{\epsilon_{n_1} + \epsilon_{n_2} + \epsilon_{n_3} - \epsilon_{n'}}
        -
        \frac{f(-\epsilon_{n})}{\epsilon_{n_1} + \epsilon_{n_2} + \epsilon_{n_3} - \epsilon_{n}}
    }{\epsilon_{n} - \epsilon_{n'}}
    \right] \nonumber\\
    &\phantom{\sum ((( ( }
    \left.
        +
        \frac{f(-\epsilon_{n_3}) f(-\epsilon_{n_2}) f(-\epsilon_{n_1})}
                          {[\epsilon_{n} - \epsilon_{n_1} - \epsilon_{n_2} - \epsilon_{n_3}]
                           [\epsilon_{n'} - \epsilon_{n_1} - \epsilon_{n_2} - \epsilon_{n_3}]}
    \right\} \, .
\end{align}

\subsubsection{Discussion}
Eqs.~\eqref{MatsubaraSummed2ndOrder},~\eqref{MatsubaraSummed3rdOrder2PI}
and~\eqref{MatsubaraSummed3rdOrder2PR} provide three relatively compact and straightforward
expressions for the second and third order contributions of the one-body propagator.
The denominator factorisation in these expressions 
is possible whenever the denominators are associated to
spanning trees differing only in the choice of equivalent lines.
At second order only one class of trees appears.
At third order, for each diagram, only three different classes appear.
Deriving general rules to get directly fully factorised
amplitudes would be interesting, in order to get optimal
algebraic formula ready to be implemented numerically.
Those rules may also mitigate the increasing number
of spanning trees with perturbative order since, in general, 
only one member of each class of trees needs to be considered.
Such refinements are beyond the scope of this article and are left for future developments.

As discussed earlier, infrared divergences may appear in the form of vanishing denominators
of separate spanning tree contributions.
We have chosen to write Eq.~\eqref{MatsubaraSummed3rdOrder2PR}
in a form which highlights the cancellation of the infrared divergence
associated to the denominator $\left[\epsilon_n - \epsilon_{n'}\right]$
for different spanning trees.
In this case, we find two denominators that differ only
in the exchange $\epsilon_n \leftrightarrow \epsilon_{n'}$. 
Expanding the term that depends on $\epsilon_{n'}$ around $\epsilon_{n}$, and taking the limit $\epsilon_{n'} \to \epsilon_{n}$, one finds
\begin{multline}
    \lim_{\epsilon_{n'} \to \epsilon_{n}}
    \left[
    \frac{
        \frac{f(-\epsilon_{n'})}{\epsilon_{n_1} + \epsilon_{n_2} + \epsilon_{n_3} - \epsilon_{n'}}
        -
        \frac{f(-\epsilon_{n})}{\epsilon_{n_1} + \epsilon_{n_2} + \epsilon_{n_3} - \epsilon_{n}}
    }{\epsilon_{n} - \epsilon_{n'}}
    \right]
    \\ =
    \frac{f'(-\epsilon_{n})}{\epsilon_{n_1} + \epsilon_{n_2} + \epsilon_{n_3} - \epsilon_{n}}
    -
    \frac{f(-\epsilon_{n})}{\left[\epsilon_{n_1} + \epsilon_{n_2} + \epsilon_{n_3} - \epsilon_{n}\right]^2}
    \ ,
\end{multline}
where $f'$ denotes the derivative of the Fermi-Dirac distribution.
Similarly, one could consider the case where $\left[\epsilon_{n} - \epsilon_{n'}\right]$
and $\left[ \epsilon_{n_1} + \epsilon_{n_2} + \epsilon_{n_3} - \epsilon_{n} \right]$
are simultaneously vanishing.
We have checked that the double limit
$\lim_{\epsilon_{n}\to(\epsilon_{n_1}+\epsilon_{n_2}+\epsilon_{n_3})} \lim_{\epsilon_{n'}\to\epsilon_{n}}$
of the Matsubara sum is well-defined also in this case.
In general, the Matsubara sum is absolutely convergent\footnote{This is due to
the fact that any independent Matsubara frequency appears at least in
two different propagators except for frequencies associated to tadpoles.
Each propagator is an $O\left(\omega_l^{-1}\right)$ function, such that the sequence
that is summed over in Eq.~\eqref{WeightFactorIntegral} is of
$O\left(\omega_l^{-2}\right)$, and the series converges absolutely.
The particular case of the tadpole, as usual, is taken care of by the regularising
$\eta$-term coming out of Feynman rules.} so that
$I\left(\mathscr{G}\right)$ is always finite.
Thus, any divergence occurring in the sum over quasiparticle energies
for a particular $I\left(\mathscr{A}\right)$ is artificial and 
necessarily cancels out when combined with contributions
from different spanning trees.

In this section, we have derived explicitly several perturbative contributions
to the contravariant one-body Green's function.
Remarkably, the Nambu-covariant formalism used in the formulation of NCPT
allows to obtain very compact expressions.
We have been able to showcase the full third order contribution
in the case of a HFB partitioning 
and an Hamiltonian with a two-body interaction.
At the same time, we have shown that the Nambu-covariant formalism
allows to easily derive the first order contribution
to the contravariant one-body Green's function with arbitrarily
high $k$-body interactions.
However, perturbative contributions are in some cases insufficient.
Relevant approximations for several physical systems,
such as strongly correlated fermions,
require non-perturbative summations of subsets of diagrams.
Such infinite summations are addressed in Part~II,
where we discuss self-consistently dressed propagators
and vertices in the Nambu-covariant formalism.

\section{Connection with standard formalisms}\label{subsec:ConnectStandardFormalism}
The contravariant many-body Green's functions and the un-oriented Feynman diagrams
are fundamental objects of the Nambu-covariant formalism.
To clarify their meaning we discuss their relation with
their counterpart appearing in the more standard Gorkov and Bogoliubov formalisms.

\subsection{Gorkov formalism}\label{subsubsec:GorkovConnection}
Let us first make the connection with the Gorkov formalism as discussed extensively,
for nuclear physics applications, in Ref.~\cite{Soma2011,Barbieri2022}.
To do so, we consider the orthonormal single-particle bases $\mathcal{B}$
and $\tilde{\mathcal{B}}$ related by the bijection $\tilde{.}$ defined by
\begin{equation}
    \tilde{.} : \ket{a} \mapsto \ket{\tilde{a}} \equiv \eta_{a} \ket{\hat{a}} \ ,
\end{equation}
where $\eta_{a}$ is a phase factor and $\hat{.}$ an involution
on the elements of $\mathcal{B}$ as defined in Ref.~\cite{Soma2011}\footnote{To avoid
a conflict of notations, the bijection between elements of $\mathcal{B}$ and
$\tilde{\mathcal{B}}$ has been renamed $\tilde{.}\,$, whereas the notation $\bar{.}$
is used in Refs.~\cite{Soma2011,Barbieri2022}. The latter would clash with our previous use of $\bar{.}$~.
The involution on $\mathcal{B}$ has been renamed
$\hat{.}\,$, while $\tilde{.}$ is used in Refs.~\cite{Soma2011,Barbieri2022}.}.
We also consider the Hamiltonian to be Hermitian and reading
\begin{equation}
    H \equiv
        \sum_{bc} T_{bc} \ a^\dagger_{b}a_{c}
        + 
        \frac{1}{(2!)^2} 
        \sum_{bcde} \bar{V}_{bcde} \ a^\dagger_{b} a^\dagger_{c} a_{e} a_{d} \ .
\end{equation}
To make the connection with the Gorkov formalism transparent,
we work in the field basis
\begin{equation}\label{OrthoGorkovFieldBasis}
    {\mathcal{B}^f}' = \set{a^{\dagger}_{b}} \cup \set{a_{\tilde{b}}}
\end{equation}
which is related to the simple field basis
\begin{equation}
    \mathcal{B}^f = \set{a^{\dagger}_{b}} \cup \set{a_{b}}
\end{equation}
by a non-canonical transformation.
Their respective Nambu fields ${\mathrm{A}'}^{\mu}$ and $\mathrm{A}^{\mu}$
are related according to the transformation of Eqs.~\eqref{NambuTransforms} 
with
\begin{equation}
    {\mathcal{W}^{(b,l_b)}}_{(c,l_c)}
    \equiv
    \begin{pmatrix}
        1 & 0 \\
        0 & \eta^{-1}_{c} \delta_{\hat{b}c}
    \end{pmatrix}_{l_b l_c} \ ,
\end{equation}
where the sub-indices $l_b l_c$ means that the $2\times2$ matrix is to be evaluated
at those Nambu indices.
Explicitly, the Nambu fields associated to ${\mathcal{B}^f}'$ read\footnote{Here, the single-particle bases $\mathcal{B}$ and $\tilde{\mathcal{B}}$ are assumed
to be orthonormal. Hence, the associated creation and annihilation operators are Hermitian conjugated to each other.}
\begin{subequations}
\begin{align}
    \mathrm{A}'_{(b, 1)} &\equiv \bar{a}_b = a^\dagger_b \ , \\
    \mathrm{A}'_{(b, 2)} &\equiv a_{\tilde{b}} \ , \\
    {\mathrm{A}'}^{(b, 1)} &= a_b \ , \\
    {\mathrm{A}'}^{(b, 2)} &= \bar{a}_{\tilde{b}} = a^\dagger_{\tilde{b}} \ .
\end{align}
\end{subequations}
In terms of the Nambu fields ${\mathrm{A}'}^{\mu}$, the Hamiltonian reads
\begin{align}
    H =
        &\frac{1}{2!} 
        \sum_{\mu_1\mu_2} t_{\mu_1\mu_2} \ {\mathrm{A}'}^{\mu_1} {\mathrm{A}'}^{\mu_2}
        \nonumber \\
        &+
        \frac{1}{4!} 
        \sum_{\mu_1\mu_2\mu_3\mu_4}
            v^{(2)}_{\mu_1\mu_2\mu_3\mu_4} \
            {\mathrm{A}'}^{\mu_1} {\mathrm{A}'}^{\mu_2}
            {\mathrm{A}'}^{\mu_3} {\mathrm{A}'}^{\mu_4}
            \ .
\end{align}
Working in the particular field basis ${\mathcal{B}^f}'$,
the relationship between the contravariant many-body Green's functions
and the many-body Green's functions as devised in Ref.~\cite{Soma2011}
straightforwardly reads
\begin{subequations}\label{CovGF_to_GorkovGF}
\begin{align}
    (-1)^k &\mathcal{G}^{(b_1,1) \dots (b_{2k},1)}(\tau_1, \dots, \tau_{2k}) \nonumber \\
        &= \mean{\mathrm{T}\left[ a_{b_1}(\tau_{1}) \dots a_{b_{2k}}(\tau_{2k}) \right]} \ , \\
    (-1)^k &\mathcal{G}^{(b_1,2) (b_2,1) \dots (b_{2k},1)}(\tau_1, \dots, \tau_{2k}) \nonumber \\
        &= \mean{\mathrm{T}\left[ a^\dagger_{\tilde{b}_1}(\tau_{1})
                                 a_{b_2}(\tau_{2}) \dots a_{b_{2k}}(\tau_{2k}) \right]} \ , \\
        &\vdots \nonumber \\
    (-1)^k &\mathcal{G}^{(b_1,2) (b_2,2) \dots (b_{2k},2)}(\tau_1, \dots, \tau_{2k}) \nonumber \\
        &= \mean{\mathrm{T}\left[ a^\dagger_{\tilde{b}_1}(\tau_{1})
                                 a^\dagger_{\tilde{b}_2}(\tau_{2})
                                 \dots a^\dagger_{\tilde{b}_{2k}}(\tau_{2k}) \right]} \ .
\end{align}
\end{subequations}
In principle, there are $2^{2k}$ different $k$-body Green's functions,
counting both anomalous and normal ones.
Using the Hermitian symmetry of the Hamiltonian and
the antisymmetry of the time-ordering, the number of independent
$k$-body Green's functions reduces to $k+1$.
It is clear that if one is interested in high-$k$ many-body Green's functions,
working with the unique tensor Green's function defined in Eq.~\eqref{DefkBodyGF}
is much more convenient than having to consider, separately,
its $k+1$ independent components.

Let us now relate the amplitude of an un-oriented diagram $\mathscr{G}$
to the amplitudes associated to Gorkov diagrams as defined in Ref.~\cite{Soma2011}.
It is straightforward to verify that a given un-oriented diagram $\mathscr{G}$
with \emph{fixed} Nambu indices on its line, reduces to one oriented diagram
of the Gorkov diagrammatic, up to a prefactor.
Indeed, we clearly see from Eqs.~\eqref{GenericFeynmanAmplitudeTime} and
Eqs.~\eqref{CovGF_to_GorkovGF} that fixing Nambu indices amounts to
fixing the kind of Gorkov propagators (normal or anomalous)
appearing in the amplitude of the diagram.
Then, as detailed in~\ref{subsec:antisymVerticesElmts},
the totally antisymmetric vertex $v^{(2)}_{[\lambda_1\lambda_2\lambda_3\lambda_4]}$,
at fixed Nambu indices, reduces to a particular matrix element of type
$\bar{V}_{bcde}$.
This is clearly seen from Eq.~\eqref{ExplAntisym2Bvert}.
Eventually, the amplitude of an un-oriented Feynman diagram with fixed Nambu indices
is a sum of a product of standard matrix elements of the potential
and normal/anomalous propagators, all contracted on single-particle indices
according to the topology of the un-oriented diagram.
Fixing Nambu indices on an un-oriented
diagram thus leads to a contribution proportional to the amplitude of a Gorkov diagram
whose topology is the same as the un-oriented diagram and whose orientation is dictated
by the fixed Nambu indices.
This particular Gorkov diagram is said to be associated to the fixing of Nambu indices.

One should be careful that fixing Nambu indices, however, \emph{does not}
give directly the amplitude of a Gorkov diagram.
Several sets of fixed Nambu indices can have the same associated Gorkov diagram.
Only when summing over all the sets of Nambu indices associated to a given Gorkov diagram
one will recover its full amplitude.
A proper proof of this
statement is beyond the scope of this article.
The most delicate point is to check that the right symmetry factor is recovered
when summing over all sets of fixed Nambu indices associated to one
Gorkov diagram\footnote{This point is subtle because several sets of Nambu indices
might lead to several topologically equivalent Gorkov diagrams.
This is typically the case when fixing Nambu indices of equivalent lines.
This can also occur when several sets of fixed Nambu indices lead to the
same Gorkov diagram up to a permutation of vertices.}.

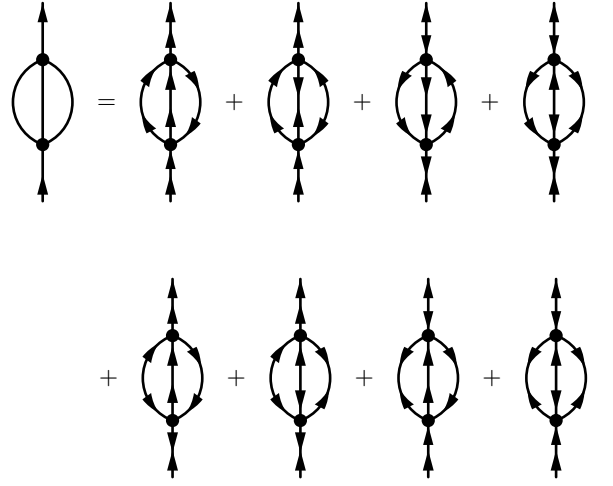
\begin{figure}[t]
\centering
\parbox{35pt}{\begin{fmffile}{2ndOrderGF_FixedExtNambu}
    \begin{fmfgraph*}(35,75)
       \fmfkeep{2ndOrderGF_FixedExtNambu}
    \fmfbottom{i}
    \fmftop{o}
    \fmf{prop_pn, tension=1.5, tag=1}{i,v1}
    \fmf{plain, tag=3}{v1,v2}
    \fmf{prop_nm, tension=1.5, tag=5}{v2,o}
    \fmffreeze
    \fmf{plain, right=0.7, tag=4}{v1,v2}
    \fmffreeze
    \fmf{plain, right=0.7, tag=2}{v2,v1}
    \fmfv{d.shape=circle,d.filled=full,d.size=2thick}{v1}
    \fmfv{d.shape=circle,d.filled=full,d.size=2thick}{v2}
\end{fmfgraph*}
\end{fmffile}}
=
\parbox{35pt}{\begin{fmffile}{2ndOrderGF_FixedNambu_1}
    \begin{fmfgraph*}(35,75)
       \fmfkeep{2ndOrderGF_FixedNambu_1}
    \fmfbottom{i}
    \fmftop{o}
    \fmf{prop_pm, tension=1.5, tag=1}{i,v1}
    \fmf{prop_pm, tag=3}{v1,v2}
    \fmf{prop_pm, tension=1.5, tag=5}{v2,o}
    \fmffreeze
    \fmf{prop_mp, right=0.7, tag=4}{v1,v2}
    \fmffreeze
    \fmf{prop_mp, right=0.7, tag=2}{v2,v1}
    \fmfv{d.shape=circle,d.filled=full,d.size=2thick}{v1}
    \fmfv{d.shape=circle,d.filled=full,d.size=2thick}{v2}
\end{fmfgraph*}
\end{fmffile}}
+
\parbox{35pt}{\begin{fmffile}{2ndOrderGF_FixedNambu_2}
    \begin{fmfgraph*}(35,75)
       \fmfkeep{2ndOrderGF_FixedNambu_2}
    \fmfbottom{i}
    \fmftop{o}
    \fmf{prop_pm, tension=1.5, tag=1}{i,v1}
    \fmf{prop_pp, tag=3}{v1,v2}
    \fmf{prop_pm, tension=1.5, tag=5}{v2,o}
    \fmffreeze
    \fmf{prop_mm, right=0.7, tag=4}{v1,v2}
    \fmffreeze
    \fmf{prop_mp, right=0.7, tag=2}{v2,v1}
    \fmfv{d.shape=circle,d.filled=full,d.size=2thick}{v1}
    \fmfv{d.shape=circle,d.filled=full,d.size=2thick}{v2}
\end{fmfgraph*}
\end{fmffile}}
+
\parbox{35pt}{\begin{fmffile}{2ndOrderGF_FixedNambu_3}
    \begin{fmfgraph*}(35,75)
       \fmfkeep{2ndOrderGF_FixedNambu_3}
    \fmfbottom{i}
    \fmftop{o}
    \fmf{prop_pp, tension=1.5, tag=1}{i,v1}
    \fmf{prop_mp, tag=3}{v1,v2}
    \fmf{prop_mm, tension=1.5, tag=5}{v2,o}
    \fmffreeze
    \fmf{prop_pm, right=0.7, tag=4}{v1,v2}
    \fmffreeze
    \fmf{prop_pm, right=0.7, tag=2}{v2,v1}
    \fmfv{d.shape=circle,d.filled=full,d.size=2thick}{v1}
    \fmfv{d.shape=circle,d.filled=full,d.size=2thick}{v2}
\end{fmfgraph*}
\end{fmffile}}
+
\parbox{35pt}{\begin{fmffile}{2ndOrderGF_FixedNambu_4}
    \begin{fmfgraph*}(35,75)
       \fmfkeep{2ndOrderGF_FixedNambu_4}
    \fmfbottom{i}
    \fmftop{o}
    \fmf{prop_pp, tension=1.5, tag=1}{i,v1}
    \fmf{prop_mm, tag=3}{v1,v2}
    \fmf{prop_mm, tension=1.5, tag=5}{v2,o}
    \fmffreeze
    \fmf{prop_pp, right=0.7, tag=4}{v1,v2}
    \fmffreeze
    \fmf{prop_pm, right=0.7, tag=2}{v2,v1}
    \fmfv{d.shape=circle,d.filled=full,d.size=2thick}{v1}
    \fmfv{d.shape=circle,d.filled=full,d.size=2thick}{v2}
\end{fmfgraph*}
\end{fmffile}}

\vspace{1cm}

\hspace{1.5cm}+
\parbox{35pt}{\begin{fmffile}{2ndOrderGF_FixedNambu_5}
    \begin{fmfgraph*}(35,75)
       \fmfkeep{2ndOrderGF_FixedNambu_5}
    \fmfbottom{i}
    \fmftop{o}
    \fmf{prop_pp, tension=1.5, tag=1}{i,v1}
    \fmf{prop_pm, tag=3}{v1,v2}
    \fmf{prop_pm, tension=1.5, tag=5}{v2,o}
    \fmffreeze
    \fmf{prop_mp, right=0.7, tag=4}{v1,v2}
    \fmffreeze
    \fmf{prop_mm, right=0.7, tag=2}{v2,v1}
    \fmfv{d.shape=circle,d.filled=full,d.size=2thick}{v1}
    \fmfv{d.shape=circle,d.filled=full,d.size=2thick}{v2}
\end{fmfgraph*}
\end{fmffile}}
+
\parbox{35pt}{\begin{fmffile}{2ndOrderGF_FixedNambu_6}
    \begin{fmfgraph*}(35,75)
       \fmfkeep{2ndOrderGF_FixedNambu_6}
    \fmfbottom{i}
    \fmftop{o}
    \fmf{prop_pp, tension=1.5, tag=1}{i,v1}
    \fmf{prop_mm, tag=3}{v1,v2}
    \fmf{prop_pm, tension=1.5, tag=5}{v2,o}
    \fmffreeze
    \fmf{prop_pp, right=0.7, tag=4}{v1,v2}
    \fmffreeze
    \fmf{prop_mm, right=0.7, tag=2}{v2,v1}
    \fmfv{d.shape=circle,d.filled=full,d.size=2thick}{v1}
    \fmfv{d.shape=circle,d.filled=full,d.size=2thick}{v2}
\end{fmfgraph*}
\end{fmffile}}
+
\parbox{35pt}{\begin{fmffile}{2ndOrderGF_FixedNambu_7}
    \begin{fmfgraph*}(35,75)
       \fmfkeep{2ndOrderGF_FixedNambu_7}
    \fmfbottom{i}
    \fmftop{o}
    \fmf{prop_pm, tension=1.5, tag=1}{i,v1}
    \fmf{prop_pm, tag=3}{v1,v2}
    \fmf{prop_mm, tension=1.5, tag=5}{v2,o}
    \fmffreeze
    \fmf{prop_mp, right=0.7, tag=4}{v1,v2}
    \fmffreeze
    \fmf{prop_pp, right=0.7, tag=2}{v2,v1}
    \fmfv{d.shape=circle,d.filled=full,d.size=2thick}{v1}
    \fmfv{d.shape=circle,d.filled=full,d.size=2thick}{v2}
\end{fmfgraph*}
\end{fmffile}}
+
\parbox{35pt}{\begin{fmffile}{2ndOrderGF_FixedNambu_8}
    \begin{fmfgraph*}(35,75)
       \fmfkeep{2ndOrderGF_FixedNambu_8}
    \fmfbottom{i}
    \fmftop{o}
    \fmf{prop_pm, tension=1.5, tag=1}{i,v1}
    \fmf{prop_mm, tag=3}{v1,v2}
    \fmf{prop_mm, tension=1.5, tag=5}{v2,o}
    \fmffreeze
    \fmf{prop_pp, right=0.7, tag=4}{v1,v2}
    \fmffreeze
    \fmf{prop_pp, right=0.7, tag=2}{v2,v1}
    \fmfv{d.shape=circle,d.filled=full,d.size=2thick}{v1}
    \fmfv{d.shape=circle,d.filled=full,d.size=2thick}{v2}
\end{fmfgraph*}
\end{fmffile}}
\vspace{0.5cm}
\caption{Gorkov diagrams factorised in the second order un-oriented diagram for
a given set of fixed external Nambu indices. The fixed Nambu indices are represented
by external arrows on the original un-oriented diagram.}
\label{fig:Cov_To_Gorkov_2ndOrder}
\end{figure}

As a consequence of this
statement, an un-oriented diagram with fixed external Nambu indices
equals the sum of all Gorkov diagrams obtained by orienting the diagram in
a compatible way with the fixed external Nambu indices.
We have explicitly checked this for the first and second order
expansion of the one-body Green's function. We refer to this relation
as a factorisation of Gorkov diagrams.
An example of such factorisation at second order 
is shown in
Fig.~\ref{fig:Cov_To_Gorkov_2ndOrder}.
Let us denote the number of Gorkov diagrams factorised in a given
un-oriented diagram $\mathscr{G}$ (at fixed external Nambu indices)
by $F_{\text{Gorkov}}(\mathscr{G})$.
A naive upper bound on the number of diagrams factorised
consists in counting all possible orientations, i.e.\
\begin{equation}
    F_{\text{Gorkov}}(\mathscr{G}) \leq 2^{2(I+k)} \ ,
\end{equation}
where $I$ and $2k$ are respectively the number of internal and external lines of
$\mathscr{G}$.
We can refine this upper bound by taking into account the fact that the potential is
particle-number conserving.
Let us assume that only interactions up to $k_{\text{max}}$-body are considered.
As a consequence, only orientations where $k_i$ incoming and $k_i$ outgoing
arrows occur at each $k_i$-body vertex give a non-zero contribution.
Instead of counting orientations of lines we count orientations
of vertices which are limited to $\binom{2k_i}{k_i}$ instead of $2^{2k_i}$.
For an un-oriented diagram $\mathscr{G}_n$ (at fixed external Nambu indices)
with $n$ vertices, we get the refined upper bound
\begin{equation}
    F_{\text{Gorkov}}(\mathscr{G}_n)
        \leq \binom{2k_{\text{max}}}{k_{\text{max}}}^{n} \ .
\end{equation}
\emph{We estimate that the number of factorised Gorkov diagrams grows exponentially with $n$.}
Let us stress that this estimate does not take into account
the topology of the diagram.
For example, two different orientations of the vertices might lead to
the same Gorkov diagram up to a permutation of equivalent lines.
Taking into account such double counting would partially reduce the estimate,
nevertheless we expect the exponential growth with the number of vertices
to hold in general.

\subsection{Bogoliubov formalism}\label{subsubsec:BogoConnection}
We now consider the connection between the Nambu-covariant formalism and
the Bogoliubov formalism discussed in Refs.~\cite{Signoracci2015,Duguet2016}.
We focus on the perturbative version of the formalisms, 
namely NCPT and Bogoliubov many-body perturbation theory
(BMBPT)~\cite{Tichai2018a}.
The latter relies on quasiparticle creation
and annihilation operators related to
single-particle creation and annihilation operators obtained by
the unitary Bogoliubov transformation
\begin{subequations}\label{UnitaryBogoTransfo}
\begin{align}
    \beta_{k} &\equiv
        \sum_{b} U^*_{bk} a_b + V^*_{bk} a^\dagger_b \ , \\
    \beta^{\dagger}_{k} &\equiv
        \sum_{b} U_{bk} a_b + V_{bk} a^\dagger_b \ ,
\end{align}
\end{subequations}
where quasiparticle states are indexed over $k$ indices and
where single-particle states are those of an orthonormal basis $\mathcal{B}$
indexed over $b$ indices.
The Bogoliubov formalism developed in Refs.~\cite{Signoracci2015,Duguet2016,Tichai2018a}
focuses mainly on the case where the Hamiltonian is partitioned as
\begin{subequations}
\begin{align}
    H &\equiv H_0 + H_1 \ , \\
    H_0 &\equiv H^{00} + \bar{H}^{11} \ , \\
    H_1 &\equiv H^{20} + \Breve{H}^{11} + H^{02} \nonumber \\
             &\phantom{\equiv \ }
                + H^{40}
                + H^{31} + H^{22} + H^{13}
                + H^{04} \ ,
\end{align}
\end{subequations}
where $H^{00}$ is proportional to the identity and $\bar{H}^{11}$ is diagonal, i.e.\
\begin{subequations}
\begin{align}
    H^{00} &\propto \mathbbm{1}_{\mathscr{F}} \ , \\
    \bar{H}^{11} &\equiv \sum_{k} E_k  \beta^\dagger_k \beta_k \ ,
\end{align}
\end{subequations}
with $E_k > 0$. For such partitioning the unperturbed propagator contains only
normal components (with respect to quasiparticle creation
and annihilation operators).
We also use the shorthand notation $H^{ij}$ to denote
an operator involving $i$ quasiparticle creation operators and
$j$ quasiparticle annihilation operators, i.e.\ for example
\begin{equation}
    H^{31} \equiv
        \frac{1}{3!} \sum_{k_1k_2k_3k_4}
            H^{31}_{k_1k_2k_3k_4}
            \beta^\dagger_{k_1} \beta^\dagger_{k_2} \beta^\dagger_{k_3} \beta_{k_4} \ .
\end{equation}
This reformulation allows to exchange anomalous lines in Gorkov diagrammatics
with anomalous vertices in Bogoliubov diagrammatics~\cite{Nozieres1964}.
Here, by anomalous vertices we mean vertices where the number of
incoming lines does not necessarily match the number of outgoing lines.

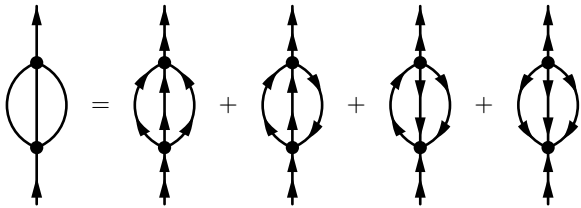
\begin{figure}[t]
\centering
\parbox{35pt}{\begin{fmffile}{2ndOrderGF_FixedExtNambu}
    \begin{fmfgraph*}(35,75)
       \fmfkeep{2ndOrderGF_FixedExtNambu}
    \fmfbottom{i}
    \fmftop{o}
    \fmf{prop_pn, tension=1.5, tag=1}{i,v1}
    \fmf{plain, tag=3}{v1,v2}
    \fmf{prop_nm, tension=1.5, tag=5}{v2,o}
    \fmffreeze
    \fmf{plain, right=0.7, tag=4}{v1,v2}
    \fmffreeze
    \fmf{plain, right=0.7, tag=2}{v2,v1}
    \fmfv{d.shape=circle,d.filled=full,d.size=2thick}{v1}
    \fmfv{d.shape=circle,d.filled=full,d.size=2thick}{v2}
\end{fmfgraph*}
\end{fmffile}}
=
\parbox{35pt}{\begin{fmffile}{2ndOrderGF_Bogo_1}
    \begin{fmfgraph*}(35,75)
       \fmfkeep{2ndOrderGF_Bogo_1}
    \fmfbottom{i}
    \fmftop{o}
    \fmf{prop_pm, tension=1.5, tag=1}{i,v1}
    \fmf{prop_pm, tag=3}{v1,v2}
    \fmf{prop_pm, tension=1.5, tag=5}{v2,o}
    \fmffreeze
    \fmf{prop_pm, right=0.7, tag=4}{v1,v2}
    \fmffreeze
    \fmf{prop_mp, right=0.7, tag=2}{v2,v1}
    \fmfv{d.shape=circle,d.filled=full,d.size=2thick}{v1}
    \fmfv{d.shape=circle,d.filled=full,d.size=2thick}{v2}
\end{fmfgraph*}
\end{fmffile}}
+
\parbox{35pt}{\begin{fmffile}{2ndOrderGF_Bogo_2}
    \begin{fmfgraph*}(35,75)
       \fmfkeep{2ndOrderGF_Bogo_2}
    \fmfbottom{i}
    \fmftop{o}
    \fmf{prop_pm, tension=1.5, tag=1}{i,v1}
    \fmf{prop_pm, tag=3}{v1,v2}
    \fmf{prop_pm, tension=1.5, tag=5}{v2,o}
    \fmffreeze
    \fmf{prop_mp, right=0.7, tag=4}{v1,v2}
    \fmffreeze
    \fmf{prop_mp, right=0.7, tag=2}{v2,v1}
    \fmfv{d.shape=circle,d.filled=full,d.size=2thick}{v1}
    \fmfv{d.shape=circle,d.filled=full,d.size=2thick}{v2}
\end{fmfgraph*}
\end{fmffile}}
+
\parbox{35pt}{\begin{fmffile}{2ndOrderGF_Bogo_3}
    \begin{fmfgraph*}(35,75)
       \fmfkeep{2ndOrderGF_Bogo_3}
    \fmfbottom{i}
    \fmftop{o}
    \fmf{prop_pm, tension=1.5, tag=1}{i,v1}
    \fmf{prop_mp, tag=3}{v1,v2}
    \fmf{prop_pm, tension=1.5, tag=5}{v2,o}
    \fmffreeze
    \fmf{prop_mp, right=0.7, tag=4}{v1,v2}
    \fmffreeze
    \fmf{prop_mp, right=0.7, tag=2}{v2,v1}
    \fmfv{d.shape=circle,d.filled=full,d.size=2thick}{v1}
    \fmfv{d.shape=circle,d.filled=full,d.size=2thick}{v2}
\end{fmfgraph*}
\end{fmffile}}
+
\parbox{35pt}{\begin{fmffile}{2ndOrderGF_Bogo_4}
    \begin{fmfgraph*}(35,75)
       \fmfkeep{2ndOrderGF_Bogo_4}
    \fmfbottom{i}
    \fmftop{o}
    \fmf{prop_pm, tension=1.5, tag=1}{i,v1}
    \fmf{prop_mp, tag=3}{v1,v2}
    \fmf{prop_pm, tension=1.5, tag=5}{v2,o}
    \fmffreeze
    \fmf{prop_mp, right=0.7, tag=4}{v1,v2}
    \fmffreeze
    \fmf{prop_pm, right=0.7, tag=2}{v2,v1}
    \fmfv{d.shape=circle,d.filled=full,d.size=2thick}{v1}
    \fmfv{d.shape=circle,d.filled=full,d.size=2thick}{v2}
\end{fmfgraph*}
\end{fmffile}}
\vspace{0.5cm}
\caption{Bogoliubov diagrams factorised in the second order un-oriented diagram for
a given set of fixed external Nambu indices. The fixed Nambu indices are represented
by external arrows on the original un-oriented diagram.
Pay attention that diagrams containing directed cycles are not discarded
at non-vanishing temperatures, contrary to a common selection rule
derived in the zero temperature formalism of Ref.~\cite{Duguet2016}.}
\label{fig:Cov_To_Bogo_2ndOrder}
\end{figure}

To connect NCPT and BMBPT,
we start from the field basis
$\mathcal{B}^f = \set{a^\dagger_b} \cup \set{a_b}$,
where $a^\dagger_b$ and $a_b$ are the same single-particle creation and annihilation
operators used in
Eqs.~\eqref{UnitaryBogoTransfo}. We then perform the same unitary
Bogoliubov transformation as in Eqs.~\eqref{UnitaryBogoTransfo}
to perform a change of field basis from $\mathcal{B}^f$ to
quasiparticle and Nambu indices.
The new Nambu fields are thus defined by Eqs.~\eqref{NambuTransforms}, 
with the transformation
\begin{equation}
    {\mathcal{W}^{(b,l_b)}}_{(k,l_k)}
    \equiv
    \begin{pmatrix}
        U_{bk} & V^*_{bk} \\
        V_{bk} & U^*_{bk}
    \end{pmatrix}_{l_b l_k} \ .
\end{equation}
In this case, the new Nambu fields read\footnote{Since the single-particle basis 
$\mathcal{B}$ is orthonormal and the Bogoliubov transformation is unitary,
the creation/annihilation operators associated to the new Nambu fields
are Hermitian conjugated to each other. For more details on unitary
Bogoliubov transformations see App.~A of Part~II.}
\begin{subequations}
\begin{align}
    \mathrm{A}'_{(k, 1)} &\equiv \bar{\beta}_k = \beta^\dagger_k \ , \\
    \mathrm{A}'_{(k, 2)} &\equiv \beta_{k} \ , \\
    {\mathrm{A}'}^{(k, 1)} &= \beta_k \ , \\
    {\mathrm{A}'}^{(k, 2)} &= \bar{\beta}_{k} = \beta^\dagger_{k} \ ,.
\end{align}
\end{subequations}
The connection between BMBPT and NCPT diagrammatics, 
expressed in terms of Nambu fields ${\mathrm{A}'}^{\mu}$,
follows in a similar fashion as for the Gorkov case.
The main difference is that, this time, only normal lines occur
and anomalous vertices are allowed. Many-body Green's functions
are related to the contravariant ones with fixed Nambu indices.
Un-oriented diagrams with fixed external Nambu indices are then again
related to the sum of all possible orientations. We refer
to this relationship between diagrams as a factorisation of Bogoliubov diagrams.
An example of such factorisation is given in
Fig.~\ref{fig:Cov_To_Bogo_2ndOrder}.
Similarly to the Gorkov case, we denote the number of Bogoliubov diagrams
factorised in a given un-oriented diagram $\mathscr{G}$
by $F_{\text{Bogoliubov}}(\mathscr{G})$.
This time, the upper bound on $F_{\text{Bogoliubov}}(\mathscr{G})$
which takes into account the fact that only normal lines are allowed reads
\begin{equation}
    F_{\text{Bogoliubov}}(\mathscr{G}) \leq 2^{I} \ ,
\end{equation}
where $I$ is the number of internal lines of $\mathscr{G}$.
\emph{We estimate that the number of factorised Bogoliubov diagrams grows exponentially with $I$.}
Let us stress again that this estimate does not take into account
the topology of the diagram.
For example, two different orientations of the lines could lead to
the same Bogoliubov diagram up to a permutation of equivalent lines.
Taking into account such double counting would reduce the estimate. Again,
we expect the exponential growth with the number of internal lines
to hold, provided the number of vertices also grows\footnote{Heuristically,
the growth of the number of vertices is essential to avoid saturation of vertices,
i.e.\ to avoid being in a situation where any additional internal line
is also an additional equivalent line. This worst case scenario is avoided
if we assume to have only up to $k_{\text{max}}$-body interaction since
at some point adding new internal lines will require adding new vertices.}.

\subsection{Nambu-invariance and truncations}\label{subsubsec:discussion}
Let us consider a many-body approximation defined as a truncation on the set
of un-oriented diagrams contributing to the many-body Green's functions.
Since any diagram is a tensor of the same nature as the Green's function
to which it contributes, the approximated Green's functions are guaranteed
to conserve their tensor character.
Let $\mathcal{G}^{\mu_1\dots\mu_{2k}}_{\text{MB}}(\tau_1, \dots, \tau_{2k})$ be
the approximated $k$-body Green's functions, and $O$ be a $k$-body observable
such that
\begin{equation}
    O =
    \sum_{\mu_1 \dots \mu_{2k}}
      o_{\mu_1 \dots \mu_{2k}} \
      \mathrm{A}^{\mu_1} \dots \mathrm{A}^{\mu_{2k}} \ .
\end{equation}
The approximated expectation value of the observable is defined as
\begin{equation}
    \mean{O}_{\text{MB}} \equiv (-1)^{k}
        \sum_{\mu_1\dots\mu_{2k}}
            o_{\mu_1 \dots \mu_{2k}} \
            \mathcal{G}^{\mu_1 \dots \mu_{2k}}_{\text{MB}}(0^{+\dots+}, \dots, 0) \ ,
\end{equation}
where the zero-time limit is to be understood as taken while keeping
$\tau_1 > \dots > \tau_{2k}$.
As long as $\mathcal{G}^{\mu_1\dots\mu_{2k}}_{\text{MB}}(\tau_1, \dots, \tau_{2k})$
is a $(2k,0)$-tensor at any fixed set of times,
the approximated expectation value $\mean{O}_{\text{MB}}$
will be a type $(0,0)$ (or scalar) tensor.
Consequently, the approximated expectation value is not only single-particle
invariant but also \emph{Nambu invariant} which includes, as a sub-case,
the invariance with respect to any Bogoliubov transformation.

This invariance property of the approximated observables is \emph{not} guaranteed
when considering many-body approximations defined by a set of
Gorkov or Bogoliubov diagrams.
In general, Gorkov or Bogoliubov diagrammatic contributions 
that differ only by their orientations are reshuffled when performing
a Bogoliubov transformation or, more generally, when performing a change of
field basis. In other words, the invariance under transformations of an approximated observable
is only recovered when the set of diagrams defining the approximation
contains \emph{all} the possible orientations of each diagram it contains.
As it was discussed in Secs.~\ref{subsubsec:GorkovConnection}
and~\ref{subsubsec:BogoConnection}, such a many-body approximation can be expressed
as a truncation on the set of un-oriented diagrams obtained from NCPT
and $\mean{O}_{\text{MB}}$ is a scalar tensor.
This is the case, for example, of the recent and successful many-body approximations
referred to as BMBPT($n$) in Ref.~\cite{Tichai2018a}, which include
all Bogoliubov diagrams containing up to $n$ vertices.

Ensuring that the many-body approximation considered is Nambu invariant is not only
desirable from a purely theoretical point of view. We expect that invariance will also
have an impact on numerical efficiency.
In the case of truncated model space calculations, Nambu invariance ensures
the crucial property that the complete basis set limit is independent from
the field basis chosen to implement the calculations.
Such basis independence opens the way towards further optimisations.
For instance, the field basis can be chosen to maximise the reliability and speed
of convergence towards the complete basis set limit.
Such basis optimisation strategies have been studied at length in quantum
chemistry, see Ref.~\cite{Nagy2017} for a recent review. 
Basis optimisation was recently shown
to be of great importance for successful large-scale calculations of
nuclear structure observables~\cite{Caprio2012,Tichai2019,Hoppe2021}.
So far, only invariance with respect to the choice of single-particle basis
has been considered, with a focus on symmetry-conserving calculations.
Nambu-invariant many-body approximations could benefit greatly
from further prospects in the optimisation of the field basis,
although it is not yet immediate to give \emph{a priori} arguments
for preferring a certain field basis over another.
An optimised field basis $\mathcal{B}^f$ might be of Gorkov type
(i.e.\ taking advantage of particle-number conservation at vertices
by using a field basis based on single-particle states),
of Bogoliubov type (i.e.\ taking advantage of a diagonal propagator by using
a field basis based on quasiparticle states) or, actually, anything else.
The virtue of the NCPT developed in this article is that it remains unbiased
with respect to the chosen field basis $\mathcal{B}^f$,
allowing for a large range of potential optimisation strategies.

Last but not least, let us briefly discuss the diagrammatic factorisation
mentioned in Secs.~\ref{subsubsec:GorkovConnection}
and~\ref{subsubsec:BogoConnection} and its potential impact on
numerical efficiency.
At first sight, factorisation may appear to be an artifice recasting several
standard diagrams into a single un-oriented diagram at the price of
doubling the dimension of the one-body space, from $\mathscr{H}_1$
to the one of $\mathscr{H}^f$. 
Working with un-oriented diagrams
has, nonetheless, two \emph{a piori} main advantages.
The first benefit is formal: one can work with tensors with a greater
degree of symmetry than in other approaches, such as
$v^{(k)}_{[\mu_1 \dots \mu_{2k}]}$, which is totally antisymmetric.
This suggests that the increase in size of the model space can be
(at least partly) mitigated by
the additional symmetries
in the evaluation of the tensor network.
The second benefit is specific to numerical implementations on
massively parallel hardware: a greater degree of parallelisation of
the floating-point operations is exposed in NCPT equations, without increasing
data movements.
This is a general consequence of replacing several different
tensor contractions by a unique one, with larger dimensions. 
Parallelising floating-point operations without increasing
data movements is key to increase efficiency of GPU-based
architectures~\cite{Dongarra2014}\footnote{To emphasise the importance
of reducing data movements, let us stress that the reading (writing)
bandwidth between CPUs and GPUs ranges, for a modern standard,
from $16$ GB/s (for $8$ reading (writing) lanes of a $4^{\text{th}}$
generation PCI Express) to $150$ GB/s (for $48$ reading (writing) lanes
of $2^{\text{nd}}$ generation NVlink), while the computational power of a GPU
ranges between $1$ Tflop/s and $10$ Tflop/s (e.g.\ for a NVIDIA K20
and V100 GPU).}. 
Therefore, we expect that trading a smaller set of tensor networks
for a larger model space will
increase the gain obtained by using accelerator hardware, such as GPUs.
We stress that accelerators are becoming more and more important
in the numerical evaluation of tensor networks, thanks to
the rapidly growing software infrastructures deploying more efficient
and easier-to-use algorithms~\cite{Dongarra2014,Gates2019}.
Taking advantage of these developments to their fullest potential is a point that
should not be neglected.
We are aware that this discussion is purely qualitative, and
the practical benefits remain speculative at this point.
Ultimately, one would have to develop quantitative studies which will likely
depend on the many-body problem
and the computational resources at stake.

\section{Conclusions}\label{sec:conclusions}

Since the development of a microscopic theory of superconductivity
by Bardeen, Cooper and Schrieffer~\cite{Bardeen1957a,Bardeen1957b}
several reformulations of symmetry-broken many-body theory have occurred.
Nambu's original reformulation was based 
on so-called Nambu fields~\cite{Nambu1960}, and used the fact that
these fields respect the usual canonical anticommutation rules which, 
in the notation of this paper, read
\begin{equation}
    \Set{ \mathrm{A}^\mu , \mathrm{A}_\nu } = \delta_{\mu\nu} \ .
\end{equation}
This property allowed Nambu to develop a perturbation theory based on
oriented Feynman diagrams free of anomalous lines, but where
propagators are $2\times2$ matrices.
Whether working with a matrix propagator
or with separate
normal and anomalous propagators, 
a commonly used shorthand notation has been to
denote by un-oriented diagrams the sum of their oriented versions,
compatible with the Feynman rules to be employed.
In such approaches, the non-trivial dependence of the amplitudes on
Nambu indices precludes a clear factorisation of the different orientations
of a given diagram.
To the best of our knowledge, a reformulation of Feynman diagrammatics
verifying the requirement that Nambu indices only impact the value
of a diagram via their contractions in a tensor network was first
introduced by De Dominicis and Martin~\cite{DeDominicis1964a,DeDominicis1964b}.
The key element for the success of this formulation
is that it expresses the many-body problem solely in terms of contravariant Nambu fields, 
$\mathrm{A}^{\mu}$, despite having
\begin{equation}
    \Set{ \mathrm{A}^\mu , \mathrm{A}^\nu } = g^{\mu\nu} \neq \delta_{\mu\nu} \ .
\end{equation}
Later on, diagrammatics with the same property were independently
reintroduced by Kleinert~\cite{Kleinert1981,Kleinert1982}
and Haussmann~\cite{Haussmann1993,Haussmann1999}.
These approaches are all based on diagrams with un-oriented lines
and totally antisymmetric vertices, going beyond the traditional Hugenholtz
antisymmetrisation.

In the present work, previous reformulations have been taken further
by extending them to a general Hamiltonian expressed in a general field basis.
The success of such reformulation has been shown to be underpinned by the
algebraic structure of Nambu tensors, detailed in Sec.~\ref{sec:tensor}.
The covariance and contravariance of Nambu tensors was defined
to be with respect to any change of basis of $\mathscr{H}^f$. These
include, as a sub-case,
changes of single-particle basis; unitary and non-unitary Bogoliubov
transformations; and an additional set of non-canonical linear
transformations, i.e.\ transformations modifying the metric $g_{\mu\nu}$.
The perturbation theory resulting from working in the Nambu-covariant formalism,
dubbed Nambu-Covariant Perturbation Theory, deals with two- and
many-body interactions; it is conceived explicitly at non-zero temperature;
and it eventually results in simplified un-oriented Feynman diagrammatics
with fully antisymmetrised interaction vertices.

In Secs.~\ref{subsec:FeynRulesTime} and~\ref{subsec:FeynRulesEnergy},
we give the Feynman rules that provide an expansion of the (contravariant) 
many-body Green's functions.
The un-oriented Feynman diagrams appearing in the perturbative expansion 
decompose the many-body Green's functions in terms of Nambu tensors.
Let us also stress two particular subtleties appearing in NCPT.
The first, which did not appear in previous
works~\cite{DeDominicis1964a,DeDominicis1964b,Kleinert1981,Kleinert1982,Haussmann1993,Haussmann1999},
is the partial antisymmetrisation of interaction vertices, which is required when tadpoles are
contracted over them.
This is necessary unless we assume the components $v^{(k)}_{\mu_1\dots\mu_{2k}}$
to be totally antisymmetric from the start. Although such decomposition of the
Hamiltonian is in general possible, it might not be optimal for 
numerical applications.
The second subtlety 
is the almost, but not quite, direct connection between the NCPT diagrammatics
for fixed external Nambu indices, and the more standard Gorkov
and Bogoliubov diagrammatics, discussed in
Secs.~\ref{subsubsec:GorkovConnection} and~\ref{subsubsec:BogoConnection}.

Although the developments of NCPT in
Sec.~\ref{sec:covPT} are restricted to the single-reference case,
we work in a general enough setting that could be
of interest for further extensions.
In particular, we leave open the possibility of working with a field basis
built upon a non-orthogonal single-particle basis, as in
Eq.~\eqref{BasicExampleFieldBasis}. Our formalism can also deal with non-Hermitian
unperturbed Hamiltonians.
These two advantages (a non-orthogonal basis, diagonalising a
non-Hermitian unperturbed Hamiltonian) should be useful 
in applications
to certain flavours of multi-reference perturbation theory.
For instance, let us mention the multi-configuration perturbation theory
(MCPT)~\cite{Rolik2003} which has recently been applied to open-shell
nuclei~\cite{Tichai2018b}; or the non-orthogonal configuration interaction
with second-order perturbation theory built on top (NOCI-PT2), which has
been recently developed and applied in quantum chemistry~\cite{Burton2020}. 
Our general setting may also be useful in
the developments of projected Bogoliubov many-body perturbation theory
(PBMBPT)~\cite{Duguet2016}.
In this case, at zero temperature, spontaneously broken symmetries are restored
by mixing several single-reference calculations over vacua which differ
by a non-unitary Bogoliubov transformation. While such developments lie beyond the scope 
of this work,
it would be interesting to study how the Nambu-covariant
formalism could be used to reformulate the above approaches
and see if any formal simplifications or numerical optimisations arise.
Beyond formal and numerical improvements, the ability to handle
non-Hermitian Hamiltonians and their non-orthogonal basis is also
important phenomenologically for applications to open quantum systems.
For example, the description of nuclear reactions
relies crucially on effective Hamiltonians
arising from a Feshbach projection~\cite{Feshbach1962}.
These effective Hamiltonians are, in general, non-Hermitian.
We refer the reader to Ref.~\cite{Ashida2020} for a recent review on 
non-Hermitian Hamiltonians and their
physical applications.

To conclude, the focus of this work has been on the perturbative
aspects of many-body theory and how they can be formulated
in a Nambu-covariant fashion.
In the case of a physical system made of strongly correlated
fermions, perturbative approximations may be insufficient.
One avenue to solve this problem consists in dressing propagators
and vertices using advanced many-body techniques in order to effectively
sum infinite sets of diagrams.
In Part~II of this work, we consider infinite summations of diagrams
obtained via self-consistently dressed propagators and vertices
in a Nambu-covariant fashion~\cite{part2}. The mere fact that these additional
considerations are possible is proof of the versatility and potentiality
of this new Nambu-covariant formalism.

\section*{Acknowledgments}
The authors thank J.~W.~T.~Keeble for proofreading the manuscript.
M.~D.\ would like to acknowledge useful discussions and comments
with the nuclear theory group of CEA-Saclay on an earlier version
of this work.

This work is supported by STFC, through Grants Nos 
ST/L005743/1 and ST/P005314/1; 
by the Spanish MICINN
through the ``Ram\'on y Cajal" program with grant RYC2018-026072 and
the ``Unit of Excellence Mar\'ia de Maeztu 2020-2023" award to the Institute of Cosmos Sciences (CEX2019-000918-M).
TRIUMF receives federal funding via a contribution agreement with the National Research Council of Canada.

\appendix

\section{Matrix elements}\label{app:MatElts}

In this appendix, we relate the standard expression of a $k$-body operator
to its fully covariant representation. In practical applications,
one will need to perform a transformation from the
standard operator matrix elements into the fully covariant tensor
coordinates appearing in the Nambu-covariant formalism.
For a fixed single-particle basis, this transformation should be a
one-off (pre-processing) step before fully-fledged
many-body steps are developed. 

For a given choice of single-particle basis $\mathcal{B}$,
a $k$-body operator $O$ reads, in terms of the associated creation and annihilation operators,
\begin{equation}\label{kbody_Gorkov}
  O \equiv \sum_{\substack{b_1 \dots b_k \\ c_1 \dots c_k}}
              o_{b_1 \dots b_k c_1 \dots c_k} \ 
              \bar{a}_{b_1} \dots \bar{a}_{b_k}
              a_{c_k} \dots a_{c_1} \ ,
\end{equation}
where $o_{b_1 \dots b_k c_1 \dots c_k}$ are complex numbers.
Similarly, for a given field basis $\mathcal{B}^f$ of $\mathscr{H}^f$,
the operator reads, in terms of a mixed $(k,k)$-tensor of coordinates
${o^{\mu_1 \dots \mu_{k}}}_{\mu_{k+1} \dots \mu_{2k}}$,
\begin{align}
    O &=
         \sum_{\mu_1\dots \mu_{2k}} 
            {o^{\mu_1 \dots \mu_{k}}}_{\mu_{k+1} \dots \mu_{2k}} \  
            \mathrm{A}_{\mu_1} \dots \mathrm{A}_{\mu_k}
            \mathrm{A}^{\mu_{k+1}} \dots \mathrm{A}^{\mu_{2k}} \ . 
\end{align}
For applications to NCPT, it is more convenient to work with
the fully covariant $(0,2k)$-tensor of coordinates $o_{\mu_1 \dots \mu_{2k}}$, 
which verify
\begin{subequations}\label{kbody_CovRep}
\begin{align}
    O &=
         \sum_{\mu_1\dots \mu_{2k}} 
            o_{\mu_1 \dots \mu_{2k}} \  
            \mathrm{A}^{\mu_1} \dots \mathrm{A}^{\mu_{2k}} \ .
\end{align}
\end{subequations}

Conveniently choosing Nambu fields as given in Eqs.~\eqref{CovNambuFieldsDef}
and using the associated metric given in Eq.~\eqref{basicExampleMetric},
Eq.~\eqref{kbody_Gorkov} can be re-expressed as
\begin{equation}
  O = \sum_{\substack{b_1 \dots b_k \\ c_1 \dots c_k}}
              o_{b_1 \dots b_k c_k \dots c_1} \ 
              \mathrm{A}_{(b_1,1)} \dots \mathrm{A}_{(b_k,1)}
              \mathrm{A}^{(c_1,1)} \dots \mathrm{A}^{(c_k,1)} \ .
\end{equation}
Therefore, we can easily relate $o_{b_1 \dots b_k c_1 \dots c_k}$ to the mixed
representation matrix elements,  
${o^{\mu_1 \dots \mu_{k}}}_{\mu_{k+1} \dots \mu_{2k}}$, according to
\begin{multline}
  {o^{(b_1,l_1) \dots (b_k,l_k)}}_{(c_1,m_1) \dots (c_k,m_k)} = %\\ 
    o_{b_1 \dots b_k c_k \dots c_1} \ 
    E^{11}_{l_1m_1} \dots E^{11}_{l_km_k}  \ , \label{kBody_MixedNambuGorkov_Relation}
\end{multline}
where the family $E^{ij}$ denotes the canonical basis of $2\times2$ matrices, i.e.\
\begin{equation}
    E^{ij}_{kl} \equiv \delta_{ik} \delta_{jl} \ .
\end{equation}
Let us recall that the metric in $\mathcal{B}^f$ reads simply
\begin{equation}
    g_{\mu\nu} = \delta_{\mu\bar{\nu}} \ .
\end{equation}
Consequently,
the fully covariant coordinates $o_{\mu_1 \dots \mu_{2k}}$ are 
related to the original matrix elements
by the expression
\begin{multline}
    o_{(b_1,l_1) \dots (b_k,l_k) (c_1,m_1) \dots (c_k,m_k)} = %\\
        o_{b_1 \dots b_k c_k \dots c_1} \ 
            E^{21}_{l_1m_1} \dots E^{21}_{l_km_k}  \ .
    \label{kBody_CovNambuGorkov_Relation}
\end{multline}
Notice that in Eqs.~\eqref{kBody_MixedNambuGorkov_Relation}
and~\eqref{kBody_CovNambuGorkov_Relation}, the components of the tensor
are separable between Nambu indices and single-particle indices.

\section{Antisymmetrisation of vertices}\label{app:AntisymVertices}
In this appendix, we detail how totally and partially antisymmetric
vertices arise in the Feynman rules given in
Sec.~\ref{subsec:FeynRulesTime}. As an example,
we then work out explicit expressions of antisymmetrised vertices
in terms of standard single-particle matrix elements
in the case of a two-body interaction.

\subsection{Vertex factorisation}\label{subsec:FactorUnsym}
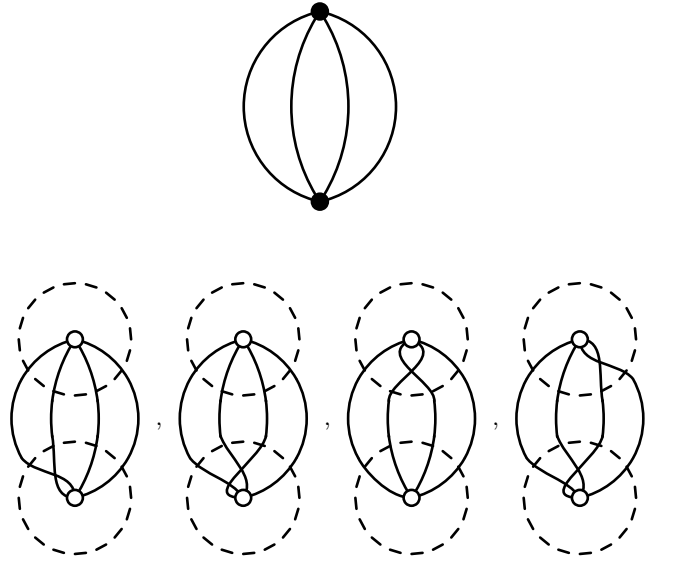
\begin{figure}[t]
  \centering
  \hspace{-0.2cm}\parbox{60pt}{\begin{fmffile}{2ndOrder_lnZ_TwoBody_Symm}
    \begin{fmfgraph*}(60,120)
       \fmfkeep{2ndOrder_lnZ_TwoBody_Symm}
    \fmfbottom{i}
    \fmftop{o}
    \fmf{phantom, tension=3, tag=1}{i,v1}
    \fmf{phantom, tag=3}{v1,v2}
    \fmf{phantom, tension=3, tag=5}{v2,o}
    \fmffreeze
    \fmf{plain, right=0.3, tag=4}{v1,v2}
    \fmf{plain, right=0.3, tag=2}{v2,v1}
    \fmf{plain, right=0.8, tag=4}{v1,v2}
    \fmf{plain, right=0.8, tag=2}{v2,v1}
    \fmfv{d.shape=circle,d.filled=full,d.size=3thick}{v1}
    \fmfv{d.shape=circle,d.filled=full,d.size=3thick}{v2}
\end{fmfgraph*}
\end{fmffile}}

\vspace{0.25cm}

\parbox{55pt}{\begin{fmffile}{2ndOrder_lnZ_TwoBody1_UnSymm}
    \begin{fmfgraph*}(50,100)
       \fmfkeep{2ndOrder_lnZ_TwoBody1_UnSymm}
    \fmfbottom{i}
    \fmftop{o}
    \fmf{phantom, tension=3}{i,v1}
    \fmf{phantom}{v1,v2}
    \fmf{phantom, tension=3}{v2,o}
    \fmffreeze
    \fmf{phantom, right=0.8, tag=1}{v2,v1}
    \fmf{phantom, right=0.3, tag=2}{v2,v1}
    \fmf{phantom, right=0.3, tag=3}{v1,v2}
    \fmf{phantom, right=0.8, tag=4}{v1,v2}
    \fmfv{d.shape=circle,d.filled=empty,d.size=3thick}{v1}
    \fmfv{d.shape=circle,d.filled=empty,d.size=3thick}{v2}
    \fmfipath{circlet,circleb,dcirclet,dcircleb}
    \fmfiset{circlet}{fullcircle scaled 1w shifted vloc(__v2)}
    \fmfiset{circleb}{fullcircle scaled 1w shifted vloc(__v1)}
    \fmfiset{dcirclet}{fullcircle scaled 0.85w shifted vloc(__v2)}
    \fmfiset{dcircleb}{fullcircle scaled 0.85w shifted vloc(__v1)}
    \fmfipath{p[]}
    \fmfiset{p1}{vpath1(__v1,__v2)}
    \fmfiset{p2}{vpath2(__v1,__v2)}
    \fmfiset{p3}{vpath3(__v1,__v2)}
    \fmfiset{p4}{vpath4(__v1,__v2)}
    \fmfi{dashes}{dcirclet}
    \fmfi{dashes}{dcircleb}
    \fmfi{plain}{subpath (15length(p1)/50,35length(p1)/50) of p1}
    \fmfi{plain}{subpath (19length(p2)/50,31length(p2)/50) of p2}
    \fmfi{plain}{subpath (19length(p3)/50,31length(p3)/50) of p3}
    \fmfi{plain}{subpath (15length(p4)/50,35length(p4)/50) of p4}
    % Top part
    \fmfipair{ct[]}
    \fmfiequ{ct1}{p1 intersectionpoint circlet}
    \fmfiequ{ct2}{p2 intersectionpoint circlet}
    \fmfiequ{ct3}{p3 intersectionpoint circlet}
    \fmfiequ{ct4}{p4 intersectionpoint circlet}
    \fmfipath{t[]}
    \fmfiset{t1}{vloc(__v2){dir -166.9} .. ct1}
    \fmfiset{t2}{vloc(__v2){dir -123} .. ct2}
    \fmfiset{t3}{vloc(__v2){dir -57} .. ct3}
    \fmfiset{t4}{vloc(__v2){dir -13.1} .. ct4}
    \fmfi{plain}{t1}
    \fmfi{plain}{t2}
    \fmfi{plain}{t3}
    \fmfi{plain}{t4}
    % Bottom part
    \fmfipair{cb[]}
    \fmfiequ{cb1}{p1 intersectionpoint circleb}
    \fmfiequ{cb2}{p2 intersectionpoint circleb}
    \fmfiequ{cb3}{p3 intersectionpoint circleb}
    \fmfiequ{cb4}{p4 intersectionpoint circleb}
    \fmfipath{b[]}
    \fmfiset{b1}{vloc(__v1){dir 180} .. cb2{dir 100}}
    \fmfiset{b2}{vloc(__v1){dir 90} .. cb1{dir 130}}
    \fmfiset{b3}{vloc(__v1){dir 57} .. cb3}
    \fmfiset{b4}{vloc(__v1){dir 13.1} .. cb4}
    \fmfi{plain}{b1}
    \fmfi{plain}{b2}
    \fmfi{plain}{b3}
    \fmfi{plain}{b4}
\end{fmfgraph*}
\end{fmffile}}
,
\parbox{55pt}{\begin{fmffile}{2ndOrder_lnZ_TwoBody2_UnSymm}
    \begin{fmfgraph*}(50,100)
       \fmfkeep{2ndOrder_lnZ_TwoBody2_UnSymm}
    \fmfbottom{i}
    \fmftop{o}
    \fmf{phantom, tension=3}{i,v1}
    \fmf{phantom}{v1,v2}
    \fmf{phantom, tension=3}{v2,o}
    \fmffreeze
    \fmf{phantom, right=0.8, tag=1}{v2,v1}
    \fmf{phantom, right=0.3, tag=2}{v2,v1}
    \fmf{phantom, right=0.3, tag=3}{v1,v2}
    \fmf{phantom, right=0.8, tag=4}{v1,v2}
    \fmfv{d.shape=circle,d.filled=empty,d.size=3thick}{v1}
    \fmfv{d.shape=circle,d.filled=empty,d.size=3thick}{v2}
    \fmfipath{circlet,circleb,dcirclet,dcircleb}
    \fmfiset{circlet}{fullcircle scaled 1w shifted vloc(__v2)}
    \fmfiset{circleb}{fullcircle scaled 1w shifted vloc(__v1)}
    \fmfiset{dcirclet}{fullcircle scaled 0.85w shifted vloc(__v2)}
    \fmfiset{dcircleb}{fullcircle scaled 0.85w shifted vloc(__v1)}
    \fmfipath{p[]}
    \fmfiset{p1}{vpath1(__v1,__v2)}
    \fmfiset{p2}{vpath2(__v1,__v2)}
    \fmfiset{p3}{vpath3(__v1,__v2)}
    \fmfiset{p4}{vpath4(__v1,__v2)}
    \fmfi{dashes}{dcirclet}
    \fmfi{dashes}{dcircleb}
    \fmfi{plain}{subpath (15length(p1)/50,35length(p1)/50) of p1}
    \fmfi{plain}{subpath (19length(p2)/50,31length(p2)/50) of p2}
    \fmfi{plain}{subpath (19length(p3)/50,31length(p3)/50) of p3}
    \fmfi{plain}{subpath (15length(p4)/50,35length(p4)/50) of p4}
    % Top part
    \fmfipair{ct[]}
    \fmfiequ{ct1}{p1 intersectionpoint circlet}
    \fmfiequ{ct2}{p2 intersectionpoint circlet}
    \fmfiequ{ct3}{p3 intersectionpoint circlet}
    \fmfiequ{ct4}{p4 intersectionpoint circlet}
    \fmfipath{t[]}
    \fmfiset{t1}{vloc(__v2){dir -166.9} .. ct1}
    \fmfiset{t2}{vloc(__v2){dir -123} .. ct2}
    \fmfiset{t3}{vloc(__v2){dir -57} .. ct3}
    \fmfiset{t4}{vloc(__v2){dir -13.1} .. ct4}
    \fmfi{plain}{t1}
    \fmfi{plain}{t2}
    \fmfi{plain}{t3}
    \fmfi{plain}{t4}
    % Bottom part
    \fmfipair{cb[]}
    \fmfiequ{cb1}{p1 intersectionpoint circleb}
    \fmfiequ{cb2}{p2 intersectionpoint circleb}
    \fmfiequ{cb3}{p3 intersectionpoint circleb}
    \fmfiequ{cb4}{p4 intersectionpoint circleb}
    \fmfipath{b[]}
    \fmfiset{b1}{vloc(__v1){dir 180} .. cb3{dir 80}}
    \fmfiset{b2}{vloc(__v1){dir 123} .. cb1{dir 130}}
    \fmfiset{b3}{vloc(__v1){dir 57} .. cb2{dir 110}}
    \fmfiset{b4}{vloc(__v1){dir 13.1} .. cb4}
    \fmfi{plain}{b1}
    \fmfi{plain}{b2}
    \fmfi{plain}{b3}
    \fmfi{plain}{b4}
\end{fmfgraph*}
\end{fmffile}}
,
\parbox{55pt}{\begin{fmffile}{2ndOrderGF_TwoBody3_UnSymm}
    \begin{fmfgraph*}(50,100)
       \fmfkeep{2ndOrder_lnZ_TwoBody3_UnSymm}
    \fmfbottom{i}
    \fmftop{o}
    \fmf{phantom, tension=3}{i,v1}
    \fmf{phantom}{v1,v2}
    \fmf{phantom, tension=3}{v2,o}
    \fmffreeze
    \fmf{phantom, right=0.8, tag=1}{v2,v1}
    \fmf{phantom, right=0.3, tag=2}{v2,v1}
    \fmf{phantom, right=0.3, tag=3}{v1,v2}
    \fmf{phantom, right=0.8, tag=4}{v1,v2}
    \fmfv{d.shape=circle,d.filled=empty,d.size=3thick}{v1}
    \fmfv{d.shape=circle,d.filled=empty,d.size=3thick}{v2}
    \fmfipath{circlet,circleb,dcirclet,dcircleb}
    \fmfiset{circlet}{fullcircle scaled 1w shifted vloc(__v2)}
    \fmfiset{circleb}{fullcircle scaled 1w shifted vloc(__v1)}
    \fmfiset{dcirclet}{fullcircle scaled 0.85w shifted vloc(__v2)}
    \fmfiset{dcircleb}{fullcircle scaled 0.85w shifted vloc(__v1)}
    \fmfipath{p[]}
    \fmfiset{p1}{vpath1(__v1,__v2)}
    \fmfiset{p2}{vpath2(__v1,__v2)}
    \fmfiset{p3}{vpath3(__v1,__v2)}
    \fmfiset{p4}{vpath4(__v1,__v2)}
    \fmfi{dashes}{dcirclet}
    \fmfi{dashes}{dcircleb}
    \fmfi{plain}{subpath (15length(p1)/50,35length(p1)/50) of p1}
    \fmfi{plain}{subpath (19length(p2)/50,31length(p2)/50) of p2}
    \fmfi{plain}{subpath (19length(p3)/50,31length(p3)/50) of p3}
    \fmfi{plain}{subpath (15length(p4)/50,35length(p4)/50) of p4}
    % Top part
    \fmfipair{ct[]}
    \fmfiequ{ct1}{p1 intersectionpoint circlet}
    \fmfiequ{ct2}{p2 intersectionpoint circlet}
    \fmfiequ{ct3}{p3 intersectionpoint circlet}
    \fmfiequ{ct4}{p4 intersectionpoint circlet}
    \fmfipath{t[]}
    \fmfiset{t1}{vloc(__v2){dir -166.9} .. ct1}
    \fmfiset{t2}{vloc(__v2){dir -155} .. ct3{dir -95}}
    \fmfiset{t3}{vloc(__v2){dir -25} .. ct2{dir -85}}
    \fmfiset{t4}{vloc(__v2){dir -13.1} .. ct4}
    \fmfi{plain}{t1}
    \fmfi{plain}{t2}
    \fmfi{plain}{t3}
    \fmfi{plain}{t4}
    % Bottom part
    \fmfipair{cb[]}
    \fmfiequ{cb1}{p1 intersectionpoint circleb}
    \fmfiequ{cb2}{p2 intersectionpoint circleb}
    \fmfiequ{cb3}{p3 intersectionpoint circleb}
    \fmfiequ{cb4}{p4 intersectionpoint circleb}
    \fmfipath{b[]}
    \fmfiset{b1}{vloc(__v1){dir 180} .. tension 1 and 2 .. cb1}
    \fmfiset{b2}{vloc(__v1){dir 123} .. cb2}
    \fmfiset{b3}{vloc(__v1){dir 57} .. cb3}
    \fmfiset{b4}{vloc(__v1){dir 13.1} .. cb4}
    \fmfi{plain}{b1}
    \fmfi{plain}{b2}
    \fmfi{plain}{b3}
    \fmfi{plain}{b4}
\end{fmfgraph*}
\end{fmffile}}
,
\parbox{55pt}{\begin{fmffile}{2ndOrderGF_TwoBody4_UnSymm}
    \begin{fmfgraph*}(50,100)
       \fmfkeep{2ndOrder_lnZ_TwoBody4_UnSymm}
    \fmfbottom{i}
    \fmftop{o}
    \fmf{phantom, tension=3}{i,v1}
    \fmf{phantom}{v1,v2}
    \fmf{phantom, tension=3}{v2,o}
    \fmffreeze
    \fmf{phantom, right=0.8, tag=1}{v2,v1}
    \fmf{phantom, right=0.3, tag=2}{v2,v1}
    \fmf{phantom, right=0.3, tag=3}{v1,v2}
    \fmf{phantom, right=0.8, tag=4}{v1,v2}
    \fmfv{d.shape=circle,d.filled=empty,d.size=3thick}{v1}
    \fmfv{d.shape=circle,d.filled=empty,d.size=3thick}{v2}
    \fmfipath{circlet,circleb,dcirclet,dcircleb}
    \fmfiset{circlet}{fullcircle scaled 1w shifted vloc(__v2)}
    \fmfiset{circleb}{fullcircle scaled 1w shifted vloc(__v1)}
    \fmfiset{dcirclet}{fullcircle scaled 0.85w shifted vloc(__v2)}
    \fmfiset{dcircleb}{fullcircle scaled 0.85w shifted vloc(__v1)}
    \fmfipath{p[]}
    \fmfiset{p1}{vpath1(__v1,__v2)}
    \fmfiset{p2}{vpath2(__v1,__v2)}
    \fmfiset{p3}{vpath3(__v1,__v2)}
    \fmfiset{p4}{vpath4(__v1,__v2)}
    \fmfi{dashes}{dcirclet}
    \fmfi{dashes}{dcircleb}
    \fmfi{plain}{subpath (15length(p1)/50,35length(p1)/50) of p1}
    \fmfi{plain}{subpath (19length(p2)/50,31length(p2)/50) of p2}
    \fmfi{plain}{subpath (19length(p3)/50,31length(p3)/50) of p3}
    \fmfi{plain}{subpath (15length(p4)/50,35length(p4)/50) of p4}
    % Top part
    \fmfipair{ct[]}
    \fmfiequ{ct1}{p1 intersectionpoint circlet}
    \fmfiequ{ct2}{p2 intersectionpoint circlet}
    \fmfiequ{ct3}{p3 intersectionpoint circlet}
    \fmfiequ{ct4}{p4 intersectionpoint circlet}
    \fmfipath{t[]}
    \fmfiset{t1}{vloc(__v2){dir -166.9} .. ct1}
    \fmfiset{t2}{vloc(__v2){dir -123} .. ct2}
    \fmfiset{t3}{vloc(__v2){dir -90} .. ct4{dir -50}}
    \fmfiset{t4}{vloc(__v2){dir 0} .. ct3{dir -80}}
    \fmfi{plain}{t1}
    \fmfi{plain}{t2}
    \fmfi{plain}{t3}
    \fmfi{plain}{t4}
    % Bottom part
    \fmfipair{cb[]}
    \fmfiequ{cb1}{p1 intersectionpoint circleb}
    \fmfiequ{cb2}{p2 intersectionpoint circleb}
    \fmfiequ{cb3}{p3 intersectionpoint circleb}
    \fmfiequ{cb4}{p4 intersectionpoint circleb}
    \fmfipath{b[]}
    \fmfiset{b1}{vloc(__v1){dir 180} .. cb3{dir 80}}
    \fmfiset{b2}{vloc(__v1){dir 123} .. cb1{dir 130}}
    \fmfiset{b3}{vloc(__v1){dir 57} .. cb2{dir 110}}
    \fmfiset{b4}{vloc(__v1){dir 13.1} .. cb4}
    \fmfi{plain}{b1}
    \fmfi{plain}{b2}
    \fmfi{plain}{b3}
    \fmfi{plain}{b4}
\end{fmfgraph*}
\end{fmffile}}
\vspace{0.25cm}
\caption{Top: a second-order diagram $\mathscr{G}^Z_2$ contributing to
$\ln{\frac{Z}{Z_0}}$ which is expressed with fully antisymmetric vertices
(filled dots). Bottom: four examples out of the 24 distinct diagrams in
$\text{Sym}(\mathscr{G}^Z_2)$ expressed in terms of
un-symmetrised vertices (empty dots). All diagrams are distinct only
by their part within the dashed circles where half-lines are permuted.}
\label{fig:TwoBodyUnsymDiagEx}
\end{figure}

Let us consider the Feynman rules that are obtained in terms of
un-symmetrised vertices. These can be obtained following, for instance, Chap.~$5$
of Ref.~\cite{Blaizot1986} by using Wick's theorem and recasting
time-ordered integrals.
Compared to Sec.~\ref{subsec:FeynRulesTime}, the only difference lies in the
symmetry factor, which includes only considerations from permutation of vertices;
and in the expressions of the vertices, which are not symmetrised.
For a perturbation theory defined by the partitioning of Eq.~\eqref{GeneralPartitionPT},
and a diagram $\mathscr{G}_n$ with $n$ vertices, the generic Feynman amplitude 
reads
\begin{multline}
    \mathcal{A}_{\text{Unsymm}}^{\mu_1 \dots \mu_{2k}}(\tau_{\mu_1}, \dots, \tau_{\mu_{2k}})
    =
    \frac{(-1)^{n+L}}{S} \\
    \times \sum_{\lambda\dots\lambda}
        \frac{v^{(k_1)}_{\lambda \dots \lambda}}{(2k_1)!}
        \dots
        \frac{v^{(k_n)}_{\lambda \dots \lambda}}{(2k_n)!}
        \int^{\beta}_{0} \mathrm{d}\tau_1 \dots \mathrm{d}\tau_n \
        \prod_{e \in I} -\mathcal{G}^{(0)\lambda\lambda}(\tau_{i},\tau_{j}) \\
        \times \prod_{e \in E_{\text{in}}} -\mathcal{G}^{(0)\lambda\mu}(\tau_{i},\tau_{\mu})
        \prod_{e \in E_{\text{out}}} -\mathcal{G}^{(0)\mu\lambda}(\tau_{\mu},\tau_{j})
        \ ,
\end{multline}
where we use the same generic notation as in Eq.~\eqref{GenericFeynmanAmplitudeTime}.
As mentioned earlier, there is no contribution to the symmetry factor coming from
equivalent lines nor tadpoles and
we work directly with the un-symmetrised vertices
$\frac{1}{(2k)!}v^{(k)}_{\mu_1 \dots \mu_{2k}}$ appearing in the Hamiltonian.
We represent these un-symmetrised vertices with an empty dot,
in order to distinguish them from fully antisymmetrised ones that are
represented with solid dots.
Fig.~\ref{fig:TwoBodyUnsymDiagEx} shows the example of a second-order 
diagram in the expansion of $\ln{\frac{Z}{Z_0}}$
with fully antisymmetric vertices (top)
as well as several corresponding diagrams
but with un-symmetrised vertices (bottom).

At this point, in a standard oriented diagrammatic,
we would obtain its Hugenholtz version by
using the antisymmetry of matrix elements
with respect to exchange of outgoing and, separately, of incoming half-lines.
This antisymmetry can be assumed without loss of generality thanks
to the canonical anticommutation rules of creation and annihilation operators,
namely Eqs.~\eqref{CAR_sp1} and \eqref{CAR_sp2}.
To do so, we first define the set of oriented diagrams
$\text{Hug}(\mathscr{G})$ made out of oriented diagrams
$\mathscr{G}'$ differing from an oriented diagram $\mathscr{G}$
by permutations of incoming half-lines attached to a same vertex
and (separately) of outgoing half-lines attached to a same vertex.
Thanks to the assumed antisymmetry of matrix elements and
the minus sign rule, the amplitude associated to any oriented diagram
$\mathscr{G}' \in \text{Hug}(\mathscr{G})$ is the same as the one
of $\mathscr{G}$, referred to as $\mathcal{A}_{\text{Unsymm}}$.
Consequently, we can discard any contribution from oriented diagrams
in $\text{Hug}(\mathscr{G})$ so long as we associate 
the amplitude
\begin{equation}
    \mathcal{A}_{\text{Hug}}
        \equiv \text{Card}\left(\text{Hug}(\mathscr{G})\right)
                \times \mathcal{A}_{\text{Unsymm}} 
\end{equation}
to  $\mathscr{G}$, 
where $\text{Card}\left( \ . \ \right)$ denotes the cardinal of a set.
In practice, the Hugenholtz diagrammatic contributions are obtained directly
from modified Feynman rules, which take into account the notion of
equivalent lines.
For more details on the Hugenholtz version of oriented diagrammatics
we refer to standard textbooks on quantum many-body theory
such as Ref.~\cite{Blaizot1986}.

\begin{figure}[t]
  \centering
\parbox{60pt}{\begin{fmffile}{2ndOrderGF_TwoBody5_UnSymm}
    \begin{fmfgraph*}(50,100)
       \fmfkeep{2ndOrder_lnZ_TwoBody5_UnSymm}
    \fmfbottom{i}
    \fmftop{o}
    \fmf{phantom, tension=3}{i,v1}
    \fmf{phantom}{v1,v2}
    \fmf{phantom, tension=3}{v2,o}
    \fmffreeze
    \fmf{phantom, right=0.8, tag=1}{v2,v1}
    \fmf{phantom, right=0.3, tag=2}{v2,v1}
    \fmf{phantom, right=0.3, tag=3}{v1,v2}
    \fmf{phantom, right=0.8, tag=4}{v1,v2}
    \fmfv{d.shape=circle,d.filled=empty,d.size=3thick}{v1}
    \fmfv{d.shape=circle,d.filled=empty,d.size=3thick}{v2}
    \fmfipath{circlet,circleb,dcirclet,dcircleb}
    \fmfiset{circlet}{fullcircle scaled 1w shifted vloc(__v2)}
    \fmfiset{circleb}{fullcircle scaled 1w shifted vloc(__v1)}
    \fmfiset{dcirclet}{fullcircle scaled 0.85w shifted vloc(__v2)}
    \fmfiset{dcircleb}{fullcircle scaled 0.85w shifted vloc(__v1)}
    \fmfipath{p[]}
    \fmfiset{p1}{vpath1(__v1,__v2)}
    \fmfiset{p2}{vpath2(__v1,__v2)}
    \fmfiset{p3}{vpath3(__v1,__v2)}
    \fmfiset{p4}{vpath4(__v1,__v2)}
    \fmfi{dashes}{dcirclet}
    \fmfi{dashes}{dcircleb}
    \fmfi{plain}{subpath (15length(p1)/50,35length(p1)/50) of p1}
    \fmfi{plain}{subpath (19length(p2)/50,31length(p2)/50) of p2}
    \fmfi{plain}{subpath (19length(p3)/50,31length(p3)/50) of p3}
    \fmfi{plain}{subpath (15length(p4)/50,35length(p4)/50) of p4}
    % Top part
    \fmfipair{ct[]}
    \fmfiequ{ct1}{p1 intersectionpoint circlet}
    \fmfiequ{ct2}{p2 intersectionpoint circlet}
    \fmfiequ{ct3}{p3 intersectionpoint circlet}
    \fmfiequ{ct4}{p4 intersectionpoint circlet}
    \fmfipath{t[]}
    \fmfiset{t1}{vloc(__v2){dir -166.9} .. ct1}
    \fmfiset{t2}{vloc(__v2){dir -155} .. ct3{dir -95}}
    \fmfiset{t3}{vloc(__v2){dir -25} .. ct2{dir -85}}
    \fmfiset{t4}{vloc(__v2){dir -13.1} .. ct4}
    \fmfi{plain}{t1}
    \fmfi{plain}{t2}
    \fmfi{plain}{t3}
    \fmfi{plain}{t4}
    % Bottom part
    \fmfipair{cb[]}
    \fmfiequ{cb1}{p1 intersectionpoint circleb}
    \fmfiequ{cb2}{p2 intersectionpoint circleb}
    \fmfiequ{cb3}{p3 intersectionpoint circleb}
    \fmfiequ{cb4}{p4 intersectionpoint circleb}
    \fmfipath{b[]}
    \fmfiset{b1}{vloc(__v1){dir 180} .. tension 1 and 2 .. cb1}
    \fmfiset{b2}{vloc(__v1){dir 155} .. cb3{dir 95}}
    \fmfiset{b3}{vloc(__v1){dir 25} .. cb2{dir 85}}
    \fmfiset{b4}{vloc(__v1){dir 13.1} .. cb4}
    \fmfi{plain}{b1}
    \fmfi{plain}{b2}
    \fmfi{plain}{b3}
    \fmfi{plain}{b4}
\end{fmfgraph*}
\end{fmffile}}
\hspace{0.5cm}
$\Longleftrightarrow$
\hspace{0.5cm}
\parbox{60pt}{\begin{fmffile}{2ndOrderGF_TwoBody6_UnSymm}
    \begin{fmfgraph*}(50,100)
       \fmfkeep{2ndOrder_lnZ_TwoBody6_UnSymm}
    \fmfbottom{i}
    \fmftop{o}
    \fmf{phantom, tension=3}{i,v1}
    \fmf{phantom}{v1,v2}
    \fmf{phantom, tension=3}{v2,o}
    \fmffreeze
    \fmf{phantom, right=0.8, tag=1}{v2,v1}
    \fmf{phantom, right=0.3, tag=2}{v2,v1}
    \fmf{phantom, right=0.3, tag=3}{v1,v2}
    \fmf{phantom, right=0.8, tag=4}{v1,v2}
    \fmfv{d.shape=circle,d.filled=empty,d.size=3thick}{v1}
    \fmfv{d.shape=circle,d.filled=empty,d.size=3thick}{v2}
    \fmfipath{circlet,circleb,dcirclet,dcircleb}
    \fmfiset{circlet}{fullcircle scaled 1w shifted vloc(__v2)}
    \fmfiset{circleb}{fullcircle scaled 1w shifted vloc(__v1)}
    \fmfiset{dcirclet}{fullcircle scaled 0.85w shifted vloc(__v2)}
    \fmfiset{dcircleb}{fullcircle scaled 0.85w shifted vloc(__v1)}
    \fmfipath{p[]}
    \fmfiset{p1}{vpath1(__v1,__v2)}
    \fmfiset{p2}{vpath2(__v1,__v2)}
    \fmfiset{p3}{vpath3(__v1,__v2)}
    \fmfiset{p4}{vpath4(__v1,__v2)}
    \fmfi{dashes}{dcirclet}
    \fmfi{dashes}{dcircleb}
    \fmfi{plain}{subpath (15length(p1)/50,35length(p1)/50) of p1}
    \fmfi{plain}{subpath (19length(p2)/50,31length(p2)/50) of p2}
    \fmfi{plain}{subpath (19length(p3)/50,31length(p3)/50) of p3}
    \fmfi{plain}{subpath (15length(p4)/50,35length(p4)/50) of p4}
    % Top part
    \fmfipair{ct[]}
    \fmfiequ{ct1}{p1 intersectionpoint circlet}
    \fmfiequ{ct2}{p2 intersectionpoint circlet}
    \fmfiequ{ct3}{p3 intersectionpoint circlet}
    \fmfiequ{ct4}{p4 intersectionpoint circlet}
    \fmfipath{t[]}
    \fmfiset{t1}{vloc(__v2){dir -166.9} .. ct1}
    \fmfiset{t2}{vloc(__v2){dir -123} .. ct2}
    \fmfiset{t3}{vloc(__v2){dir -57} .. ct3}
    \fmfiset{t4}{vloc(__v2){dir -13.1} .. ct4}
    \fmfi{plain}{t1}
    \fmfi{plain}{t2}
    \fmfi{plain}{t3}
    \fmfi{plain}{t4}
    % Bottom part
    \fmfipair{cb[]}
    \fmfiequ{cb1}{p1 intersectionpoint circleb}
    \fmfiequ{cb2}{p2 intersectionpoint circleb}
    \fmfiequ{cb3}{p3 intersectionpoint circleb}
    \fmfiequ{cb4}{p4 intersectionpoint circleb}
    \fmfipath{b[]}
    \fmfiset{b1}{vloc(__v1){dir 180} .. tension 1 and 2 .. cb1}
    \fmfiset{b2}{vloc(__v1){dir 123} .. cb2}
    \fmfiset{b3}{vloc(__v1){dir 57} .. cb3}
    \fmfiset{b4}{vloc(__v1){dir 13.1} .. cb4}
    \fmfi{plain}{b1}
    \fmfi{plain}{b2}
    \fmfi{plain}{b3}
    \fmfi{plain}{b4}
\end{fmfgraph*}
\end{fmffile}}
\vspace{0.25cm}
\caption{Example of two different sets of permutations of half-lines
which gives the same un-oriented diagram.}
\label{fig:TwoBodyUnsymDiagExDbleCnting}
\end{figure}
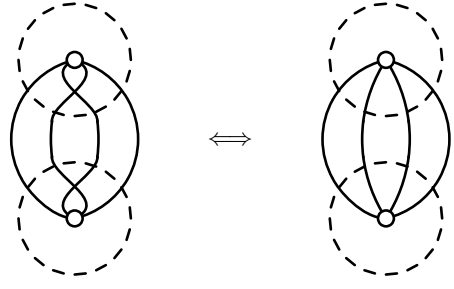

In the case of NCPT, the same procedure cannot be carried
out so easily, because the tensors $v^{(k)}_{\mu_1 \dots \mu_{2k}}$ are
in general \emph{not} totally antisymmetric and the Nambu fields
do not anticommute, namely
\begin{equation}
    \Set{ \mathrm{A}^\mu , \mathrm{A}^\nu } = g^{\mu\nu} \neq 0 \ .
\end{equation}
To prove Eq.~\eqref{GenericFeynmanAmplitudeTime},
which generalises the Hugenholtz antisymmetrisation, we use a different approach. 
We first define the set $\text{Sym}(\mathscr{G}_n)$
of un-oriented diagrams with $n$ un-symmetrised vertices obtained from a given
un-oriented diagram $\mathscr{G}_n$ (also with un-symmetrised vertices)
by any permutation of half-lines attached to a same vertex.
Examples of un-oriented diagrams with un-symmetrised vertices which are obtained
by this procedure are given at the bottom of Fig.~\ref{fig:TwoBodyUnsymDiagEx}.  
Let us emphasise that, compared to the previous Hugenholtz factorisation,
we consider a larger set of permutations where \emph{all} half-lines
of a given vertex can be permuted.
Without the antisymmetry of $v^{(k)}_{\mu_1 \dots \mu_{2k}}$,
the amplitude associated to un-oriented diagrams in
$\text{Sym}(\mathscr{G}_n)$ are not equal and we distinguish them
by specifying a set of permutations (one for each vertex)
denoted by $\sigma_1 \dots \sigma_n$. The sum of amplitudes of
un-oriented diagrams in $\text{Sym}(\mathscr{G}_n)$ defines
$\mathcal{A}^{\mu_1 \dots \mu_{2k}}(\tau_{\mu_1}, \dots, \tau_{\mu_{2k}})$
and reads
\begin{multline}\label{DefSumOverPermutation}
    \mathcal{A}^{\mu_1 \dots \mu_{2k}}(\tau_{\mu_1}, \dots, \tau_{\mu_{2k}})
    =  \frac{1}{\prod_{l=2}^{l_{\text{max}}}(l!)^m} \\
    \sum_{\sigma_1 \dots \sigma_n \in S_{2k_1} \times \dots \times S_{2k_n}}
    \mathcal{A}_{\text{Unsymm}}^{\mu_1 \dots \mu_{2k}}(\tau_{\mu_1}, \dots, \tau_{\mu_{2k}})
    [\sigma_1 \dots \sigma_n] \ ,
\end{multline}
where $\mathcal{A}_{\text{Unsymm}}^{\mu_1 \dots \mu_{2k}}(\tau_{\mu_1}, \dots, \tau_{\mu_{2k}})[\sigma_1 \dots \sigma_n]$ denotes the amplitude
associated to the un-oriented diagram obtained from $\mathscr{G}_n$
by applying the half-line permutations $\sigma_1 \dots \sigma_n$
to its $n$ vertices. The symmetry factor depends on equivalent lines, in such a way
that it compensates for the double counting of un-oriented diagrams in
$\text{Sym}(\mathscr{G}_n)$ when considering all permutations of half-lines
$\sigma_1 \dots \sigma_n \in S_{2k_1} \times \dots \times S_{2k_n}$.
Such double-counting is illustrated in
Fig.~\ref{fig:TwoBodyUnsymDiagExDbleCnting} where the same
un-oriented diagram is obtained with two different sets of permutations.
For simplicity, we are assuming here that the un-oriented diagram
$\mathscr{G}_n$ is free of tadpoles. Following the labelling convention
for vertices defined in Sec.~\ref{subsec:FeynRulesTime},
the amplitude $\mathcal{A}_{\text{Unsymm}}^{\mu_1 \dots \mu_{2k}}(\tau_{\mu_1}, \dots, \tau_{\mu_{2k}})[\sigma_1 \dots \sigma_n]$
of an un-oriented diagram obtained from $\mathscr{G}_n$ with the permutations
$\sigma_1 \dots \sigma_n$ reads
\begin{multline}
    \mathcal{A}_{\text{Unsymm}}^{\mu_1 \dots \mu_{2k}}(\tau_{\mu_1}, \dots,
        \tau_{\mu_{2k}})[\sigma_1 \dots \sigma_n]
    =
    \frac{(-1)^{n+L}}{S} \\
    \times \sum_{\lambda \dots \lambda}
        \epsilon(\sigma_1)
        \frac{v^{(k_1)}_{\lambda \dots \lambda}}{(2k_1)!}
        \dots
        \epsilon(\sigma_n)
        \frac{v^{(k_n)}_{\lambda \dots \lambda}}{(2k_n)!}
        \\
    \times
        \int^{\beta}_{0} \mathrm{d}\tau_1 \dots \mathrm{d}\tau_n \
        \prod_{e \in I} -\mathcal{G}^{(0)\lambda_\sigma\lambda_\sigma}(\tau_{i},\tau_{j}) \\
        \times \prod_{e \in E_{\text{in}}}
            -\mathcal{G}^{(0)\lambda_\sigma\mu}(\tau_{i},\tau_{\mu})
        \prod_{e \in E_{\text{out}}}
            -\mathcal{G}^{(0)\mu\lambda_\sigma}(\tau_{\mu},\tau_{j})
        \ ,
\end{multline}
where the signature of a permutation $\epsilon(\sigma_i)$
comes from extra crossing due to the permutations of half-lines
of the $i^{\text{th}}$ vertex and $\lambda_{\sigma}$ denotes
generically the modified global index, due to the permutations
of half-lines of vertices, on which a propagator is contracted.
Since the $\lambda$'s are dummy indices, we can perform the change 
of variables $\lambda_{\sigma_i} \leftarrow \lambda$
for indices attached to the $i^{\text{th}}$ vertex
(which amounts to renaming the half-lines of the $i^{\text{th}}$ vertex)
which leads to
\begin{multline}\label{AntisymmChangedLabellingLambdas}
    \mathcal{A}_{\text{Unsymm}}^{\mu_1 \dots \mu_{2k}}(\tau_{\mu_1}, \dots,
        \tau_{\mu_{2k}})[\sigma_1 \dots \sigma_n]
    =
    \frac{(-1)^{n+L}}{S} \\
    \times \sum_{\lambda \dots \lambda}
        \epsilon(\sigma_1)
        \frac{v^{(k_1)}_{\lambda_{\sigma^{-1}_1} \dots \lambda_{\sigma^{-1}_1}}}{(2k_1)!}
        \dots
        \epsilon(\sigma_n)
        \frac{v^{(k_n)}_{\lambda_{\sigma^{-1}_n} \dots \lambda_{\sigma^{-1}_n}}}{(2k_n)!}
        \\
    \times
        \int^{\beta}_{0} \mathrm{d}\tau_1 \dots \mathrm{d}\tau_n \
        \prod_{e \in I} -\mathcal{G}^{(0)\lambda\lambda}(\tau_{i},\tau_{j}) \\
        \times
        \prod_{e \in E_{\text{in}}} -\mathcal{G}^{(0)\lambda\mu}(\tau_{i},\tau_{\mu})
        \prod_{e \in E_{\text{out}}} -\mathcal{G}^{(0)\mu\lambda}(\tau_{\mu},\tau_{j})
        \ ,
\end{multline}
where $v^{(k_i)}_{\lambda_{\sigma^{-1}_i} \dots \lambda_{\sigma^{-1}_i}}$ denotes the
value $v^{(k_i)}_{\lambda \dots \lambda}$ where the indices have been permuted
according to $\sigma^{-1}_i$.
Plugging Eq.~\eqref{AntisymmChangedLabellingLambdas} into
Eq.~\eqref{DefSumOverPermutation}, and performing the change of variables
$\sigma^{-1}_i \leftarrow \sigma_i$ we obtain
the amplitude associated to $\text{Sym}(\mathscr{G}_n)$, namely
\begin{multline}\label{PluggedInSumOverPermutation}
    \mathcal{A}^{\mu_1 \dots \mu_{2k}}(\tau_{\mu_1}, \dots, \tau_{\mu_{2k}})
    =  \\
    \frac{1}{\prod_{l=2}^{l_{\text{max}}}(l!)^m}
    \sum_{\sigma_1 \dots \sigma_n \in S_{2k_1} \times \dots \times S_{2k_n}}
    \frac{(-1)^{n+L}}{S} \\
    \times \sum_{\lambda \dots \lambda}
        \epsilon(\sigma_1)
        \frac{v^{(k_1)}_{\lambda_{\sigma_1} \dots \lambda_{\sigma_1}}}{(2k_1)!}
        \dots
        \epsilon(\sigma_n)
        \frac{v^{(k_n)}_{\lambda_{\sigma_n} \dots \lambda_{\sigma_n}}}{(2k_n)!}
        \\
    \times
        \int^{\beta}_{0} \mathrm{d}\tau_1 \dots \mathrm{d}\tau_n \
        \prod_{e \in I} -\mathcal{G}^{(0)\lambda\lambda}(\tau_{i},\tau_{j}) \\
        \times \prod_{e \in E_{\text{in}}} -\mathcal{G}^{(0)\lambda\mu}(\tau_{i},\tau_{\mu})
        \prod_{e \in E_{\text{out}}} -\mathcal{G}^{(0)\mu\lambda}(\tau_{\mu},\tau_{j})
        \ ,
\end{multline}
where we have used the identity $\epsilon(\sigma^{-1}) = \epsilon(\sigma)$.
The key point here is that the propagators appearing in
Eq.~\eqref{PluggedInSumOverPermutation} do not explicitly depend on
the permutations $\sigma_1, \dots \sigma_n$, which allows us to factor
them out of the sum over $\sigma_1, \dots \sigma_n$.
Assuming we can permute the sums over $\lambda$'s and over $\sigma$'s,
we obtain
\begin{multline}\label{GenericFeynmanAmplitudeTimeProof}
    \mathcal{A}^{\mu_1 \dots \mu_{2k}}(\tau_{\mu_1}, \dots, \tau_{\mu_{2k}})
    = 
    \frac{(-1)^{n+L}}{S \times \prod_{l=2}^{l_{\text{max}}}(l!)^m} \\
    \times \sum_{\lambda \dots \lambda}
        \left(
            \frac{1}{(2k_1)!}
            \sum_{\sigma_1 \in S_{2k_1}}
            \epsilon(\sigma_1)
            v^{(k_1)}_{\lambda_{\sigma_1} \dots \lambda_{\sigma_1}}
        \right) \phantom{
                    \sum_{\sigma_1}
                    \epsilon(\sigma_1)
                    v^{(k_1)}_{\lambda_{\sigma_1} \dots \lambda_{\sigma_1}}
                }
        \\
        \times \dots \times
        \left(
            \frac{1}{(2k_n)!}
            \sum_{\sigma_n \in S_{2k_n}}
            \epsilon(\sigma_n)
            v^{(k_n)}_{\lambda_{\sigma_n} \dots \lambda_{\sigma_n}}
        \right)
        \\
    \times
        \int^{\beta}_{0} \mathrm{d}\tau_1 \dots \mathrm{d}\tau_n \
        \prod_{e \in I} -\mathcal{G}^{(0)\lambda\lambda}(\tau_{i},\tau_{j}) \\
        \times
        \prod_{e \in E_{\text{in}}} -\mathcal{G}^{(0)\lambda\mu}(\tau_{i},\tau_{\mu})
        \prod_{e \in E_{\text{out}}} -\mathcal{G}^{(0)\mu\lambda}(\tau_{\mu},\tau_{j})
\end{multline}
which is exactly Eq.~\eqref{GenericFeynmanAmplitudeTime} in the case where there
are no tadpoles.
Therefore, instead of summing amplitudes associated to un-oriented diagrams
with un-symmetrised vertices, we sum over the distinct 
$\text{Sym}(\mathscr{G}_n)$ which are faithfully represented by %the so-called
un-oriented diagrams with the fully antisymmetric vertices defined in
Eq.~\eqref{FullyAntisymVertexSym}.

For an un-oriented diagram $\mathscr{G}_n$
(with un-symmetrised vertices) that contains $p_1, \dots, p_n$ tadpoles on its vertices,
the derivation is similar.
The only difference is in Eq.~\eqref{DefSumOverPermutation}, which
we replace by
\begin{multline}
    \mathcal{A}^{\mu_1 \dots \mu_{2k}}(\tau_{\mu_1}, \dots, \tau_{\mu_{2k}})
    =  \frac{\prod^{n}_{i=1} 2^{p_i}}{2^T} 
    \frac{\prod^{n}_{i=1} p_i!}{\prod_{l=2}^{l_{\text{max}}}(l!)^m} \\
    \sum_{\substack{\sigma_1 \dots \sigma_n \\
            \in (S_{2k_1}/S^{p_1}_2 \times S_{p_1}) 
                \\\times \dots 
                \times (S_{2k_n}/S^{p_n}_2 \times S_{p_n})
        }}
        \mathcal{A}_{\text{Unsymm}}^{\mu_1 \dots \mu_{2k}}
        (\tau_{\mu_1}, \dots, \tau_{\mu_{2k}})[\sigma_1 \dots \sigma_n] \ ,
\end{multline}
where $S_{2k}/S^{p}_2 \times S_p$ denotes the set
of permutations that do not permute half-lines inside tadpoles nor
several tadpoles between them.
The term $p_i!$ appears in the numerator to compensate for the same term
in the denominator coming from the $p_i$-tuple of equivalent lines making
the $p_i$ tadpoles on the $i^{\text{th}}$ vertex.
There is indeed no need to compensate double-counting
in this case since the set of permutations, $S_{2k}/S^{p}_2 \times S_p$,
is already restricted to not contain any permutation
exchanging tadpoles on a given vertex.
The remaining factor does not modify the amplitude, because
the total number of tadpoles is $T = p_1 + \dots + p_n$ so that
\begin{equation}
    \frac{\prod^{n}_{i=1} 2^{p_i}}{2^T} = 1 \ .
\end{equation}
Similarly to the case without tadpoles,
Eq.~\eqref{GenericFeynmanAmplitudeTimeProof} eventually reads
\begin{multline}\label{GenericFeynmanAmplitudeTimeProofTadpole}
    \mathcal{A}^{\mu_1 \dots \mu_{2k}}(\tau_{\mu_1}, \dots, \tau_{\mu_{2k}})
    = 
    \frac{(-1)^{n+L}}{S \times 2^T \prod_{l=2}^{l_{\text{max}}}(l!)^m} \\
    \times \sum_{\lambda \dots \lambda}
        \left(
            \frac{2^{p_1} p_1 !}{(2k_1)!}
            \sum_{\sigma_1 \in S_{2k_1}/S^{p_1}_2 \times S_{p_1}}
            \epsilon(\sigma_1)
            v^{(k_1)}_{\lambda_{\sigma_1} \dots \lambda_{\sigma_1}}
        \right) \phantom{
                    v^{(k_1)}_{\lambda_{\sigma_1} \dots \lambda_{\sigma_1}}
                }
        \\
        \times \dots \times
        \left(
            \frac{2^{p_n} p_n !}{(2k_n)!}
            \sum_{\sigma_n \in S_{2k_n}/S^{p_n}_2 \times S_{p_n}}
            \epsilon(\sigma_n)
            v^{(k_n)}_{\lambda_{\sigma_n} \dots \lambda_{\sigma_n}}
        \right)
        \\
    \times
        \int^{\beta}_{0} \mathrm{d}\tau_1 \dots \mathrm{d}\tau_n \
        \prod_{e \in I} -\mathcal{G}^{(0)\lambda\lambda}(\tau_{i},\tau_{j}) \\
        \times
        \prod_{e \in E_{\text{in}}} -\mathcal{G}^{(0)\lambda\mu}(\tau_{i},\tau_{\mu})
        \prod_{e \in E_{\text{out}}} -\mathcal{G}^{(0)\mu\lambda}(\tau_{\mu},\tau_{j})
        \ ,
\end{multline}
where the partial antisymmetrisation of vertices appears explicitly as defined
in Eq.~\eqref{PartialAntisymVertexSym}.

\subsection{Antisymmetrised interaction}\label{subsec:antisymVerticesElmts}
We now proceed to explain how the fully antisymmetrised vertices of NCPT
can be obtained from standard interaction matrix elements. 
Let us consider the perturbation theory defined
by the partition
\begin{subequations}
  \begin{align}
    H &\equiv H_0 + H_1 \ , \\
    H_0 &\equiv
      \frac{1}{2} \sum_{\mu\nu} U_{\mu\nu} \mathrm{A}^\mu \mathrm{A}^\nu \ , \\
    H_1 &\equiv 
        \frac{1}{4!}
        \sum_{\mu_1 \mu_2 \mu_3 \mu_4}
            v^{(2)}_{\mu_1 \mu_2 \mu_3 \mu_4} \
            \mathrm{A}^{\mu_1} \mathrm{A}^{\mu_2}
            \mathrm{A}^{\mu_3} \mathrm{A}^{\mu_4} \ ,
    \label{TwoBodyPertNambu}
\end{align}
\end{subequations}
where only a pure two-body interaction contributes to the perturbation,
and where the Nambu fields $\mathrm{A}^{\mu}$ are left to be specified.
Let $\mathcal{B}$ be a single-particle basis.
In the standard Gorkov formalism, the perturbation $H_1$
is expressed in terms of $\bar{V}_{bcde}$ such that, in the basis $\mathcal{B}$,
\begin{equation}
    H_1 = \frac{1}{(2!)^2} 
        \sum_{bcde} \bar{V}_{bcde} \ \bar{a}_{b} \bar{a}_{c} a_{e} a_{d} \ .
        \label{TwoBodyPertSp}
\end{equation}
Without loss of generality, $\bar{V}_{bcde}$ is assumed to be partially
antisymmetric, i.e.\ 
\begin{equation}\label{Antiymmetries_barV}
    \bar{V}_{bcde} =  - \bar{V}_{cbde} = - \bar{V}_{bced} = \bar{V}_{cbed} \ . 
\end{equation}
It is convenient to work with the canonical field basis
$\mathcal{B}^f = \set{\bar{a}_b} \cup \set{a_b}$, i.e.\ 
% $\mathcal{B}^f = \mathcal{B} \cup \bar{\mathcal{B}}$, i.e.\ 
with the contravariant Nambu fields
\begin{subequations}
\begin{align}
    \mathrm{A}^{(b, 1)} &= a_b \ , \\
    \mathrm{A}^{(b, 2)} &= \bar{a}_{b} \ .
\end{align}
\end{subequations}
As discussed in~\ref{app:MatElts}, the relation between
$v^{(2)}_{\mu_1 \mu_2 \mu_3 \mu_4}$ and $\bar{V}_{bcde}$ reads
\begin{equation}\label{TwoBody_NambuCov_SpGorkov}
    \frac{1}{3!} v^{(2)}_{(b,l_b) (c,l_c) (d,l_d) (e,l_e)} =
         \bar{V}_{b c e d} \, 
            E^{21}_{l_b l_e} E^{21}_{l_c l_d}  \ ,
\end{equation}
where the $\frac{1}{3!}$ factor is due to a different normalisation in
Eq.~\eqref{TwoBodyPertNambu} and Eq.~\eqref{TwoBodyPertSp}.

In~\ref{subsec:FactorUnsym} we have shown why only
the totally antisymmetric part of vertices contributes to the amplitude
of an un-oriented Feynman diagram. In the case of vertices with tadpoles,
a certain partial antisymmetrisation occurs instead.
Regarding the totally antisymmetric part of $v^{(2)}_{\mu_1 \mu_2 \mu_3 \mu_4}$,
we use the symmetries in Eq.~\eqref{Antiymmetries_barV} to simplify the $4!$ terms
into a sum of only $6$ terms, namely
\begin{align}\label{ExplAntisym2Bvert}
v^{(2)}_{[(b,l_b) (c,l_c) (d,l_d) (e,l_e)]} &=
          \bar{V}_{bced} \, E^{21}_{l_b l_e} E^{21}_{l_c l_d}
        + \bar{V}_{decb} \, E^{21}_{l_d l_c} E^{21}_{l_e l_b}
    \nonumber \\
    &\phantom{} 
        - \bar{V}_{bdec} \, E^{21}_{l_b l_e} E^{21}_{l_d l_c}
        - \bar{V}_{cedb} \, E^{21}_{l_c l_d} E^{21}_{l_e l_b}
    \nonumber \\
    &\phantom{} 
        + \bar{V}_{bedc} \, E^{21}_{l_b l_d} E^{21}_{l_e l_c}
        + \bar{V}_{cdeb} \, E^{21}_{l_c l_e} E^{21}_{l_d l_b}
    \ .
\end{align}
Regarding partial antisymmetric parts of $v^{(2)}_{\mu_1 \mu_2 \mu_3 \mu_4}$,
we consider as an example $v^{(2)}_{[\mu_1 \dot{\mu}_2 \dot{\mu}_3 \mu_4]}$
which appears in the first order expansion of the propagator given in
Eq.~\eqref{1stOrderkmax2Summed}.
Similarly to Eq.~\eqref{ExplAntisym2Bvert}, we obtain a sum of $4$ terms namely
\begin{align}\label{ExplPartialAntisym2Bvert}
v^{(2)}_{[(b,l_b) \dot{(c,l_c)} \dot{(d,l_d)} (e,l_e)]} &=
      2 \bar{V}_{bced} \, E^{21}_{l_b l_e} E^{21}_{l_c l_d}
    \nonumber \\
&\phantom{}
    - 2 \bar{V}_{cedb} \, E^{21}_{l_c l_d} E^{21}_{l_e l_b}
    \nonumber \\
&\phantom{}
    + \bar{V}_{bedc} \, E^{21}_{l_b l_d} E^{21}_{l_e l_c}
    + \bar{V}_{cdeb} \, E^{21}_{l_c l_e} E^{21}_{l_d l_b}
    \ .
\end{align}
The above expressions of the totally and partially antisymmetric vertices
are simple because of our choice of basis.
We stress that this choice is arbitrary and might not be optimal for some 
specific applications. One may thus want to perform
a change of field basis. This is very easily done thanks to the 
Nambu covariance of both $v^{(2)}_{[\mu_1 \mu_2 \mu_3 \mu_4]}$
and $v^{(2)}_{[\mu_1 \dot{\mu}_2 \dot{\mu}_3 \mu_4]}$.

Let us mention two common cases where the expressions simplify further.
The first case consists in assuming the interaction potential to be Hermitian
and having real-valued matrix elements.
In this case, one can work with an orthonormal single-particle basis
$\mathcal{B}$ so that the Hermitian property reads
\begin{equation}\label{HermitianMatrixElts}
        \bar{V}^*_{bcde} = \bar{V}_{debc} \ .
\end{equation}
Combined with the realness assumption, the matrix elements verify
\begin{equation}\label{HermitianANDrealMatrixElts}
        \bar{V}_{bcde} = \bar{V}_{debc} \ .
\end{equation}
With this,
the totally and partially antisymmetric parts of the vertices simplify, respectively, to
\begin{subequations} \label{HermitianOnlyMatEls}
\begin{align}
v^{(2)}_{[(b,l_b) (c,l_c) (d,l_d) (e,l_e)]} &=
        \bar{V}_{bced}
          ( E^{21}_{l_b l_e} E^{21}_{l_c l_d}
            + E^{21}_{l_d l_c} E^{21}_{l_e l_b} )
    \nonumber \\
    &\phantom{} 
        - \bar{V}_{bdec}
            ( E^{21}_{l_b l_e} E^{21}_{l_d l_c}
              + E^{21}_{l_c l_d} E^{21}_{l_e l_b} )
    \nonumber \\
    &\phantom{} 
        + \bar{V}_{bedc}
            ( E^{21}_{l_b l_d} E^{21}_{l_e l_c}
              + E^{21}_{l_c l_e} E^{21}_{l_d l_b} )
    \ , \\
v^{(2)}_{[(b,l_b) \dot{(c,l_c)} \dot{(d,l_d)} (e,l_e)]} &=
      2 \bar{V}_{bced} \, E^{21}_{l_b l_e} E^{21}_{l_c l_d}
    \nonumber \\
&\phantom{}
    - 2 \bar{V}_{cedb} \, E^{21}_{l_c l_d} E^{21}_{l_e l_b}
    \nonumber \\
&\phantom{}
    + \bar{V}_{bedc}
        ( E^{21}_{l_b l_d} E^{21}_{l_e l_c}
          + E^{21}_{l_c l_e} E^{21}_{l_d l_b} )
    \ .
\end{align}
\end{subequations}

The second common simplification arises if we assume the potential to be Hermitian
and time-reversal invariant. In this case, we again take $\mathcal{B}$
to be orthonormal and the potential matrix elements verify
Eq.~\eqref{HermitianMatrixElts}.
To simplify the expressions of the antisymmetrised potential, it is now convenient
to work in the field basis
${\mathcal{B}^f}' = \set{a^\dagger_b} \cup \set{a_{\tilde{b}}}$,
where $a_{\tilde{b}}$ are the annihilation operators
associated to the orthonormal single-particle basis $\tilde{\mathcal{B}}$
obtained from $\mathcal{B}$ by applying the time-reversal operator.
In this case, time-reversal invariance
implies
\begin{equation}\label{TimeReversalMatrixElts}
    \bar{V}^*_{bcde} = \bar{V}_{\tilde{b}\tilde{c}\tilde{d}\tilde{e}} \ ,
\end{equation}
where $\bar{V}_{\tilde{b}\tilde{c}\tilde{d}\tilde{e}}$ denotes the matrix elements
of the potential in the $\tilde{\mathcal{B}}$ single-particle basis.
Combined with the Hermitian property of Eq.~\eqref{HermitianMatrixElts}, the matrix elements verify
\begin{equation}\label{HermitianANDTimeReversalMatrixElts}
    \bar{V}_{bcde} = \bar{V}_{\tilde{d}\tilde{e}\tilde{b}\tilde{c}} \ .
\end{equation}
Then, performing a non-canonical change of field basis to go from
$\mathcal{B}^f$ to ${\mathcal{B}^f}'$ and using
Eq.~\eqref{HermitianANDTimeReversalMatrixElts}, we obtain
\begin{subequations}
\begin{align}
v^{(2)}_{[(b,l_b) (c,l_c) (d,l_d) (e,l_e)]} &=
        \bar{V}_{\tilde{b}\tilde{c}ed}
          ( E^{21}_{l_b l_e} E^{21}_{l_c l_d}
            + E^{21}_{l_d l_c} E^{21}_{l_e l_b} )
    \nonumber \\
    &\phantom{} 
        - \bar{V}_{\tilde{b}\tilde{d}ec}
            ( E^{21}_{l_b l_e} E^{21}_{l_d l_c}
              + E^{21}_{l_c l_d} E^{21}_{l_e l_b} )
    \nonumber \\
    &\phantom{} 
        + \bar{V}_{\tilde{b}\tilde{e}dc}
            ( E^{21}_{l_b l_d} E^{21}_{l_e l_c}
              + E^{21}_{l_c l_e} E^{21}_{l_d l_b} )
    \ , \\
v^{(2)}_{[(b,l_b) \dot{(c,l_c)} \dot{(d,l_d)} (e,l_e)]} &=
      2 \bar{V}_{\tilde{b}\tilde{c}ed} \, E^{21}_{l_b l_e} E^{21}_{l_c l_d}
    \nonumber \\
&\phantom{}
    - 2 \bar{V}_{\tilde{c}\tilde{e}db} \, E^{21}_{l_c l_d} E^{21}_{l_e l_b}
    \nonumber \\
&\phantom{}
    + \bar{V}_{\tilde{b}\tilde{e}dc}
        ( E^{21}_{l_b l_d} E^{21}_{l_e l_c}
          + E^{21}_{l_c l_e} E^{21}_{l_d l_b} )
    \ .
\end{align}
\end{subequations}
These expressions are very similar to Eqs.~\eqref{HermitianOnlyMatEls}, 
but now include matrix elements between single-particle states
and their time-reversal.

\section{Gaudin's rules for a general diagram}\label{app:SumGeneralGraph}
In Sec.~\ref{subsec:MatsubaraSum}, we discussed Gaudin's summation rules
for evaluating the Matsubara frequency sums appearing in the algebraic expression 
of diagrams in the energy representation. To be concise, we focused on
connected diagrams without tadpoles nor external lines. 
We now proceed to discuss the required extensions to evaluate
the sums for any diagram.

\subsection{Extension to tadpoles}\label{subsec:tadpoleGaudin}
Let $\mathscr{G}$ be a connected un-oriented Feynman diagram without external lines.
If $\mathscr{G}$ contains tadpoles, we can still apply the same set of
Gaudin's summation rules described in Sec.~\ref{subsec:MatsubaraSum},
so long as we perform an intermediate pre-processing step. 
This step consists simply in analytically performing the sum
over Matsubara frequencies stemming from each tadpole.
For any given tadpole, the Matsubara sum is performed explicitly by applying the identity
\begin{equation}\label{tadpoleIdentity}
    \frac{1}{\beta}
    \sum_{\omega_t} -\mathcal{G}^{(0)\mu_t\nu_t}(\omega_t) e^{-i \omega_t \eta}
    =
    \sum_{n_t} f(-\epsilon_{n_t}) \ X^{(n_t) \mu_t} \bar{X}^{(n_t) \nu_t} \ ,
\end{equation}
where $\omega_t$ are Matsubara frequencies;
$\epsilon_{n_t}$, quasiparticle energies; 
and $\mu_t$ and $\nu_t$, global
indices associated to the tadpole $t$.

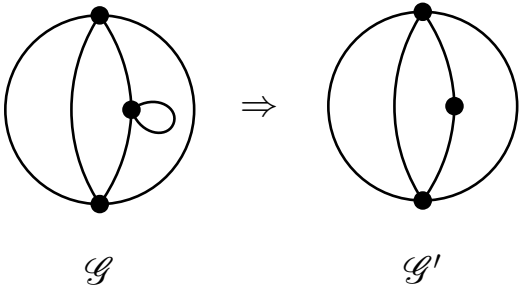
\begin{figure}[t]
  \centering
  \parbox{100pt}{\begin{fmffile}{3rdOrderGF_Tadpole}
    \begin{fmfgraph*}(100,100)
       \fmfkeep{3rdOrderGF_Tadpole}
    \fmfbottom{i}
    \fmftop{o}
    \fmf{phantom, tension=10}{i,v1}
    \fmf{phantom, tension=10}{v2,o}
    \fmf{plain, right=1, tag=1}{v1,v2}
    \fmf{plain, left=1, tag=4}{v1,v2}
    \fmffreeze
    \fmf{plain, left=0.3, tag=3}{v1,v2}
    \fmf{plain, tension=1.5}{vmid,vmid}
    \fmf{plain, right=0.15, tag=2}{v1,vmid,v2}
    \fmffreeze
    \fmfshift{(0.12w,0)}{vmid}
    \fmfv{d.shape=circle,d.filled=full,d.size=3thick}{v1}
    \fmfv{d.shape=circle,d.filled=full,d.size=3thick}{v2}
    \fmfv{d.shape=circle,d.filled=full,d.size=3thick}{vmid}
    \fmflabel{\Large$\mathscr{G}$}{i}
\end{fmfgraph*}
\end{fmffile}} \Large$\Rightarrow$
  \parbox{100pt}{\begin{fmffile}{3rdOrderGF_TadpoleStripped}
  \begin{fmfgraph*}(100,100)
    \fmfkeep{3rdOrderGF_TadpoleStripped}
    \fmfbottom{i}
    \fmftop{o}
    \fmf{phantom, tension=10}{i,v1}
    \fmf{phantom, tension=10}{v2,o}
    \fmf{plain, right=1, tag=1}{v1,v2}
    \fmf{plain, left=1, tag=4}{v1,v2}
    \fmffreeze
    \fmf{plain, left=0.3, tag=3}{v1,v2}
    \fmf{plain, right=0.15, tag=2}{v1,vmid,v2}
    \fmffreeze
    \fmfshift{(0.12w,0)}{vmid}
    \fmfv{d.shape=circle,d.filled=full,d.size=3thick}{v1}
    \fmfv{d.shape=circle,d.filled=full,d.size=3thick}{v2}
    \fmfv{d.shape=circle,d.filled=full,d.size=3thick}{vmid}
    \fmflabel{\Large$\mathscr{G}'$}{i}
\end{fmfgraph*}
\end{fmffile}}
\vspace{0.5cm}
\caption{Example of a $3^{\text{rd}}$ order diagram $\mathscr{G}$
containing a tadpole (left), together with its associated diagram $\mathscr{G}'$
obtained by stripping tadpoles (right).}
\label{fig:tadpoleCaseExtension}
\end{figure}

Having performed the tadpole Matsubara sums, we still need to evaluate the remaining 
contributions in the sum.
One can do this by applying Gaudin's summation rules to the diagram $\mathscr{G}'$, 
obtained from $\mathscr{G}$ by stripping all of its tadpole lines. Explicitly, the result of
the Matsubara sums in $\mathscr{G}$, $I\left( \mathscr{G} \right)$, is related to 
the full sums in $\mathscr{G'}$, $I\left( \mathscr{G'} \right)$, by 
\begin{equation}\label{GaudinExtensionTadpole}
    I\left( \mathscr{G} \right)
        =
        \prod_{t \in T} 
            \left( 
              \sum_{n_t} f(-\epsilon_{n_t}) \ X^{(n_t) \mu_t} \bar{X}^{(n_t) \nu_t}
            \right)
        \times
        I\left( \mathscr{G}' \right) \ ,
\end{equation}
where $t$ indexes all the tadpoles $T$ in $\mathscr{G}$ 
and $n_t$ indexes quasiparticle energies associated to the tadpole $t$.
We illustrate the transformation from a diagram $\mathscr{G}$ to a tadpole-free 
diagram $\mathscr{G}'$ in Fig.~\ref{fig:tadpoleCaseExtension}.
With Eq.~\eqref{GaudinExtensionTadpole}, Gaudin's summation rules
are extended to any connected diagram without external lines.

\subsection{Extension to external lines}\label{subsec:extGaudin}
Let us now consider a connected diagram $\mathscr{G}$ with $2k$ external lines.
We consider, without loss of generality, that $\mathscr{G}$
does not contain any tadpole. If it does, one can use the pre-processing step and 
results in the previous subsection to evaluate their contribution.

We can incorporate the presence of external lines in Gaudin's summation rules 
to the price of some minor modifications, which we enumerate now. 
Rules $1.$ and $2.$ of Sec.~\ref{subsec:MatsubaraSum} are still valid,
but with the additional restriction that spanning trees,
as well as their complementary diagrams, must only be made of internal lines.

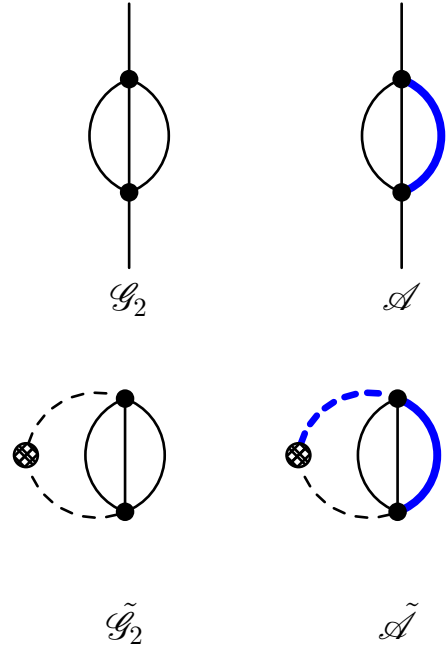
\begin{figure}[t]
  \centering
  \vspace{0.7cm}

  \parbox{100pt}{\begin{fmffile}{2ndOrderGF_large}
    \begin{fmfgraph*}(100,100)
       \fmfkeep{2ndOrderGF_large}
    \fmfbottom{i}
    \fmftop{o}
    \fmf{plain, tension=1.5, tag=1}{i,v1}
    \fmf{plain, tag=3}{v1,v2}
    \fmf{plain, tension=1.5, tag=5}{v2,o}
    \fmffreeze
    \fmf{plain, right=0.7, tag=4}{v1,v2}
    \fmffreeze
    \fmf{plain, right=0.7, tag=2}{v2,v1}
    \fmfv{d.shape=circle,d.filled=full,d.size=3thick}{v1}
    \fmfv{d.shape=circle,d.filled=full,d.size=3thick}{v2}
    \fmflabel{\Large$\mathscr{G}_2$}{i}
\end{fmfgraph*}
\end{fmffile}}
\parbox{100pt}{\begin{fmffile}{2ndOrderGF_Tree1_fmflabel}
    \begin{fmfgraph*}(100,100)
       \fmfkeep{2ndOrderGF_Tree1_fmflabel}
    \fmfbottom{i}
    \fmftop{o}
    \fmf{plain, tension=1.5, tag=1}{i,v1}
    \fmf{plain, tag=3}{v1,v2}
    \fmf{plain, tension=1.5, tag=5}{v2,o}
    \fmffreeze
    \fmf{plain, right=0.7, tag=4, fore=blue, width=3thin}{v1,v2}
    \fmffreeze
    \fmf{plain, right=0.7, tag=2}{v2,v1}
    \fmfv{d.shape=circle,d.filled=full,d.size=3thick}{v1}
    \fmfv{d.shape=circle,d.filled=full,d.size=3thick}{v2}
    \fmflabel{\Large$\mathscr{A}$}{i}
\end{fmfgraph*}
\end{fmffile}}

\vspace{0.7cm}

  \parbox{100pt}{\begin{fmffile}{2ndOrderGF_large_modified}
    \begin{fmfgraph*}(75,100)
       \fmfkeep{2ndOrderGF_large_modified}
    \fmfbottom{i}
    \fmftop{o}
    \fmfleft{l}
    \fmf{phantom, tension=1.5, tag=1}{i,v1}
    \fmf{plain, tag=3}{v1,v2}
    \fmf{phantom, tension=1.5, tag=5}{v2,o}
    \fmffreeze
    \fmf{plain, right=0.7, tag=4}{v1,v2}
    \fmffreeze
    \fmf{plain, right=0.7, tag=2}{v2,v1}
    \fmffreeze
    \fmf{dashes, left=0.5}{v1,l,v2}
    \fmfv{d.shape=circle,d.filled=full,d.size=3thick}{v1}
    \fmfv{d.shape=circle,d.filled=full,d.size=3thick}{v2}
    \fmfv{d.shape=circle,d.filled=hatched,d.size=4.5thick}{l}
    \fmflabel{\Large$\tilde{\mathscr{G}}_2$}{i}
\end{fmfgraph*}
\end{fmffile}}
\parbox{100pt}{\begin{fmffile}{2ndOrderGF_Tree1_fmflabel_modified}
    \begin{fmfgraph*}(75,100)
       \fmfkeep{2ndOrderGF_Tree1_fmflabel_modified}
    \fmfbottom{i}
    \fmftop{o}
    \fmfleft{l}
    \fmf{phantom, tension=1.5, tag=1}{i,v1}
    \fmf{plain, tag=3}{v1,v2}
    \fmf{phantom, tension=1.5, tag=5}{v2,o}
    \fmffreeze
    \fmf{plain, right=0.7, tag=4, fore=blue, width=3thin}{v1,v2}
    \fmffreeze
    \fmf{plain, right=0.7, tag=2}{v2,v1}
    \fmffreeze
    \fmf{dashes, left=0.5}{v1,l}
    \fmf{dashes, left=0.5, fore=blue, width=2.5thin}{l,v2}
    \fmfv{d.shape=circle,d.filled=full,d.size=3thick}{v1}
    \fmfv{d.shape=circle,d.filled=full,d.size=3thick}{v2}
    \fmfv{d.shape=circle,d.filled=hatched,d.size=4.5thick}{l}
    \fmflabel{\Large$\tilde{\mathscr{A}}$}{i}
\end{fmfgraph*}
\end{fmffile}}
\vspace{0.75cm}
\caption{Top left: a second order diagram $\mathscr{G}_2$ with two external lines.
Top right: one spanning tree $\mathscr{A}$ of $\mathscr{G}_2$.
Bold blue lines represent lines belonging to $\mathscr{A}$.
Bottom left: the associated diagram $\tilde{\mathscr{G}}_2$.
The hatched vertex represents $v_{\text{fix}}$.
Dashed lines represent the contractions to $v_{\text{fix}}$.
Bottom right: the modified spanning tree $\tilde{\mathscr{A}}$.
The dashed bold blue line represents $e_{\text{fix}}$.}
\label{fig:externalCaseExtension}
\end{figure}

Rule $3.a.$ remains the same.
The generalisation of rule $3.b.$ is slightly more involved.
As discussed in Sec.~\ref{subsec:FeynRulesEnergy}, the global
conservation of energy fixes one Matsubara frequency among the $2k$ (external) possible frequencies.
The line associated to the energy fixed by global conservation of energy
is denoted as $e_\text{fix}$. 
Let $\mathscr{A}$ be a spanning tree (made only of internal lines) of $\mathscr{G}$.
The modified $3.b.$ rule reads
\begin{itemize}
    \item [$3.b.'$] Consider the diagram $\tilde{\mathscr{G}}$ made of $\mathscr{G}$
    plus an additional vertex $v_{\text{fix}}$ with all external lines
    contracted to it.
    Consider the spanning tree $\tilde{\mathscr{A}}$ of $\tilde{\mathscr{G}}$
    made of $\mathscr{A}$, $v_{\text{fix}}$ and $e_\text{fix}$.
    The denominator associated to line $a$ of $\mathscr{A}$
    is the one obtained by the original rule $3.b.$ when applied to line $a$
    of $\tilde{\mathscr{A}}$ in $\tilde{\mathscr{G}}$.
\end{itemize}
Modified diagrams and spanning trees stemming from rule $3.b.'$ are illustrated
in Fig.~\ref{fig:externalCaseExtension}.

Finally, we need to consider the extension of rule $3.c$. In this case, it suffices to distinguish
between internal and external lines, so that rule $3.c.$ is replaced by
\begin{itemize}
    \item [$3.c.'$] For each \emph{internal} line $e$ in $\mathscr{G}$
    multiply by a factor $X^{(n_e) \mu_e} \bar{X}^{(n_e) \nu_e}$.
    For each \emph{external} line, multiply simply by a propagator
    as in the Feynman rules.
\end{itemize}
    If the diagram must be expressed in terms of spectroscopic amplitudes only,
    we just have to replace all external propagators by their spectral representation
    using Eqs.~\eqref{SpectralRepresentation} and~\eqref{QuadraSpFunctionEnergy}.

\subsection{Extension to disconnected diagrams}\label{subsec:connectedGaudin}
With the extensions given in~\ref{subsec:tadpoleGaudin} and~\ref{subsec:extGaudin}, Gaudin's summation rules can now be applied to any connected diagram. For completeness, we now turn to the case of disconnected diagrams. Let $\mathscr{G}$ be a disconnected diagram, with $\Gamma_1 \dots \Gamma_C$ its $C$ distinct connected parts.
Since the sub-diagrams $\Gamma_1 \dots \Gamma_C$ do not share any vertices nor lines, the Matsubara frequencies of each connected parts are necessarily independent.
The global Matsubara sum can thus be factorised, so that
\begin{equation}
    I(\mathscr{G}) = \prod^{C}_{i=1} I(\Gamma_i) \ .
\end{equation}
Since each $\Gamma_i$ is a connected diagram, we can apply Gaudin's summation rules
(or their extension if $\Gamma_i$ contains external lines or tadpoles)
to compute each term $I(\Gamma_i)$.

With the tadpole, external line and disconnected extensions just discussed,
Gaudin's summation rules introduced in Sec.~\ref{subsec:MatsubaraSum}
can be applied to any diagram.

%% If you have bibdatabase file and want bibtex to generate the
%% bibitems, please use
%%
\bibliographystyle{elsarticle-num} 
\bibliography{biblio}

%% else use the following coding to input the bibitems directly in the
%% TeX file.

%%\begin{thebibliography}{00}

%% \bibitem[Author(year)]{label}
%% For example:

%% \bibitem[Aladro et al.(2015)]{Aladro15} Aladro, R., Martín, S., Riquelme, D., et al. 2015, \aas, 579, A101

%%\end{thebibliography}
\end{document}